\documentclass[prb,aps,english,twocolumn,notitlepage,superscriptaddress]{revtex4-2}

\usepackage{graphicx}
\usepackage{dcolumn}
\usepackage{rotating}
\usepackage{lipsum}
\usepackage{xcolor}
\usepackage{bm}
\usepackage[T1]{fontenc}  
\usepackage{multirow,amsmath,amssymb}
 \usepackage{babel} 
 \usepackage{txfonts}
 \usepackage{epsfig,epsf,psfrag} 
\usepackage{graphicx} 
\usepackage{pslatex}
\usepackage{epic,eepic} 
\usepackage{color,pstcol}
\usepackage{pstricks} 
\usepackage{fancyhdr} 
\usepackage{tabularx}
\usepackage{physics}
\usepackage{url}
\usepackage{listings}
\usepackage{color} 
\usepackage{caption}
\usepackage{ulem}
\usepackage{balance}
\usepackage{subcaption}
\usepackage{float}
\usepackage[utf8]{inputenc}
\usepackage{comment}

\usepackage{hyperref}
\hypersetup{
 colorlinks=true,
 linkcolor=blue,
 anchorcolor = blue,
 citecolor = blue,
 filecolor = blue,
 urlcolor = blue
}
\usepackage{dsfont}

\graphicspath{{Images/}}

\def \be {\begin{equation}} 
\def \ee {\end{equation}}

\begin{document}

\title{$\Delta_T$ Noise as a Robust Diagnostic for Chiral, Helical and Trivial Edge Modes}
\author{Sachiraj Mishra}%
\email{sachiraj29mishra@gmail.com}

\author{Colin Benjamin}%
\email{colin.nano@gmail.com}
\affiliation{School of Physical Sciences, National Institute of Science Education and Research, HBNI, Jatni-752050, India}
\affiliation{Homi Bhabha National Institute, Training School Complex, AnushaktiNagar, Mumbai, 400094, India }

\begin{abstract}
In this article, we demonstrate that $\Delta_T$ noise provides a sensitive, practical probe for distinguishing chiral edge modes from topological helical and trivial (non-topological) helical edge transport. Measured under zero-current conditions, $\Delta_T$ noise reveals contrasts that conventional conductance measurements typically miss. Crucially, $\Delta_T$ noise requires no external energy input in the form of an applied voltage bias, yet encodes the same intrinsic information that shot noise yields in the zero-temperature, finite-bias limit, without the distorting effects of Joule heating. This absence of bias-induced heating makes $\Delta_T$ noise both more precise and more reliable than conventional shot-noise approaches. 
\end{abstract}
\maketitle
\section{Introduction} 
The observation of the quantum Hall (QH) effect, manifested experimentally as precisely quantized conductance carried by unidirectional, chiral edge channels, was a defining moment in condensed-matter physics, revealing how topology can dictate macroscopic transport properties in a two-dimensional electron gas under strong magnetic fields~\cite{PhysRevLett.45.494}. Later, materials engineered to host strong spin-orbit coupling opened a new chapter: heterostructures such as HgTe/CdTe were shown to support counterpropagating, spin-polarized edge states instead of purely chiral ones, giving rise to the quantum spin Hall (QSH) effect~\cite{doi:10.1126/science.1133734, konig2007quantum, roth2009nonlocal}. The QSH phase is especially important because it requires no external magnetic field and affords direct control over the spin texture of the propagating modes, an attribute with clear implications for spintronics and quantum information technologies.

Topological protection endows both chiral and topological helical edge modes with extraordinary resilience to nonmagnetic disorder: their existence and low-loss transport are rooted in global invariants (e.g., Chern numbers or $Z_2$ indices) rather than microscopic details. Crucially, however, not every helical-looking channel is topological. Experiments have identified “trivial” helical edge states, non-topological conductors that mimic some signatures of topological edges, for example in InAs/GaSb systems~\cite{nichele2016edge, PhysRevMaterials.4.104201}. Because conventional two-terminal or multiterminal conductance measurements are often insensitive to the underlying topological class, they cannot unambiguously distinguish chiral, topological helical, and trivial helical transport. This limitation motivates the search for alternative, topology-sensitive probes capable of clearly discriminating these regimes.

In Ref.~\cite{PhysRevB.108.115301} we showed that quantum noise, encompassing both thermal fluctuations and shot noise, provides a powerful diagnostic for distinguishing topological from trivial edge transport at finite temperatures. By computing auto- and cross-correlations in multiterminal geometries, we identified noise fingerprints that are distinct for chiral, topological helical, and trivial helical edge modes; these signatures persist when extended to finite frequencies. The analyses in Ref.~\cite{PhysRevB.108.115301}, however, were carried out under finite voltage bias, so the measured quantum noise inevitably contained contributions produced in the presence of net charge current.

Here we pursue a complementary strategy by focusing on $\Delta_T$ noise, the contribution to current fluctuations driven solely by a temperature bias when the net charge current is zero. Unlike conventional shot noise, which requires a finite bias and therefore a flowing current, $\Delta_T$ noise can be measured at zero voltage and thus isolates purely non-equilibrium thermal contributions without the background induced by charge transport. This feature is highly advantageous: removing finite-current backgrounds automatically eliminates many spurious effects and simplifies interpretation. Perhaps more importantly, avoiding an applied voltage also suppresses bias-induced dissipation (notably Joule heating) that can alter the electronic environment and obscure intrinsic quantum signatures \cite{lumbroso2018electronic, sivre2019electronic}.

{Thermal noise and $\Delta_T$ noise differ fundamentally in their origin. Thermal noise arises solely at finite temperature and vanishes at zero temperature. It is proportional to the conductance and is an equilibrium contribution. It carries essentially the same information as the conductance itself. In mesoscopic systems, however, conductance alone cannot reliably distinguish between trivial and topological edge mode transport. For example, by default, in four-terminal quantum Hall measurements, trivial and topological helical edge modes generate similar conductance and thermal noise profiles, as demonstrated in Ref.~\cite{nichele2016edge} for InAs/GaSb devices, which makes it difficult to distinguish between topological and trivial edge mode transport. This motivates the need for a probe that is sensitive to the underlying topology, and quantum shot noise, both at zero temperature and at finite temperature, along with $\Delta_T$ noise fulfills this role.}

The last few years have seen intense theoretical and experimental interest in $\Delta_T$ noise across a variety of mesoscopic platforms \cite{PhysRevLett.127.136801, PhysRevLett.125.086801, PhysRevB.105.195423, PhysRevB.107.075409, PhysRevB.107.245301, PhysRevLett.125.106801, PhysRevB.107.155405, melcer2022absent, popoff2022scattering, lumbroso2018electronic, shein2022electronic, sivre2019electronic, mishra2024andreevreflectionmediateddeltat, PhysRevResearch.7.023321, k95y-7zrb, mishra2025negativespindeltatnoise, shein2024delta, PhysRevB.108.245427}. Experimental demonstrations range from atomic-scale molecular junctions \cite{lumbroso2018electronic, shein2024delta} to mesoscopic quantum circuits \cite{shein2022electronic} and tunneling devices \cite{sivre2019electronic}. On the theoretical side, it has been shown generally that the $\Delta_T$ contribution is bounded by the thermal component in typical mesoscopic conductors \cite{PhysRevLett.127.136801}, and 
$\Delta_T$ has been proposed as a probe of fractional charge and anyonic statistics in fractional quantum Hall systems \cite{PhysRevLett.125.086801, PhysRevB.105.195423, PhysRevB.107.075409, PhysRevB.107.245301, PhysRevB.108.245427}. More recent work, some of it from our group, has extended these concepts to hybrid superconducting devices, revealing nontrivial interplay between Andreev reflection and $\Delta_T$ noise \cite{mishra2024andreevreflectionmediateddeltat, k95y-7zrb, mishra2025negativespindeltatnoise}. In particular, Ref.~\cite{k95y-7zrb} demonstrates an important practical application: $\Delta_T$ noise can be used to distinguish Yu–Shiba–Rusinov bound states from Majorana bound states in realistic device settings.

Taken together, these developments motivate our present study: by exploiting the zero-current character and reduced dissipation of $\Delta_T$ noise, we construct an experimentally accessible, robust probe capable of discriminating chiral, topological helical, and trivial helical edge transport.

A variety of alternative approaches have been proposed to separate trivial edge modes from topological ones. For example, Hanbury Brown-Twiss (HBT) correlations in presence of voltage probes have been shown to yield positive correlations for chiral edge modes, negative correlations for helical ones, and a reemergence of positive correlations for trivial helical edge states~\cite{mani2017probing}. Noise measurements in Corbino structures fabricated from InAs/Ga(In)Sb heterostructures have also been employed to separate trivial from topological modes~\cite{stevens2019noise}. Moreover, electrical gating has been demonstrated as a means to differentiate topological edge states from trivial ones via changes in Hall and longitudinal resistances~\cite{PhysRevLett.115.036803}, and some theoretical proposals exist along these lines~\cite{PhysRevB.104.045144, PhysRevB.104.195307}. While each method has its merits, $\Delta_T$ noise offers a particularly direct, uncontaminated by heating current (Joule effect), and experimentally feasible universal diagnostic. 

\begingroup

We have considered two different setups in our work. The motivation for introducing two complementary measurement setups is that different configurations probe different aspects of the $\Delta_T$ noise auto and cross correlations of edge mode transport. While a single setup can provide signatures of the underlying transport mechanism, certain distinctions between chiral, helical, and topologically trivial edge modes become more transparent in one configuration than in the other. Consequently, employing two complementary setups provides a more complete and robust framework for identifying the nature of the edge states and enhances the reliability of the proposed noise-based diagnostic method.
\endgroup

The structure of the manuscript is as follows. Section~\ref{theory} presents the general framework describing quantum noise, outlines its decomposition into thermal noise-like and shot noise-like contributions, and defines $\Delta_T$ noise at vanishing charge current for a generic QH setup. Then, we present our results for $\Delta_T$ noise in four terminal QH setup along with a comparison with results for shot noise. Section~\ref{results} presents the general theory of quantum noise, outlining both the thermal and shot noise-like components, and defines $\Delta_T$ noise at vanishing charge current for a generic QSH setup. Then, we present our results of $\Delta_T$ noise in four terminal quantum spin Hall geometries, covering topological helical and trivial helical edge modes along with a comparison with results for shot noise.  Further, in Sec.~\ref{analysis}, we provide a systematic comparison of the behavior of $\Delta_T$ noise across these edge modes. Finally, in Sec.~\ref{expt}, we conclude our paper with the experimental realization of our findings. The technical derivations are provided in Appendices \ref{App_I} to \ref{App_E}. The MATHEMATICA code to calculate the $\Delta_T$ noise can be found in Ref. \cite{github}.

\section{$\Delta_T$ noise in quantum Hall setups}
\label{theory}

\subsection{Theory}

Quantum noise correlations are defined as current--current fluctuations between different terminals in a generic mesoscopic conductor. For two distinct terminals $i$ and $j$, the time-domain correlation function reads~\cite{noise}

\begin{equation} \label{eq:1}
S_{i j}(t-t') = \frac{1}{2}\Big\langle \Delta \hat{I}_{i}(t)\Delta\hat{I}_{j}(t')+\Delta\hat{I}_{j}(t')\Delta\hat{I}_{i}(t)\Big\rangle,
\end{equation}
where the current fluctuation operator is defined as $\Delta \hat{I}_{i}(t) = \hat{I}_{i}(t) - \langle \hat{I}_{i}(t)\rangle$.  

For a multiterminal setup, the current operator in terminal $i$ can be expressed as~\cite{noise}
\begin{equation} \label{eq:2}
\begin{split}
\hat{I}_{i}(t) = \frac{2e}{h} \sum_{j,l}\sum_{r,s}\int dE\, dE' \, e^{i(E-E')t/\hbar}\\
\times \hat{a}^{\dagger}_{j r}(E)\, \mathcal{A}_{j l}^{rs}(i;E,E')\, \hat{a}_{l s}(E'),
\end{split}
\end{equation}
where the indices $j,l$ denote reservoirs (distinct from $i$), and $r,s$ label transverse edge channels. Here, $\hat{a}^{\dagger}_{j r}(E)$ creates an incoming particle with energy $E$ in terminal $j$ and channel $r$, while $\hat{a}_{l s}(E')$ annihilates a particle of energy $E'$ in terminal $l$ and channel $s$. The corresponding occupation operator for carriers incident from terminal $j$ in channel $r$ is

\begin{equation} \label{eq:3}
\hat{n}_{j r}(E) = \hat{a}_{j r}^{\dagger}(E)\hat{a}_{l s}(E') \delta_{j l} \delta_{rs} \delta(E - E').
\end{equation}

The kernel $\mathcal{A}_{j l}^{rs}(i;E,E')$ appearing in Eq.~(\ref{eq:2}) is given by~\cite{noise}
\begin{equation} \label{eq:4}
\mathcal{A}_{j l}^{rs}(i;E,E') = \delta_{i j}\delta_{i l}\delta_{rs} - \sum_{\alpha} s^{\dagger}_{i j;r\alpha}(E)\, s_{i l;s\alpha}(E'),
\end{equation}
where $s_{i j;r \alpha}(E)$ is scattering amplitude corresponding to a carrier injected from terminal $j$ in channel $\alpha$ to exit via terminal $i$ in channel $r$ with energy $E$.  

The average current at contact $i$ takes the form
\begin{equation} \label{eq:5}
\begin{split}
\langle I_{i} \rangle =& \frac{2e}{h}\sum_{j} \int dE  
\Big[N_{i}\delta_{i j}-\mathrm{Tr}(s_{i j}^{\dagger}s_{i j})\Big]f_{j}(E - eV_j),
\end{split}
\end{equation}
where $f_{j}(E - eV_j)=[1+e^{(E-eV_{j})/k_B T_{j}}]^{-1}$ is the Fermi function for reservoir $j$ and $N_{j}$ denotes the number of propagating modes in $j$, whereas $s_{i j}$ is the scattering matrix amplitude scattered from terminals $j$ to $i$. $E$ is the excitation energy relative to the Fermi energy in our setup. The prefactor of 2 accounts for spin degeneracy. Here $V_{j}, T_{j}$ accounts for the applied voltage bias and temperature in terminal $j$. Combining Eqs.~(\ref{eq:2}), (\ref{eq:5}) allows us to evaluate the current fluctuations and thus determine the resulting quantum noise characteristics.  

Applying Fourier transform to Eq.~(\ref{eq:1}) gives rise to the quantum noise spectrum~\cite{noise},
\begin{equation} \label{eq:6}
\begin{split}
2\pi \delta(\omega+\Omega) S_{i j}(\omega) =& \frac{1}{2}\Big\langle 
\Delta \hat{I}_{i}(\omega)\Delta \hat{I}_{j}(\Omega) 
\\& + \Delta \hat{I}_{j}(\Omega)\Delta \hat{I}_{i}(\omega)\Big\rangle.
\end{split}
\end{equation}

Following Ref.~\cite{noise}, the general expression for the frequency-resolved noise correlations can be written as
\begin{equation} \label{eq:7}
\begin{split}
S_{i j}(\omega) = G_0\sum_{l,p}\sum_{m,n}\int dE \; 
\mathcal{A}_{l p}^{mn}(i;E,E+\hbar \omega) \\
\times \mathcal{A}_{p l}^{nm}(j;E+\hbar \omega,E) \;
\Big[f_{l}(E- eV_l)(1-f_{p}(E+\hbar \omega - eV_p)) \\
+ (1-f_{l}(E - eV_l))f_{p}(E+\hbar \omega  - eV_p)\Big].
\end{split}
\end{equation}

where $G_0 = \frac{2e^2}{h}$. In this work, we focus on the zero-frequency limit and consider, for simplicity, a single edge mode in each terminal. Under these assumptions, Eq.~(\ref{eq:7}) reduces to
\begin{equation} \label{eq:8}
\begin{split}
& S_{i j} = G_0\sum_{l,p}\int dE \; 
\mathcal{A}_{l p}(i)\,\mathcal{A}_{p l}(j) \Big[f_{l}(E - eV_l)\\&\times(1-f_{p}(E - eV_p)) + (1-f_{l}(E - eV_l))f_{p}(E - eV_p)\Big],
\end{split}
\end{equation}
where $\mathcal{A}_{j l}(i) = I_{i}\delta_{i j}\delta_{i l} - s_{i j}^{\dagger}s_{i l}$. Eq.~(\ref{eq:8}) naturally separates into two contributions: (i) a thermal noise part $S_{i j}^{\mathrm{\text{th}}}$, and (ii) a shot-noise-like part $S_{i j}^{\mathrm{sh}}$. A detailed derivation of the aforesaid terms can be found in Refs.~\cite{PhysRevB.108.115301, PhysRevB.46.12485}. Thus,
\begin{equation*}
S_{i j} = S_{i j}^{\mathrm{\text{th}}} + S_{i j}^{\mathrm{sh}}.
\end{equation*}

Starting from Eq.~(\ref{eq:8}), one can obtain the results for shot noise at both zero and finite temperatures. The shot noise in the limit of zero temperature with a nonzero voltage bias comes from the summation terms where $l \neq p$. The general expression for zero temperature shot noise cross correlation ($S_{ij}^{\text{sh}}$) and autocorrelation ($S_{ii}^{\text{sh}}$) evaluated at finite voltage bias is given as,  
\begin{subequations}\label{eq:9}
\begin{equation} 
\begin{split}
S^{\text{sh}}_{i j} =& G_0\sum_{l \neq p}\int dE \; 
\mathcal{A}_{l p}(i)\,\mathcal{A}_{p l}(j) 
\Big[\theta_{l}(eV_l-E)\\&\times(1-\theta_{p}(eV_p-E)) + (1-\theta_{l}(eV_l-E))\\&\theta_{p}(eV_p-E)\Big],
\end{split}
\end{equation}
\begin{equation}
\begin{split}
S^{\text{sh}}_{i i} =& G_0\sum_{l \neq p}\int dE \; 
\mathcal{A}_{l p}(i)\,\mathcal{A}_{p l}(i) 
 \Big[\theta_{l}(eV_l-E)\\&\times(1-\theta_{p}(eV_p-E)) + (1-\theta_{l}(eV_l-E))\\&\theta_{p}(eV_p-E)\Big].
\end{split}
\end{equation}
\end{subequations}
where, $\theta_{l}(eV_l-E) = 1 $ for $E < eV_{l}$ and zero otherwise at zero temperature.

In this work, we are interested in the $\Delta_T$ noise, which is defined as the shot noise at zero charge current, i.e., at zero voltage bias ($V_l = 0$) but finite temperature bias. For the setups considered here, which are electron-hole symmetric, the average charge current $\langle I_{i} \rangle$ vanishes in the absence of an applied voltage bias. To derive a general expression for $\Delta_T$ noise, first, one has to find the thermal noise expression at same temperature and then subtract this thermal noise from the total quantum noise as in Eq. (\ref{eq:8}). This derivation is given in Appendix \ref{App_I}. Physically, $\Delta_T$ noise characterizes current fluctuations at zero net charge flow, thereby providing a fundamentally distinct probe from the shot noise.

{The thermal noise-like contribution for cross correlation ($i \neq j$) as shown in Eq. (\ref{eq:B8}) in Appendix \ref{App_I} is given as,}
{
\begin{equation} \label{eq:B81}
S_{i j}^{\text{th}} = -2 G_0\int dE \, \Big[T_{i j}f_{j}(E)(1-f_{j}(E))
+T_{j i}f_{i}(E)(1-f_{i}(E))\Big],
\end{equation}}

{where $T_{ij}$ gives the probability for an electron originating in terminal $l$ to reach terminal $i$ and it is equal to $|s_{ij}|^2$. Similarly, the expression for thermal noise autocorrelation ($i = j$) as shown in Eq. (\ref{eq:B6}) in Appendix \ref{App_I} is given as,}
{
\begin{equation}  \label{eq:B61}
\begin{split}
S_{i i}^{\text{th}} =& 4 G_0\int dE \, f(E)(1-f(E)) [1-T_{i i}] + 2G_0 \int dE \\& \bigg(\sum_{l \neq i}T_{il}f_l(E)(1-f_l(E))-\sum_{l \neq i}T_{il}f_i(E)(1-f_i(E))\bigg)
\end{split}
\end{equation}
where $T_{ii}$ is the reflection probability into terminal $i$.}
The $\Delta_T$ noise cross-correlation ($i \neq j$) measured at zero voltage bias and finite temperature bias is again (from Appendix \ref{App_I}, Eq. (\ref{eq:A19})),
\begin{equation} \label{eq:10}
\begin{split}
\Delta_T^{i j} =& -2G_0 \sum_{l,p}\int_{-\infty}^{\infty} dE \, (f_{l}(E)-f_a(E))(f_{p}(E)-f_b(E))\\& \times \!\Big(s_{i l}^{\dagger}s_{i p}s_{j p}^{\dagger}s_{j l}\Big),
\end{split}
\end{equation}
where $f_a(E),f_b(E)$ denote energy-dependent functions. Here, $s_{il}$ describes the amplitude for an electron incident on terminal $l$ to be scattered into terminal $i$.

Similarly, the $\Delta_T$ noise autocorrelation ($i = j$) measured at zero applied voltage bias and finite temperature bias is (from Appendix \ref{App_I}, Eq. (\ref{eq:A20})),
\begin{equation} \label{eq:11}
\begin{split}
\Delta_T^{i i} = 2G_0\int_{-\infty}^{\infty} dE \Bigg(\sum_{l} T_{i l}f_l(E)^2 
- \sum_{l, p} f_{l}(E)f_{p}(E)\,\!\Big(s_{i l}^{\dagger}s_{i p}s_{i p}^{\dagger}s_{i l}\Big)\Bigg).
\end{split}
\end{equation}

\begin{figure}[H]
\centering
\includegraphics[width=1.00\linewidth]{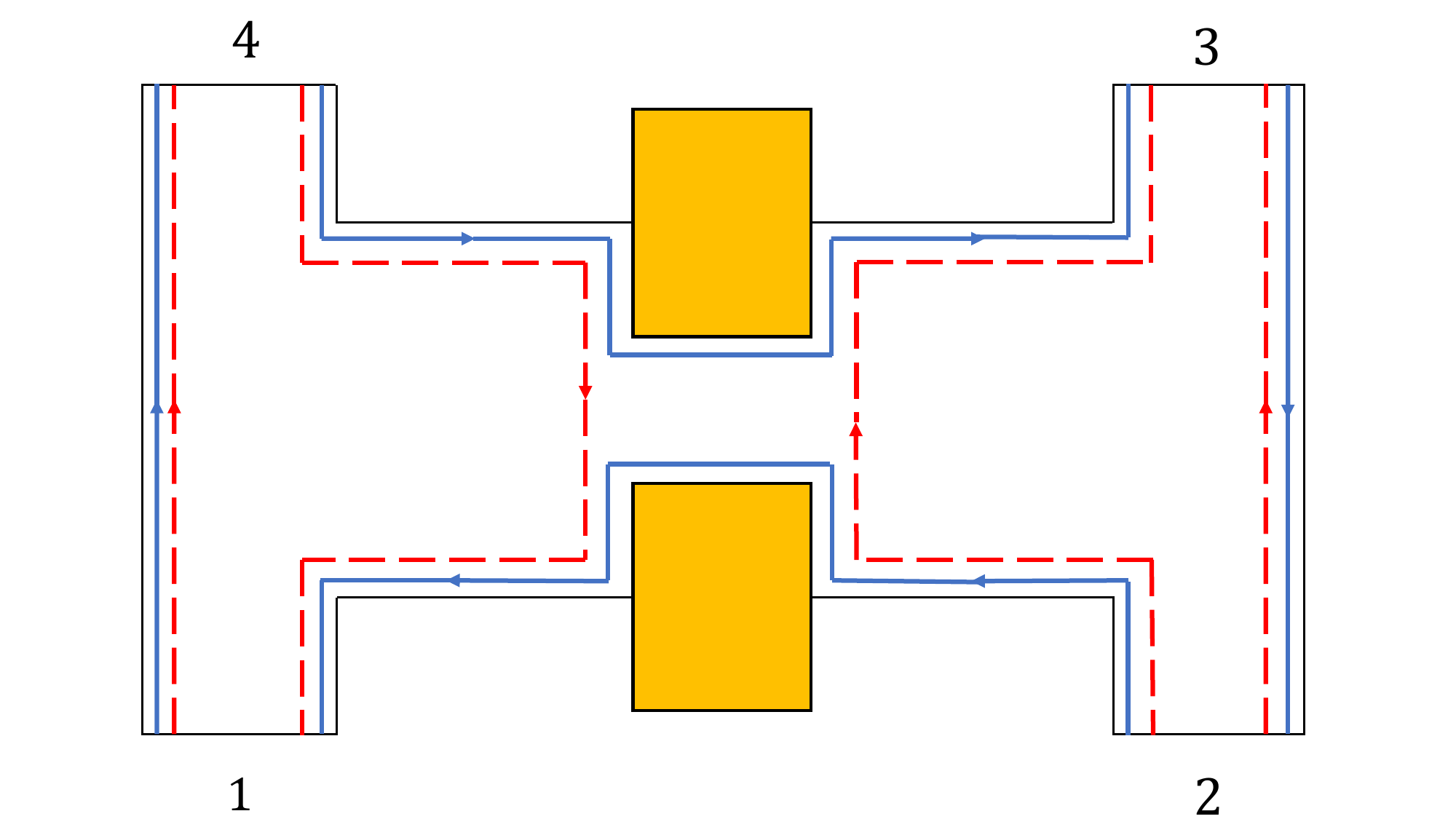}
\caption{Schematic of a four-terminal QH device featuring chiral edge channels and a narrow constriction (defined by yellow boxes). Solid lines represent edge states carrying electrons that pass through the constriction, while dashed lines indicate paths where electrons are reflected.}
\label{fig:1}
\end{figure}

We consider a four-terminal QH bar supporting chiral edge modes, as illustrated in Fig.~\ref{fig:1}. Voltage biases at zero temperature (for the calculation of shot noise) or temperature biases at zero voltage bias (for the calculation of $\Delta_T$ noise) are applied across the terminals, while the resulting quantum noise correlations are measured between any pair of terminals. 

In this work, we consider two configurations for measuring shot noise correlations:  Setup 1: $V_2 = V_3 = 0$ and $V_1 = V_4 = V$ and Setup 2: $V_2 = V_4 = 0$ and $V_1 = V_3 = V$ at zero temperature, i.e., $T_1 = T_2 = T_3 = T_4 = 0$. Similarly, we consider two configurations for $\Delta_T$ noise: Setup 1: $T_2 = T_3 = T_C$ and $T_1 = T_4 = T_H$ and Setup 2: $T_2 = T_4 = T_C$ and $T_1 = T_3 = T_H$ at zero voltage bias, i.e., $V_1 = V_2 = V_3 = V_4 = 0$. The temperature bias in either setup is $\Delta T = T_H - T_C$, with $T_H = \bar{T} + \Delta T/2$ and $T_C = \bar{T} - \Delta T/2$. Throughout this article, we focus on the linear-response regime, i.e., $\bar{T} \gg \Delta T$, the Fermi-Dirac distributions at temperature $T_H$, and zero voltage bias are $f_H(E)=[1+e^{E/k_B T_H}]^{-1}$ and at temperature $T_C$ at zero voltage bias again $f_{C}(E)=[1+e^{E/k_B T_C}]^{-1}$.

{The $s$-matrix describing the QH setup as shown in Fig. \ref{fig:1}} is {derived in Eq. (\ref{eq:C12}) in Appendix \ref{App_C1} and} therefore given as,
\begin{equation} \label{eq:12}
\begin{split}
s = \begin{pmatrix}
0 & \tau e^{i\alpha} & 0 & -i \rho e^{i\alpha}\\
0 & 0 & e^{i\chi_2} & 0\\
0 & -i \rho e^{i\alpha} & 0 & \tau e^{i\alpha}\\
e^{i\chi_1} & 0 & 0 & 0
\end{pmatrix}.
\end{split}
\end{equation}
In this geometry, the reflection probability via the constriction is defined as $\mathcal{R} = |\rho|^2$ and the transmission probability is $\mathcal{T} = |\tau|^2$. These are related to the scattering probabilities $T_{ij}=|s_{ij}|^2$ through the nonzero elements of the above $s$-matrix. For example, $T_{12}=T_{34}=\mathcal{T}$ and $T_{14}=T_{32}=\mathcal{R}$, while the remaining nonzero scattering probabilities are unity ($T_{23}=T_{41}=1$). Since the $s$-matrix is taken to be energy independent, the scattering probabilities (both $\mathcal{R}$ and $\mathcal{T}$) are energy independent. {Later, we discuss energy-dependent scattering for quantum point contact (QPC) in Sec. \ref{analysis} B.} In the following subsections, we analyze the behavior of $\Delta_T$ noise for both setups.  

\subsection{Results: $\Delta_T$ noise for chiral edge modes}

In this subsection, we analyze the cross- and auto-correlated shot noise and $\Delta_T$ noise in the four-terminal QH bar as shown in Fig. \ref{fig:1}. We begin with shot noise for setup 1, i.e., $V_2 = V_3 = 0$ and $V_1 = V_4 = V$ at zero temperature, i.e., $T_1 = T_2 = T_3 = T_4 = 0$. For this setup, the cross-correlated shot noise such as $S_{12}^{\text{sh}}$, $S_{14}^{\text{sh}}$ vanishes, while $S_{13}^{\text{sh}}$, $S_{11}^{\text{sh}}$ are finite. Using Eqs. (\ref{eq:9}) and (\ref{eq:12}), we get $S_{11}^{\text{sh}} = 2G_0 \mathcal{R} \mathcal{T} eV = -S_{13}^{\text{sh}}$. Similarly, other correlations are: $S_{24}^{\text{sh}} = S_{34}^{\text{sh}} = 0 , \quad S_{33}^{\text{sh}} =  S_{11}^{\text{sh}},$ and $S_{22}^{\text{sh}} =  S_{44}^{\text{sh}} = 0.$

For $\Delta_T$ noise, we consider setup 1: $T_2 = T_3 = T_C$ and $T_1 = T_4 = T_H$ at zero voltage bias ($V_1 = V_2 = V_3 = V_4 = 0$), the Fermi-Dirac distributions obey $f_1(E) = f_4(E) = f_H(E), f_2(E) = f_3(E) = f_C(E)$. Herein, the cross correlations $\Delta_T^{12}$ and $\Delta_T^{14}$ vanish, while $\Delta_T^{13}$ and $\Delta_T^{11}$ are finite.  Using Eq.~(\ref{eq:10}), $\Delta_T^{13}$ reduces to,  

\begin{equation} \label{eq:17}
\begin{split}
\Delta_T^{13} =& -2G_0 \sum_{l,p = 1,4}\int dE \, (f_{l}(E)-f_a(E))\\&\times(f_{p}(E)-f_b(E)) 
(s_{1 l}^{\dagger}s_{1 p}s_{3 p}^{\dagger}s_{3 l}),
\end{split}
\end{equation}
Following Refs.~\cite{PhysRevB.108.115301, PhysRevB.46.12485}, we take $f_a (E)= f_H(E)$ and $f_b (E)= f_C(E)$. Substituting the $s$-matrix from Eq.~(\ref{eq:12}), one obtains,

\begin{equation} \label{eq:18}
\Delta_T^{13} = -2G_0 \mathcal{R} \mathcal{T} \int_{-\infty}^{\infty} dE \,  (f_H(E)-f_C(E))^2.
\end{equation}
As $\mathcal{R}$ and $\mathcal{T}$ are energy independent, they are factored out of the integral. The integral, $\int_{-\infty}^{\infty} dE (f_H(E) - f_C(E))^2$, has been evaluated in Refs.~\cite{lumbroso2018electronic, mishra2024andreevreflectionmediateddeltat} in the linear-response limit, yielding,

\begin{equation}
    \int_{-\infty}^{\infty} dE \, (f_H(E) - f_C(E))^2 = k_B \bar{T} \left(\frac{\pi^2}{18} - \frac{1}{3}\right) \frac{\Delta T^2}{\bar{T}^2}.
    \label{eq:int}
\end{equation}

Thus, the closed-form result for $\Delta_T^{13}$ is,  
\begin{equation}
    \Delta_T^{13} = -2G_0 k_B \bar{T} \, \mathcal{R} \mathcal{T} \left(\frac{\pi^2}{18} - \frac{1}{3}\right) \frac{\Delta T^2}{\bar{T}^2}.
    \label{eq:15}
\end{equation}

One can also calculate the other correlations and they all vanish, both cross correlations $\Delta^{12}_T = \Delta_T^{14} = \Delta_T^{24} = 0$ and also autocorrelations $\Delta_T^{22} = \Delta_T^{44} = 0$. The auto-correlation $\Delta_T^{11}$ is finite and is $\Delta_T^{11} = -\Delta_T^{13} = \Delta_T^{33}$.

We observe that shot noise measured at zero temperature and finite voltage bias, are linear in applied voltage bias, whereas $\Delta_T$ noise are quadratic in temperature bias. {Here, we observe that the $\Delta_T$ noise and zero temperature shot noise have similar dependence on scattering probabilities. For example, $\Delta_T^{13}$ is proportional to $\mathcal{R}\mathcal{T}$ and varies quadratically with $\Delta T$. Similarly, the shot noise cross-correlation $S_{13}^{\text{sh}} $ is proportional to $G_0 \mathcal{R} \mathcal{T}$. Thus, $\Delta_T^{13}$ and $S_{13}^{\text{sh}}$ have the same sign and depend on $\mathcal{R}\mathcal{T}$, with the only difference being in their dependence on their respective biases.}

{When we calculate the thermal noise-like cross correlations $S_{13}^{\text{\text{th}}}$ and $S_{24}^{\text{\text{th}}}$, we found them to vanish. However, the cross-correlations such as $S_{12}^{\text{\text{th}}}$, $S_{14}^{\text{\text{th}}}$ and $S_{23}^{\text{\text{th}}}$ are finite. Here, $S_{14}^{\text{\text{th}}} = -2G_0(1+\mathcal{R})k_B \bar{T}$, whereas, $S_{12}^{\text{\text{th}}} = S_{23}^{\text{\text{th}}} =-2G_0 (1-\mathcal{R})k_B \bar{T}$.}

Next, when we turn to setup 2 with $V_2 = V_4 = V$ and $V_1 = V_3 = 0$ at zero temperature i.e., $T_1 = T_2 = T_3 = T_4 = 0$, we see that the cross correlations such as $S_{12}^{\text{sh}}, S_{13}^{\text{sh}}, S_{14}^{\text{sh}}, S_{23}^{\text{sh}}, S_{24}^{\text{sh}}, S_{34}^{\text{sh}}$ vanish. Similarly, the autocorrelations: ($S_{11}^{\text{sh}}, S_{22}^{\text{sh}}, S_{33}^{\text{sh}}, S_{44}^{\text{sh}}$) vanish too.

In calculating $\Delta_T$ noise in setup 2, we take $T_2 = T_4 =$ {$T_H$} and $T_1 = T_3 = T_C$ at zero voltage bias i.e., $V_1 = V_2 = V_3 = V_4 = 0$. In this case, all cross correlations, such as $ \Delta_T^{12}, \Delta_T^{13}, \Delta_T^{14}, \Delta_T^{23}, \Delta_T^{24}$, and $\Delta_T^{34}$, vanish. Furthermore, all auto-correlations, $\Delta_T^{11}, \Delta_T^{22}, \Delta_T^{33}$, and $\Delta_T^{44}$, too vanish.

\section{$\Delta_T$ noise in quantum spin Hall setups}
\label{results}

\subsection{Theory}

In analogy with QH case, the current–current correlations in a quantum spin Hall (QSH) setup can also be defined in terms of spin-resolved currents. The correlation between contacts $i$ and $j$ for spin channels $\eta$ and $\eta'$ is given by  
\begin{equation} \label{eq:48}
S_{i j}^{\eta \eta'}(t-t') = \frac{1}{2}\langle \Delta \hat{I}_{i}^{\eta}(t)\Delta\hat{I}_{j}^{\eta'}(t')+\Delta\hat{I}_{j}^{\eta'}(t')\Delta\hat{I}_{i}^{\eta}(t)\rangle,
\end{equation}
where $\Delta \hat{I}_{i}^{\eta}$ denotes the fluctuation of the spin-resolved current from its average value in contact $i$,  
\begin{equation}
\Delta \hat{I}_{i}^{\eta} = \hat{I}_{i}^{\eta} - \langle \hat{I}_{i}^{\eta}\rangle.
\end{equation}
The operator for the spin-resolved current in terminal $i$ is  
\begin{equation} \label{eq:49}
\begin{split}
\hat{I}_{i}^{\eta}(t) = \frac{e}{h}\sum_{s=1}^M \int \int dE \, dE' \, e^{i(E-E')t/\hbar} \\
\times \Big(\hat{a}_{i s}^{\eta \dagger}(E)\hat{a}_{i s}^{\eta}(E') - \hat{b}_{i s}^{\eta \dagger}(E)\hat{b}_{i s}^{\eta}(E')\Big).
\end{split}
\end{equation}
Here, $\hat{a}_{i s}^{\eta \dagger}(E)$ ($\hat{a}_{i s}^{\eta}(E)$) denotes the creation (annihilation) operator for an electron of spin $\eta$ and energy $E$ in edge channel $s$ at terminal $i$.
 Similarly, $\hat{b}_{i s}^{\eta \dagger}(E)$ and $\hat{b}_{i s}^{\eta}(E)$ correspond to the outgoing electrons. Now, the average spin-resolved current in terminal $i$ is

\begin{equation}
    \langle I_i^{\eta} \rangle = \frac{e}{h}\int dE (N_i \delta_{ij} - Tr(s_{ij}^{\eta \dagger} s_{ij}^{\eta})) f_j(E - eV_j),
\end{equation}
where $s_{ij}^{\eta} = \sum_{\rho \in \{\uparrow, \downarrow\}} s_{ij}^{\rho \eta}$, with $s_{ij}^{\rho \eta}$ representing the amplitude for an electron with spin $\eta$ in terminal $j$ to be scattered into terminal $i$ with spin $\rho$. The quantities $N_i$ and $f_j(E - eV_j)$ are defined below Eq.~(\ref{eq:5}).
Now, applying Fourier transform to Eq. (\ref{eq:48}) is gives rise to the spin-resolved quantum noise, i.e., $S_{ij}^{\eta \eta'}(\omega)$, which is given as,

\begin{equation}
\begin{split}
2\pi \delta(\omega+\Omega) S_{i j}^{\eta \eta'}(\omega) =& \frac{1}{2}\Big\langle 
\Delta \hat{I}_{i}^{\eta}(\omega)\Delta \hat{I}_{j}^{\eta}(\Omega) 
\\&+ \Delta \hat{I}_{j}^{\eta'}(\Omega)\Delta \hat{I}^{\eta}_{i}(\omega)\Big\rangle.
\end{split}
\end{equation}

Following \cite{PhysRevB.75.085328}, the general expressions for individual spin-resolved terms are given as,  
\begin{equation} \label{eq:51}
\begin{split}
&S_{i j}^{\eta \eta'} (\omega)= \frac{G_0}{2}\sum_{\rho, \rho' = \uparrow, \downarrow}\sum_{l,p}\int dE \, \mathrm{Tr}\Big[\mathcal{A}_{l p}^{\rho \rho'}(i,\eta, E, E + \hbar \omega) \\
& \times \mathcal{A}_{p l}^{\rho' \rho}(j, \eta', E + \hbar \omega, E)\Big]\Big(f_{l}(E - eV_l)\\& \times [1-f_{p}(E + \hbar \omega - eV_p)]+ [1-f_{l}(E + \hbar \omega - eV_l)]\\& \times f_{p}(E - eV_p)\Big).
\end{split}
\end{equation}

The total quantum noise correlation can then be expressed as a sum over spin contributions:  
\begin{equation} \label{eq:50}
S_{i j} (\omega) = S_{i j}^{\uparrow \uparrow} (\omega) + S_{i j}^{\uparrow \downarrow} (\omega) + S_{i j}^{\downarrow \uparrow}(\omega) + S_{i j}^{\downarrow \downarrow} (\omega).
\end{equation}

At zero frequency for single edge mode channel for each spin, Eq.~(\ref{eq:51}) reduces to,  
\begin{equation} \label{eq:52}
\begin{split}
&S_{i j}^{\eta \eta'} = \frac{G_0}{2}\sum_{\rho, \rho' = \uparrow, \downarrow}\sum_{l,p}\int dE \, \Big[\mathcal{A}_{l p}^{\rho \rho'}(i,\eta) 
\mathcal{A}_{p l}^{\rho' \rho}(j, \eta')\Big]\\&\times
\Big(f_{l}(E - eV_l)[1-f_{p}(E - eV_p)] + [1-f_{l}(E - eV_l)]\\&f_{p}(E - eV_p)\Big),
\end{split}
\end{equation}
where  
$\mathcal{A}_{l j}^{\rho \rho'}(i, \eta) = \delta_{i l}\delta_{i j}\delta_{\eta \rho}\delta_{\eta \rho'} - s_{i l}^{\eta \rho \dagger}s_{i j}^{\eta \rho'}$,
and $\eta, \eta' \in \{\uparrow, \downarrow \}$ denote the spin of the electrons. Eq.~(\ref{eq:52}) contains both thermal and shot noise-like terms. Just like QH case, here also we restrict ourselves to (1) the shot noise measured at zero temperature and finite voltage bias, (2) the $\Delta_T$ noise measured at finite temperature bias and zero voltage bias. The evaluation of shot noise at zero temperature and finite voltage bias is similar to what is done in QH case and the expression for shot noise cross correlation ($S_{ij}^{\text{sh}, ss'}$) and autocorrelation ($S_{ii}^{\text{sh}, ss'}$) for single edge mode for single spin are given as

\begin{subequations}
\begin{equation} \label{eq:53}
\begin{split}
S_{i j}^{\text{sh}, \eta \eta'} =& \frac{G_0}{2}\sum_{\rho, \rho' = \uparrow, \downarrow}\sum_{l\neq p}\int dE \quad [\mathcal{A}_{l p}^{\rho \rho'}(i, \eta)\mathcal{A}_{p l}^{\rho' \rho}(j, \eta')]\\&\times\Big[\theta_{l}(eV_l-E)(1-\theta_{p}(eV_p-E)) \\&+ (1-\theta_{l}(eV_l-E))\theta_{p}(eV_p-E)\Big]
\end{split}
\end{equation}
\begin{equation}
\begin{split}
S_{i i}^{\text{sh},\eta \eta'} =& \frac{G_0}{2}\sum_{\rho, \rho' = \uparrow, \downarrow}\sum_{l\neq p}\int dE \quad [A_{l p}^{\rho \rho'}(i, \eta)
A_{p l}^{\rho' \rho}(i, \eta')]\\&\times\Big[\theta_{l}(eV_l-E)(1-\theta_{p}(eV_p-E)) \\&+ (1-\theta_{l}(eV_l-E))\theta_{p}(eV_p-E)\Big],
\end{split}
\end{equation}
\end{subequations}
where $\theta_{l}( eV_l-E) = 1$ for $ e V_{l}<E$ and zero other wise. Once we calculate the spin-polarized shot noise, one can directly calculate the total shot noise cross correlation and auto correlation and they are given as,

\begin{equation}
\begin{split}
    S_{i j}^{\text{sh}} = &S_{i j}^{\uparrow \uparrow, \text{sh}} + S_{i j}^{\uparrow \downarrow, \text{sh}} + S_{i j}^{\downarrow \uparrow, \text{sh}} + S_{i j}^{\downarrow \downarrow, \text{sh}}, \\ S_{i i}^{\text{sh}}  = & S_{i i}^{\uparrow \uparrow, \text{sh}} + S_{i i}^{\uparrow \downarrow, \text{sh}} + S_{i i}^{\downarrow \uparrow, \text{sh}} + S_{i i}^{\downarrow \downarrow, \text{sh}}.
    \end{split}
\end{equation}

{The thermal noise correlation $S_{ij}^{\text{\text{th}}}$ or $S_{ii}^{\text{\text{th}}}$ is a sum of all the spin-spin correlations and is given as}

{
\begin{equation}
\begin{split}
    S_{i j}^{\text{\text{th}}} = &S_{i j}^{\uparrow \uparrow, \text{\text{th}}} + S_{i j}^{\uparrow \downarrow, \text{\text{th}}} + S_{i j}^{\downarrow \uparrow, \text{\text{th}}} + S_{i j}^{\downarrow \downarrow, \text{\text{th}}}, \\ S_{i i}^{\text{\text{th}}}  = & S_{i i}^{\uparrow \uparrow, \text{\text{th}}} + S_{i i}^{\uparrow \downarrow, \text{\text{th}}} + S_{i i}^{\downarrow \uparrow, \text{\text{th}}} + S_{i i}^{\downarrow \downarrow, \text{\text{th}}}.
    \end{split}
\end{equation}}

{The thermal noise cross-correlation $S_{ij}^{\text{\text{th}}}$ as derived in Eq. (\ref{eq:C5}) in Appendix \ref{App_Qn} is} 

{
\begin{equation} \label{eq:C51}
\begin{split}
S_{i j}^{\eta \eta', th} =& -G_0\int dE \Big[T_{i j}^{\eta \eta'} f_{j}(E)(1-f_{j}(E)) \\&
+ T_{j i}^{\eta \eta'} f_{i}(E)(1-f_{i}(E))\Big],
\end{split}
\end{equation}}
where $T_{ij}^{\eta\eta'}$ is the scattering probability for an electron incident from terminal $j$ with spin $\eta'$ to scatter into terminal $i$ with spin $\eta$.
{Similarly, the thermal noise auto-correlation $S_{ii}^{\text{\text{th}}}$ as derived in Eq. (\ref{eq:C9}) in Appendix \ref{App_Qn} is}
{
\begin{equation} \label{eq:C91}
\begin{split}
S_{i i}^{\eta \eta', th} =& 2 G_0\int dE \, f_{i}(E)(1-f_{i}(E)) \,
\big(\delta_{\eta \eta'} - T_{i i}^{\eta \eta'}\big)  \\& + G_0 \int dE \bigg(\sum_{l \neq i}T_{il}^{\eta \eta'}f_{l}(E) (1-f_{l}(E)) \\& - \sum_{l \neq i}T_{il}^{\eta \eta'}f_{i}(E) (1-f_{i}(E) \delta_{\eta \eta'})\bigg).
\end{split}
\end{equation}}

Similarly, cross-correlated $\Delta_T$ noise measured at zero voltage bias ($V_l = 0$) and finite temperature bias is given by,  
\begin{equation} \label{eq:57}
\Delta_T^{i j} = \Delta_T^{i j, \uparrow \uparrow} + \Delta_T^{i j, \uparrow \downarrow} + \Delta_T^{i j, \downarrow \uparrow} + \Delta_T^{i j, \downarrow \downarrow},
\end{equation}
with spin-resolved components (see, Appendix \ref{App_Qn}, Eq. (\ref{eq:C15})),
\begin{equation} \label{eq:58}
\begin{split}
\Delta_T^{i j, \eta \eta'} =& -G_0\int dE \sum_{l, p}\sum_{\rho, \rho' = \uparrow, \downarrow}(f_{l}(E)-f_a(E))\\&\times(f_{p}(E)-f_b(E))(s_{i l}^{\eta \rho \dagger}s_{i p}^{\eta \rho'}s_{j p}^{\eta' \rho' \dagger}s_{j l}^{\eta' \rho}).
\end{split}
\end{equation}
The derivation in Eq.~(\ref{eq:58}) has been found out by first getting the thermal noise-like component considering all the Fermi-Dirac distributions to be same and then later subtracting it from the total quantum noise correlation from Eq. (\ref{eq:52}).  
Following a similar procedure, the auto-correlated $\Delta_T$ noise reads, 
\begin{equation} \label{eq:63}
\Delta_T^{i i} = \Delta_T^{i i, \uparrow \uparrow} + \Delta_T^{i i, \uparrow \downarrow} + \Delta_T^{i i, \downarrow \uparrow} + \Delta_T^{i i, \downarrow \downarrow},
\end{equation}
with individual spin contributions (see, Appendix \ref{App_Qn}, Eq. (\ref{eq:C16})),
\begin{equation} \label{eq:64}
\begin{split}
&\Delta_T^{i i,\eta \eta'} = G_0\int dE \Bigg(\sum_{l \neq i} \sum_{\rho \neq \eta'}T_{il}^{\eta \rho}f_i(E)^2 \delta_{\eta \eta'} + \sum_{\rho} T_{ii}^{\eta \rho} f_i(E)^2 \delta_{\eta \eta'}\\&  \sum_{l \neq i}T_{il}^{\eta \eta'}f_l(E)^2 - \sum_{l, p} \sum_{\rho \rho'} f_l(E) f_p(E) (s_{il}^{\eta \rho \dagger} s_{ip}^{\eta \rho'} s_{ip}^{\eta' \rho' \dagger} s_{i l}^{\eta' \rho})\Bigg).
\end{split}
\end{equation}
Here, $T_{il}^{\eta \eta'}$ denotes the probability for an electron with spin $\eta'$ in terminal $l$ to be transmitted into terminal $i$ with spin $\eta$.

\begin{figure}[H]
\centering
\includegraphics[width=1.00\linewidth]{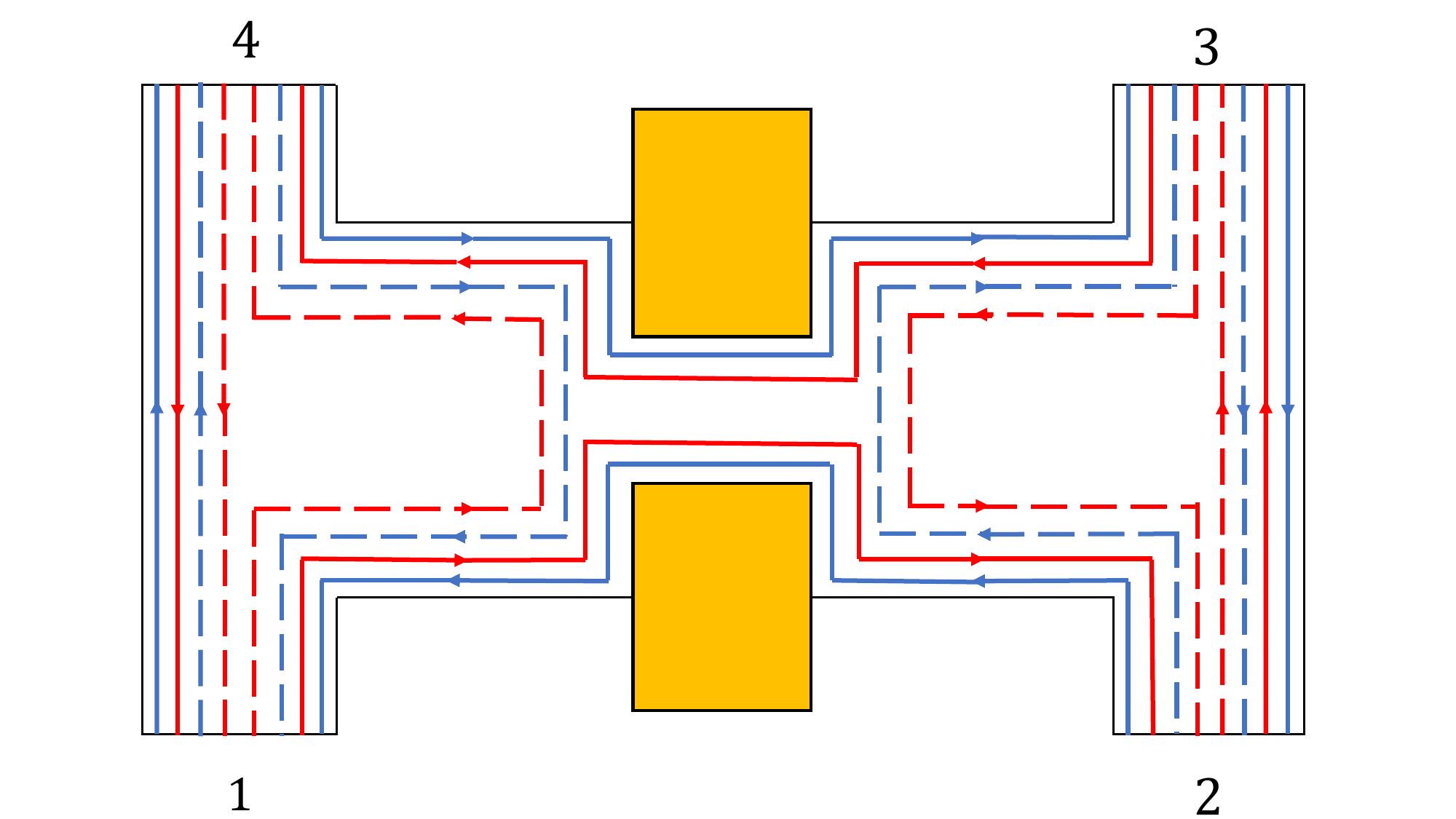}
\caption{Schematic of a four-terminal QSH device with a constriction, illustrating topological helical edge modes. Solid (dashed) blue (red) lines represent spin-up (spin-down) electrons, each can either transmit (reflect) through the constriction. The corresponding scattering matrix for this setup is provided in Eq.~(\ref{eq:12}).}
 \label{fig:2}
\end{figure}

\subsection{Results: $\Delta_T$ noise with topological helical edge modes}

We consider a four-terminal system QSH sample, exhibiting topological helical edge mode transport, as illustrated in Fig.~\ref{fig:2}. Voltage biases at zero temperature (for shot noise calculation) and temperature biases at zero voltage bias (for $\Delta_T$ noise calculation) can be applied at any contact, and the resulting current-current correlations between pairs of contacts can be measured. Similar to the QH case, we consider two setups for the shot noise measured at zero temperature and finite voltage bias, i.e., setup 1: $V_2 = V_3 = 0$, $V_1 = V_4 = V$ and setup 2: $V_2 = V_4 = 0$, $V_1 = V_3 = V$ at $T_1 = T_2 = T_3 = T_4 = 0$. Similarly, for the $\Delta_T$ noise, we again consider two setups, i.e., setup 1: $T_2 = T_3 = T_C$, $T_1 = T_4 = T_H$ and setup 2: $T_2 = T_4 = T_C$, $T_1 = T_3 = T_H$
both at $V_1 = V_2 = V_3 = V_4 = 0$.

The $s$-matrix for the setup shown in Fig.~\ref{fig:2} is derived in Eq.~(\ref{eq:C66}) of Appendix~\ref{App_C2} and is given by
\begin{equation} \label{eq:66}
\resizebox{0.89\hsize}{!}{$
s=
\begin{pmatrix}
0 & 0 & \tau e^{i\alpha} & 0 & 0 & 0 & -i\rho e^{i\alpha} & 0\\
0 & 0 & 0 & 0 & 0 & 0 & 0 & e^{-i\chi_1}\\
0 & 0 & 0 & 0 & e^{i\chi_2} & 0 & 0 & 0\\
0 & \tau e^{-i\alpha} & 0 & 0 & 0 & -i\rho e^{-i\alpha} & 0 & 0\\
0 & 0 & -i\rho e^{i\alpha} & 0 & 0 & 0 & \tau e^{i\alpha} & 0\\
0 & 0 & 0 & e^{-i\chi_2} & 0 & 0 & 0 & 0\\
e^{i\chi_1} & 0 & 0 & 0 & 0 & 0 & 0 & 0\\
0 & -i\rho e^{-i\alpha} & 0 & 0 & 0 & \tau e^{-i\alpha} & 0 & 0
\end{pmatrix}$.}
\end{equation}
In this geometry, the reflection and transmission probabilities of the quantum point contact are defined as $\mathcal{R}=|\rho|^2$ and $\mathcal{T}=|\tau|^2$, respectively. These are related to the spin-resolved scattering probabilities $T_{ij}^{\eta\eta'}=|s_{ij}^{\eta\eta'}|^2$ through the nonzero elements of the above $s$-matrix. For example, $T_{13}^{\uparrow\uparrow}=T_{31}^{\downarrow\downarrow}=T_{57}^{\downarrow\downarrow}=T_{75}^{\uparrow\uparrow}=\mathcal{R}$, whereas $T_{15}^{\uparrow\downarrow}=T_{37}^{\downarrow\uparrow}=T_{51}^{\downarrow\uparrow}=T_{73}^{\uparrow\downarrow}=\mathcal{T}$. The remaining nonzero scattering probabilities correspond to perfect transmission and are equal to unity. Since the $s$-matrix is assumed to be energy independent, both the QPC scattering probabilities $\mathcal{R}$ and $\mathcal{T}$, and hence the corresponding spin-resolved transmission probabilities $T_{ij}^{\eta\eta'}$, are also energy independent.

For each setup, the shot noise correlations and $\Delta_T$ noise correlations are computed using the scattering matrix as in Eq.~(\ref{eq:66}). For shot noise in setup 1: $V_2 = V_3 = 0, V_1 = V_4 = V$ at $T_1 = T_2 = T_3 = T_4 = 0$, we find that most of the cross correlations such as $S_{12}^{\text{sh}}$, $S_{14}^{\text{sh}}$, $S_{23}^{\text{sh}}$, and $S_{34}^{\text{sh}}$ {vanish}, whereas $S_{24}^{\text{sh}}$ is finite due to the helical nature of edge modes. Here, the crosscorrelation such as $S_{13}^{\text{sh}}$ is finite and the only contribution is from $S_{13}^{\uparrow \uparrow, \text{sh}}$, given as $S_{13}^{ \text{sh}} = - G_0 \mathcal{R} \mathcal{T} eV$, we see that it is exactly half of the QH case. This quantitative difference could distinguish the helical and chiral edge mode transport.

Similarly, the correlations such as $S_{24}^{\uparrow \uparrow, \text{sh}}, S_{24}^{\downarrow \uparrow, \text{sh}}, S_{24}^{\uparrow \downarrow, \text{sh}}$ are zero, but $S_{24}^{\downarrow \downarrow, \text{sh}}$ is finite {and given by} $S_{24}^{\downarrow \downarrow, \text{sh}} = -G_0 \mathcal{R} \mathcal{T} eV$. Therefore, the total cross {correlation} $S_{24}^{ \text{sh}}$ is
$S_{24}^{ \text{sh}} = - G_0 \mathcal{R} \mathcal{T} eV$. This result distinguishes the chiral edge mode and helical edge mode transport qualitatively. Similarly, all the autocorrelations are exactly same and they are $S_{11}^{ \text{sh}} = S_{22}^{ \text{sh}} = S_{33}^{ \text{sh}} = S_{44}^{ \text{sh}} = G_0 \mathcal{R} \mathcal{T} eV.$ Here, the finite $S_{22}^{\text{sh}}$ and $S_{44}^{\text{sh}}$ distinguish that of QH case with chiral edge modes the chiral and helical edge mode transport.

\begin{figure}[H]
\centering
\includegraphics[width=1.00\linewidth]{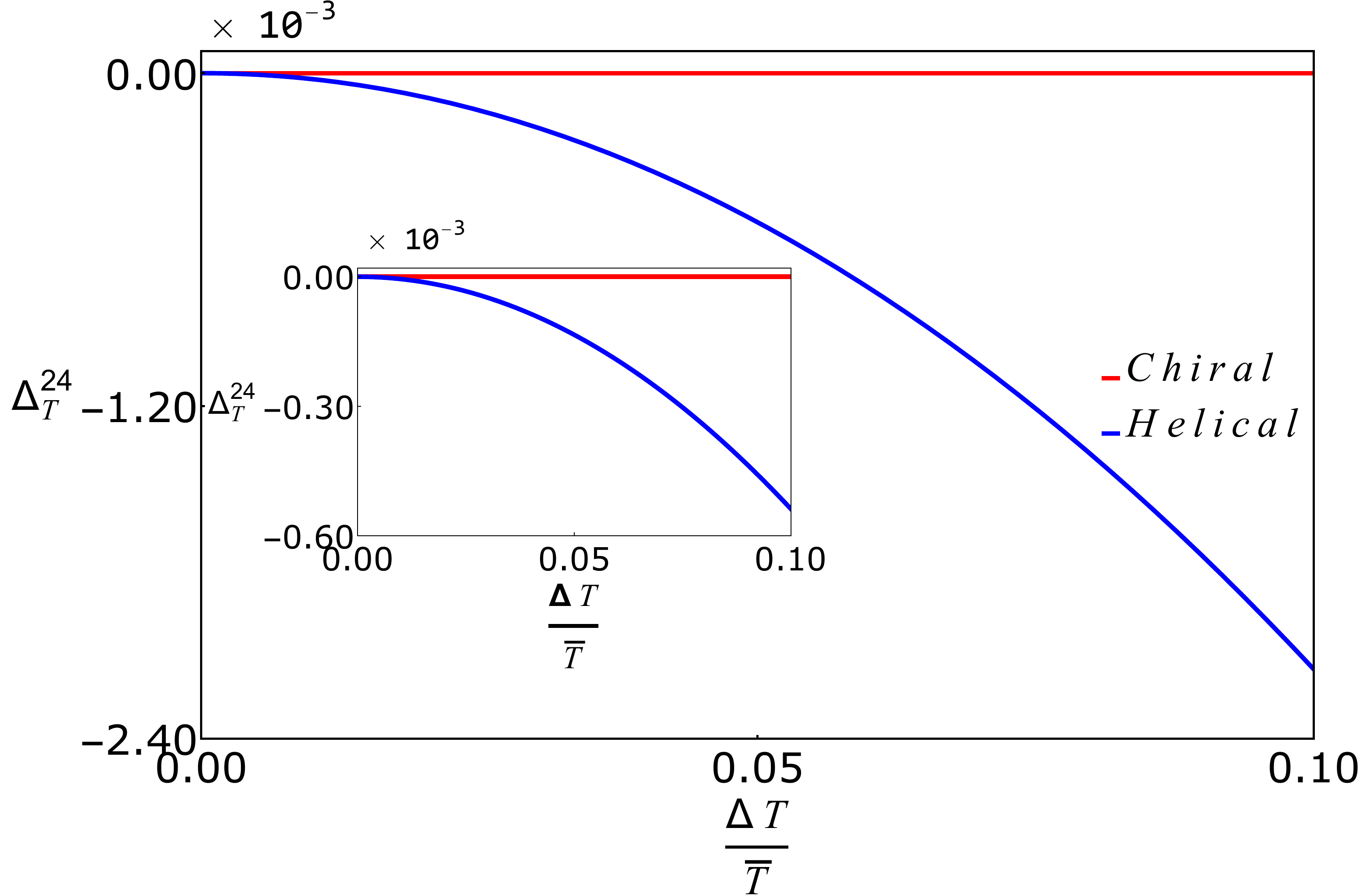}
\caption{$\Delta_T^{24}$ in units of $G_0 k_B \bar{T}$ in setup 2 ($\Delta_T^{24}$ of setup 1 in inset) at $\bar{T} = 10 K$ for $\mathcal{R} = 0.5$.}
 \label{fig:3}
\end{figure}

To calculate $\Delta_T$ noise at zero voltage bias and finite temperature bias, we consider two different setups, i.e., setup 1: $T_2 = T_3 = T_C$, $T_1 = T_4 = T_H$ and setup 2: $T_2 = T_4 = T_C$, $T_1 = T_3 = T_H$ at $V_1 = V_2 = V_3 = V_4 = 0$. Similar to the QH case, most cross correlations in setup 1, such as $\Delta_T^{12}$, $\Delta_T^{14}$, $\Delta_T^{23}$, and $\Delta_T^{34}$, vanish, while $\Delta_T^{24}$ is finite due to the helical nature of the edge modes. Specifically, the components of $\Delta_T^{24}$ satisfy $\Delta_T^{24;\uparrow\uparrow} = 0$, $\Delta_T^{24;\uparrow\downarrow} = \Delta_T^{24;\downarrow\uparrow} = 0$, and only $\Delta_T^{24;\downarrow\downarrow}$ is nonzero and contributes to $\Delta_T^{24}$ overally, with

\begin{equation}
\begin{split}
\Delta_T^{24} = 
\Delta_T^{24;\downarrow\downarrow} =& -G_0 \int dE \sum_{l,p} \sum_{\rho,\rho' = \uparrow,\downarrow} (f_l - f_a)(f_p - f_b) 
\\& \quad  \big(s_{2l}^{\downarrow \rho \dagger} s_{2p}^{\downarrow \rho' \dagger} s_{4p}^{\downarrow \rho'} s_{4l}^{\downarrow \rho}\big),
\end{split}
\end{equation}

where only terms with $l, p = 1,3$ survive. $f_a$ and $f_b$ are the energy-dependent functions, which we take {to} be $f_C$ for the calculation. Similarly, one can also derive $\Delta_T^{13}$ easily and it is exactly same as $\Delta_T^{24}$. Using the relevant scattering matrix elements from Eq.~(\ref{eq:66}), we obtain: $\Delta_T^{13} = \Delta_T^{24} =  -G_0 k_B \bar{T} \mathcal{R} \mathcal{T} \left(\frac{\pi^2}{18} - \frac{1}{3}\right) \frac{\Delta T^2}{\bar{T}^2}$. Therefore, $\Delta_T^{24}$ successfully distinguishes the chiral and helical edge mode transport, see the inset of Fig. \ref{fig:3}. 

{In QSH setup 1 as shown in Fig. \ref{fig:2}, with contacts at temperatures $T_1 = T_4 = T_H$ and $T_2 = T_3 = T_C$ and voltages at contacts $V_1 = V_2 = V_3 = V_4 = 0$, the $\Delta_T$ noise cross correlation between terminals 2 and 4 is $\Delta_T^{24} = -G_0 k_B \bar{T}\,\mathcal{R}\mathcal{T}\left(\frac{\pi^2}{18} - \frac{1}{3}\right)\frac{\Delta T^2}{\bar{T}^2}$. $\Delta_T^{24}$ is proportional to $\mathcal{R}\mathcal{T}$ and varies quadratically with $\Delta T$.}

{Instead, if we apply voltage bias $V_1 = V_4 = V$ with $V_2 = V_3 = 0$ at zero temperature $T_1 = T_2 = T_3 = T_4 = 0$, the shot noise cross-correlation $S_{24}^{\text{sh}} = -G_0\,\mathcal{R}\mathcal{T}\,eV$. Thus, $\Delta_T^{24}$ and $S_{24}^{\text{sh}}$ have the same sign and dependence on $\mathcal{R}\mathcal{T}$, with the difference being in their dependence on the respective biases: $\Delta_T^{24}$ varies quadratically with $(\Delta T)$, whereas $S_{24}^{\text{sh}}$ varies linearly with $eV$.}

{In the chiral QH setup of Fig. \ref{fig:1}, both $\Delta_T^{24}$ and $S_{24}^{\text{sh}}$ vanish. Therefore, $\Delta_T$ noise and shot noise at zero temperatures are either both finite or they both vanish.}

{A practical advantage of $\Delta_T$ noise is that it is measured under zero charge current and is therefore free from Joule heating, which can distort zero temperature shot noise, which is measured at net finite charge current. This makes $\Delta_T$ noise a more reliable experimental probe than zero temperature shot noise.}

For the autocorrelations, $\Delta_T^{11}$, $\Delta_T^{22}$, $\Delta_T^{33}$, {$\Delta_T^{44}$} remain finite, as in the QH system and they are exactly same, i.e., $\Delta_T^{11} = \Delta_T^{22} = \Delta_T^{33} = \Delta_T^{44} = G_0 k_B \bar{T} \mathcal{R} \mathcal{T} \left(\frac{\pi^2}{18} - \frac{1}{3}\right) \frac{\Delta T^2}{\bar{T}^2}.$

\begin{figure}[H]
\centering
\includegraphics[width=1.00\linewidth]{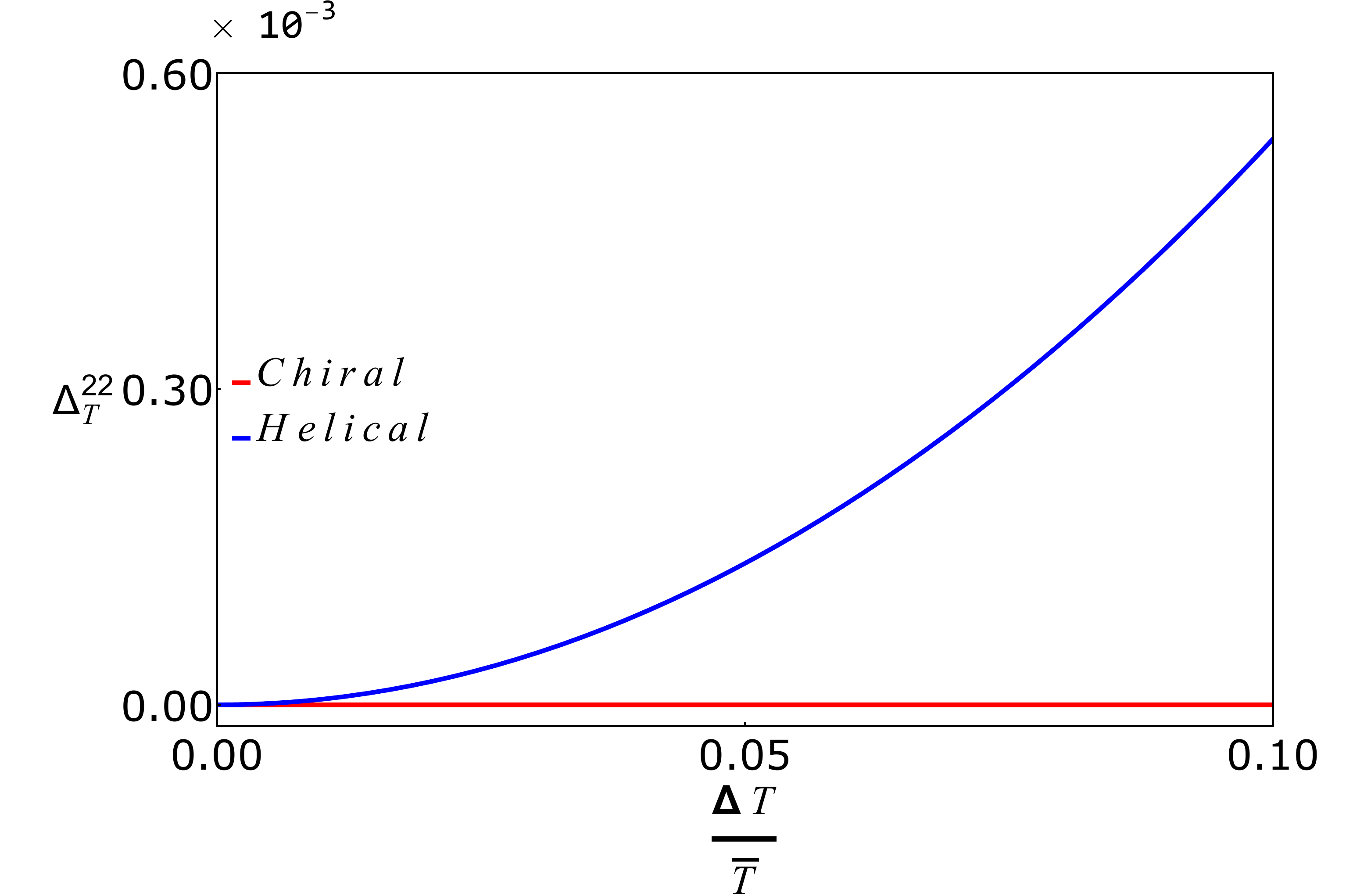}
\caption{$\Delta_T^{22}$ in units of $G_0 k_B \bar{T}$ in {setup 1} at $\bar{T} = 10 K$ for $\mathcal{R} = 0.5$.}
 \label{fig:4}
\end{figure}

Unlike the chiral case, $\Delta_T^{22}$ and $\Delta_T^{44}$ {remain finite}, providing a clear distinction between helical and chiral edge transport, see Fig. \ref{fig:4}. Thus, in setup 1, the finite values of $\Delta_T^{24}$, $\Delta_T^{22}$, and $\Delta_T^{44}$ distinguish helical modes from chiral ones.

{On the other hand, we observe that the thermal noise crosscorrelations $S_{12}^{\text{\text{th}}}$, $S_{14}^{\text{\text{th}}}$ and $S_{23}^{\text{\text{th}}}$ are exactly are exactly the same as those in the QH case with chiral edge modes, which shows the inability of the thermal noise-like contribution in distinguishing between Helical and Chiral edge modes.}

{$\Delta_T^{24}$, $\Delta_T^{22}$ are much better probes than the thermal noise-like components. As shown in Figs. \ref{fig:3} and \ref{fig:4}, we see that for chiral edge mode transport, they are exactly zero, whereas for helical edge mode transport, they are finite, which gives a clear distinction between chiral and helical edge mode transport.}

However, in setup 2 at zero temperature and finite voltage bias $V_1 = V_3 = V$, $V_2 = V_4 = 0$ at $T_1 = T_2 = T_3 = T_4 =0$, we find that $S_{24}^{\text{sh}}$ is finite and is equal to $-4G_0  \mathcal{R} \mathcal{T} eV$, while this is zero in QH case. This distinguishes the chiral and helical edge transport. In this setup, all the autocorrelations vanish.

Similarly, to calculate $\Delta_T$ noise in setup 2: $T_2$ {$= T_4 = T_H$}, {$T_1 = T_3 = T_C$} at $V_1 = V_2 = V_3 = V_4 = 0$, only $\Delta_T^{24}$ is finite and is equal to $
\Delta_T^{24} = -4G_0 k_B \bar{T} \mathcal{R} \mathcal{T} \left(\frac{\pi^2}{18} - \frac{1}{3}\right) \frac{\Delta T^2}{\bar{T}^2}$, while all other cross correlations vanish. In the corresponding QH system, $\Delta_T^{24}$ is zero, so this cross correlation provides a clear signature of helical transport, see Fig. \ref{fig:3}. All the autocorrelations vanish in this setup, consistent with the QH case.

\subsection{Results: $\Delta_T$ noise with trivial helical edge mode }

\begin{figure}[H]
\centering
\includegraphics[width=1.00\linewidth]{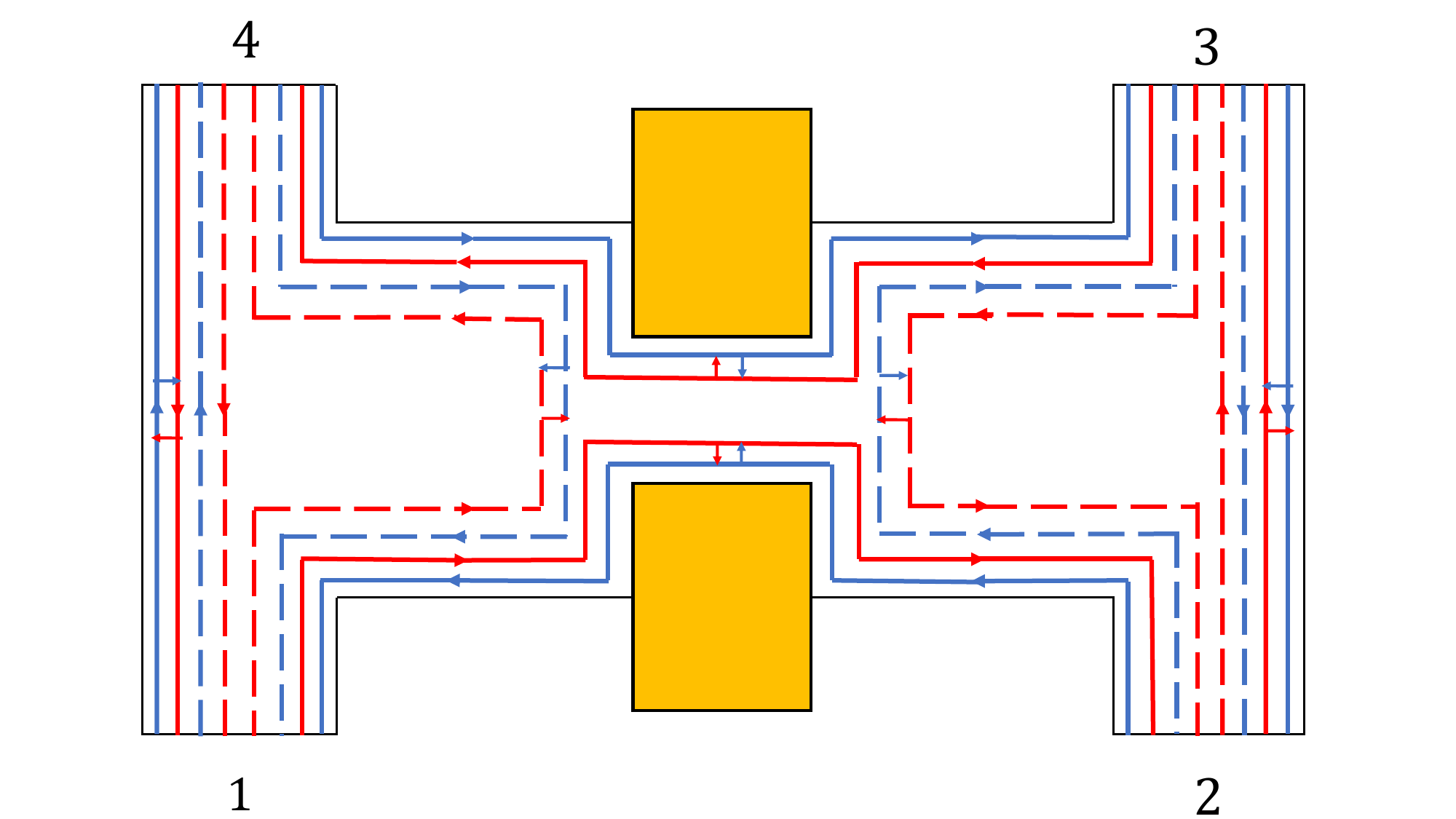}
\caption{Schematic of a four-terminal QSH device with a constriction, illustrating trivial helical edge modes. Solid (dashed) blue (red) lines represent spin-up (spin-down) electrons, each can either transmit (reflect) through the constriction. {The spin-flip scattering processes are depicted by the small arrows between edge modes.} The corresponding scattering matrix for this setup is provided in Eq.~(\ref{eq:108}).}
 \label{fig:5}
\end{figure}

Figure~\ref{fig:5} shows QSH setup with trivial or nontopological helical edge modes. In this setup, an electron with spin up can be reflected into a down-spin state, while maintaining the spin-momentum locking feature of the system~\cite{nichele2016edge}.
Consequently, the electron acquires a finite probability $P$ of reversing both its propagation direction and spin orientation due to intraedge scattering. {The spin-flip scattering processes are highlighted by small arrows in Fig. \ref{fig:5}. Since such processes are absent in the topological case, the corresponding arrows do not appear in Fig. \ref{fig:2}.} In this work, we demonstrate how the topological helical edge modes can be distinguished from trivial ones through $\Delta_T$ noise measurements, including both cross-correlations and autocorrelations. Besides the scattering at the constriction, spin-flip processes occur, leading to modifications of the scattering matrix. The $s$-matrix describing this setup, incorporating spin-flip probability $P$, is given by:
\begin{widetext}

{
\begin{equation} \label{eq:108}
s=
\begin{pmatrix}
0 & -i\sqrt{ P} e^{i\xi} & \tau e^{i \alpha}x & 0 & 0 & 0 & -i\rho e^{i \alpha}x & 0\\
-i\sqrt{ P} & 0 & 0 & 0 & 0 & 0 & 0 & x \, e^{-i\chi_1}\\
0 & 0 & 0 & -i\sqrt{ P} & x \, e^{i\chi_2} & 0 & 0 & 0\\
0 & \tau e^{-i \alpha}x & -i\sqrt{ P}e^{-i\xi} & 0 & 0 & i\rho e^{-i \alpha}x & 0 & 0\\
0 & 0 & -i\rho e^{i \alpha}x & 0 & 0 & -i\sqrt{ P}e^{i\xi} & \tau e^{i \alpha}x & 0\\
0 & 0 & 0 & x \, e^{-i\chi_2} & -i\sqrt{ P} & 0 & 0 & 0\\
x \, e^{i\chi_1} & 0 & 0 & 0 & 0 & 0 & 0 & -i\sqrt{ P}\\
0 & i\rho e^{-i \alpha}x & 0 & 0 & 0 & \tau e^{-i \alpha}x & -i\sqrt{ P}e^{-i\xi} & 0
\end{pmatrix},
\end{equation}}
\end{widetext}

where, $x = \sqrt{1 -  P}, \xi = \alpha - \theta$ with $\theta = \text{Tan}^{-1}\left(\frac{\rho}{\tau}\right)$. Note that setting $ P = 0$ reproduces the scattering matrix for topological helical edge modes given in Eq.~(\ref{eq:66}). {This $s$-matrix has been explained in Appendix \ref{App_C2}, see below Eq. (\ref{eq:C108}).}

Using the scattering matrix in Eq.~(\ref{eq:108}), we analyze shot noise and $\Delta_T$ noise in the trivial QSH setup.

Just like QSH setup with topological helical edge modes, we consider two setups for the calculation of shot noise under conditions of zero temperature with a finite applied voltage, i.e., setup 1: $V_1 = V_4 = V, V_2 = V_3 = 0$ and setup 2: $V_1 = V_3 = V, V_2 = V_4 = 0$ at zero temperature, i.e., $T_1 = T_2 = T_3 = T_4 = 0$. Similarly, for the calculation of $\Delta_T$ noise at zero voltage bias and finite temperature bias, we consider two setups, i.e., setup 1: $T_1 = T_4  = T, T_2 = T_3 = T_C$ and setup 2: $T_1 = T_3 = T, T_2 = T_4 = T_C$ at $V_1 = V_2 = V_3 = V_4 = 0$. 

In the first setup, we find that the shot noise cross correlation $S_{14}^{\text{sh}}$, which is zero for $P = 0$ (topological), is now finite, when helical edge modes are trivial because of spin-flip scattering. The same spin correlations such as $S_{14}^{\uparrow \uparrow, \text{sh}}$ and $S_{14}^{\downarrow \downarrow, \text{sh}}$ are zero, where as opposite spin correlations $S_{14}^{\uparrow \downarrow, \text{sh}}$ and $S_{14}^{\downarrow \uparrow, \text{sh}}$ are finite and they are $S^{\downarrow\uparrow;\text{sh}}_{ 14} = -4G_0  P(1 -  P) eV,
S^{\uparrow\downarrow; \text{sh}}_{14} = \mathcal{R} \, S^{\downarrow\uparrow;\text{sh}}_{ 14}$. Therefore, $S_{14}^{\text{sh}}$ is $-4G_0  P(1 -  P)(1+\mathcal{R})eV$.
This clearly distinguishes topological helical edge mode from trivial one. Similarly, $S_{12}^{\text{sh}}$ and $S_{34}^{\text{sh}}$ are finite and can clearly distinguish between two distinct helical edge modes and are given as $S_{12}^{\text{sh}} = S_{34}^{\text{sh}} = -G_0  P(1 -  P) (1-\mathcal{R}) eV.$ The autocorrelations are all finite, but cannot distinguish these helical edge modes qualitatively and they are $S_{11}^{\text{sh}} = S_{22}^{\text{sh}} = S_{33}^{\text{sh}} = S_{44}^{\text{sh}} = G_0  ( P(1 -  P) \mathcal{T} + \mathcal{R} \mathcal{T} (1 -  P)^2) eV $. However, we observe that there is a quantitative difference between the autocorrelations measured in topological helical edge mode and trivial helical edge mode, which can clearly distinguish both of them.

\begin{figure}[H]
\centering
\includegraphics[width=1.00\linewidth]{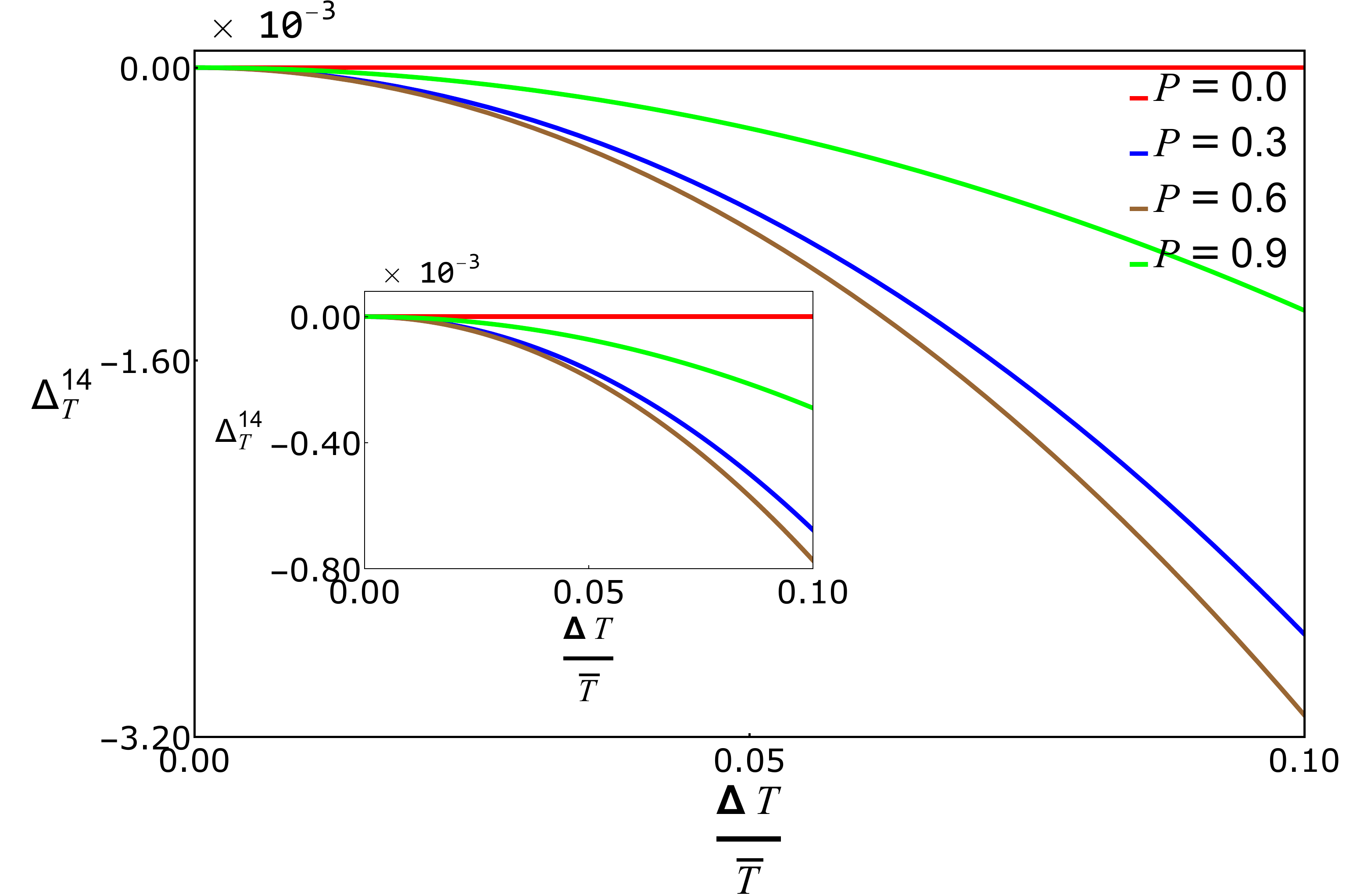}
\caption{$\Delta_T^{14}$ in units of $G_0 k_B \bar{T}$ in setup 1 ($\Delta_T^{14}$ of setup 2 in inset) at $\bar{T} = 10 K$ for $\mathcal{R} = 0.5$.}
 \label{fig:6}
\end{figure}

To calculate $\Delta_T$ noise, we consider setup 1, i.e., $T_1 = T_4 = T_H$, $T_2 = T_3 = T_C$ at zero voltage bias ($V_1 = V_2 = V_3 = V_4 = 0$), where we find that $\Delta_T^{14}$, which is zero for topological helical edge modes, becomes finite due to spin-flip scattering. Specifically, the same-spin correlations $\Delta_T^{14;\uparrow\uparrow}$ and $\Delta_T^{14;\downarrow\downarrow}$ vanish, while the opposite-spin correlations $\Delta_T^{14;\uparrow\downarrow}$ and $\Delta_T^{14;\downarrow\uparrow}$ are nonzero, with:$
\Delta_T^{14;\downarrow\uparrow} = -4G_0  P(1 -  P) \int_{-\infty}^{\infty} dE \, (f_H(E) - f_C(E))^2,
\Delta_T^{14;\uparrow\downarrow} = \mathcal{R} \, \Delta_T^{14;\downarrow\uparrow}.$ Now using Eq. (\ref{eq:int}),
we obtain the total $\Delta_T$ noise cross correlation $\Delta_T^{14} = -4G_0  P(1 -  P) (1+\mathcal{R}) k_B \bar{T} \left(\frac{\pi^2}{18} - \frac{1}{3}\right) \frac{\Delta T^2}{\bar{T}^2}$. This result shows that for topological helical edge mode ($ P = 0$), $\Delta_T^{14}$ is zero, where as for trivial helical edge mode ($ P \neq 0$), $\Delta_T^{14}$ is finite, see Fig. \ref{fig:6}. Here, we find that both $S_{14}^{\text{sh}}$ and $\Delta_T^{14}$ essentially carry the same physical information, but their dependence on external driving parameters is fundamentally different. The quantity $S_{14}^{\text{sh}}$ is governed by the applied voltage bias $eV$, whereas $\Delta_T^{14}$ is determined by the temperature bias $\Delta T$. Importantly, $S_{14}^{\text{sh}}$ is evaluated under conditions of finite charge current. The presence of this current inevitably leads to power dissipation through the Joule effect, which in turn can introduce unwanted disturbance in the setup and thereby distort the expected shot noise signal. Such distortions complicate the interpretation of the data and limit the precision with which shot noise can be used as a probe of underlying transport properties. In contrast, $\Delta_T^{14}$ is evaluated at zero average charge current. In this situation, there is no possibility of Joule heating, which means the measured signal directly reflects the fluctuations associated with the temperature gradient alone. This makes $\Delta_T^{14}$ a much cleaner and more accurate probe compared to $S_{14}^{\text{sh}}$. Therefore, $\Delta_T^{14}$ provides a distinct experimental advantage, offering higher precision and eliminating extrinsic effects that otherwise obscure the fundamental noise characteristics in voltage-driven measurements.

\begin{figure}[H]
\centering
\includegraphics[width=1.00\linewidth]{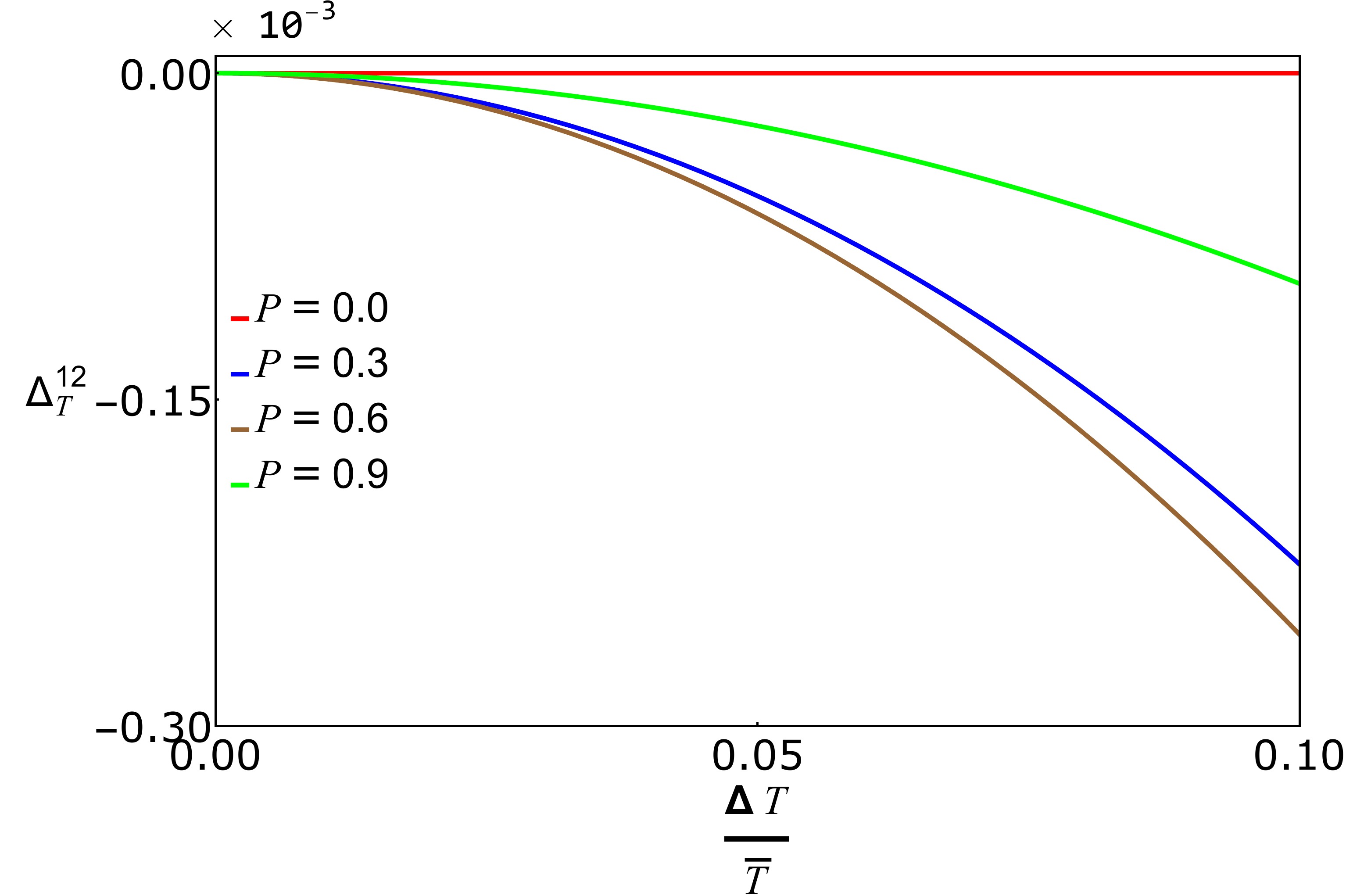}
\caption{$\Delta_T^{12}$ (= $\Delta_T^{34}$) in units of $G_0 k_B \bar{T}$ in setup 1 (we get same results in setup 2) at $\bar{T} = 10 K$ for $\mathcal{R} = 0.5$.}
 \label{fig:7}
\end{figure}

Other $\Delta_T$ noise cross correlations, such as $\Delta_T^{12}$ and $\Delta_T^{34}$, which were zero for topological edge modes, are now finite for trivial helical edge mode and are given as $\Delta_T^{12} = \Delta_T^{34} = -G_0  P(1 -  P) (1 - \mathcal{R}) k_B \bar{T} \left(\frac{\pi^2}{18} - \frac{1}{3}\right) \frac{\Delta T^2}{\bar{T}^2}$, thus distinguishing topological from trivial helical edge modes, see Fig. \ref{fig:7}.
However, the autocorrelations $\Delta_T^{11}, \Delta_T^{22}, \Delta_T^{33}, \Delta_T^{44}$ remain finite and, by themselves, do not differentiate between trivial and topological edge modes.

For setup 2, calculating shot noise, at zero temperature, we find the cross correlations $S_{12}^{\text{sh}}$ and $S_{34}^{\text{sh}}$ are same and are given as $        S_{12}^{\text{sh}} = S_{34}^{\text{sh}}= -G_0 (1 -  P) P(1 - \mathcal{R}) eV$, while
$S_{14}^{\text{sh}} = S_{23}^{\text{sh}}= -G_0 (1 -  P) P(1 + \mathcal{R}) eV$. These cross correlations help distinguish topological from trivial helical as for the former case these vanish. Similarly, the autocorrelations also can clearly distinguish topological from trivial edge mode and are all same and finite and are given as, $S_{11}^{\text{sh}} = S_{22}^{\text{sh}} = S_{33}^{\text{sh}}  =  S_{44}^{\text{sh}}=2G_0  P(1 -  P) eV.$ {On the other hand, we find that the thermal noise cross correlations, such as, $S_{12}^{\text{\text{th}}}$, $S_{14}^{\text{\text{th}}}$ and $S_{23}^{\text{\text{th}}}$ 
differ from those in the helical edge–mode case by an overall factor of 
$(1-P)$. A similar scaling also appears in the charge conductance. Although this yields a quantitative distinction in the analytical expressions for thermal noise, experimentally this is difficult to use as thermal noise alone cannot reliably distinguish between helical and trivial helical edge modes, precisely because of the same limitation encountered in charge-conductance measurements.}

\begin{figure}[H]
\centering
\includegraphics[width=1.00\linewidth]{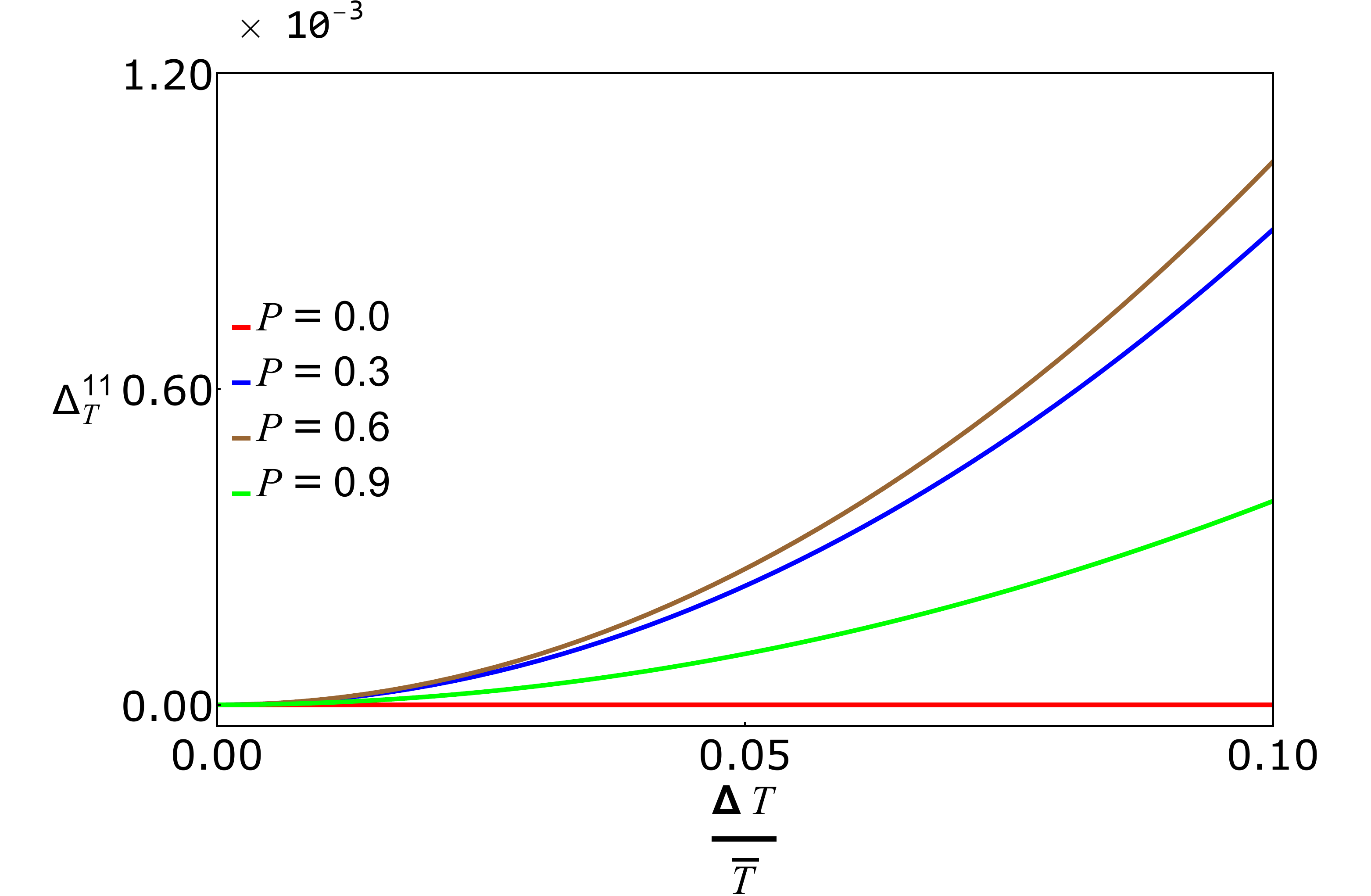}
\caption{$\Delta_T^{11} (= \Delta_T^{22} = \Delta_T^{33} = \Delta_T^{44})$ in units of $G_0 k_B \bar{T}$ in setup 2 at $\bar{T} = 10 K$ for $\mathcal{R} = 0.5$.}
 \label{fig:8}
\end{figure}

{On the other hand, the $\Delta_T$ noise cross correlations such as $\Delta_T^{14}, \Delta_T^{12}, \Delta_T^{34}$ in setup 1 and the $\Delta_T$ noise autcorrelations such as $\Delta_T^{11}, \Delta_T^{22}, \Delta_T^{33}, \Delta_T^{44}$ in setup 2 are much reliable probes as for helical, they are zero, whereas for trivial edge modes, they are finite, see Figs. \ref{fig:6}-\ref{fig:8}, making $\Delta_T$ noise an effective probe to distinguish helical from trivial edge mode transport.}

In setup 2, for $\Delta_T$ noise, we find $\Delta_T^{12} = \Delta_T^{34} = -G_0 (1 -  P)  P(1 - \mathcal{R}) k_B \bar{T} \left(\frac{\pi^2}{18} - \frac{1}{3}\right) \frac{\Delta T^2}{\bar{T}^2}$, while $\Delta_T^{14} = \Delta_T^{23} = \frac{1 + \mathcal{R}}{1 - \mathcal{R}} \Delta_T^{12}$. 
These $\Delta_T$ noise cross correlations demonstrate a clear distinction between these helical modes, see Figs. \ref{fig:6} and \ref{fig:7}. $\Delta_T$ noise autocorrelations are also finite, further differentiating the trivial from topological case, see Fig. \ref{fig:8} and are given as $\Delta_T^{11} = \Delta_T^{22} = \Delta_T^{33} = \Delta_T^{44} =2 G_0  P(1 -  P) k_B \bar{T} \left(\frac{\pi^2}{18} - \frac{1}{3}\right) \frac{\Delta T^2}{\bar{T}^2}$.

\section{Analysis}
\label{analysis}

\subsection{Zero frequency $\Delta_T$ noise}

In this section, we investigate the zero-frequency $\Delta_T$ noise cross correlations and autocorrelations to differentiate between chiral, two helical edge modes, which are topological and trivial.

We begin with setup 1, where the biasing condition is $T_1 = T_4 = T_H$ and $T_2 = T_3 = T_C$, at zero voltage bias ($V_1 = V_2 = V_3 = V_4 = 0$). A summary of the $\Delta_T$ correlations is presented in Table~\ref{Table1}. For chiral edge state transport, $\Delta_T^{12}, \Delta_T^{14}, \Delta _T^{23}$ vanish. In the case of helical edge state, these correlations also vanish. However, when trivial helical edge modes are considered, an important difference emerges: the cross correlations $\Delta_T^{12}$, $\Delta_T^{34}$, and $\Delta_T^{14}$ vary quadratically with the temperature bias $\Delta T$, whereas they vanish for topological helical modes, see Figs.~\ref{fig:6} and \ref{fig:7}. This distinction originates from finite opposite-spin contributions in case of all the cross correlations mentioned above. For example, $\Delta_T^{14;\uparrow \downarrow}$ and $\Delta_T^{14;\downarrow \uparrow}$, which arise due to spin-flip scattering with probability $P$ and are proportional to $P(1-P)$. For $P=0$, these contributions vanish identically. The relevant $s$-matrix elements, including $s_{11}^{\uparrow \downarrow}$, $s_{11}^{\downarrow \uparrow}$, $s_{14}^{\uparrow \uparrow}$, $s_{14}^{\downarrow \downarrow}$, $s_{41}^{\uparrow \uparrow}$, $s_{41}^{\downarrow \downarrow}$, $s_{44}^{\uparrow \downarrow}$, and $s_{44}^{\downarrow \uparrow}$, see Eq.~(\ref{eq:108}), play a central role in generating these correlations. Thus, $\Delta_T$ cross correlations are capable of distinguishing these two different helical edge modes, although they are ineffective in differentiating chiral from topological helical modes.  

A more revealing correlation is $\Delta_T^{24}$. For chiral edge modes, $\Delta_T^{24}$ vanishes, whereas for helical edge states, it acquires a finite value that varies quadratically with $\Delta T$, thereby providing a clear distinction between the two transport regimes, see Fig.~\ref{fig:3}. This behavior stems from the helical nature of transport: the same-spin correlation $\Delta_T^{24;\downarrow \downarrow}$ is finite, while $\Delta_T^{24;\uparrow \uparrow}$ is zero. The finite contribution arises due to the presence of the constriction and is proportional to $\mathcal{R}\mathcal{T}$, vanishing when $\mathcal{T}=1$. The relevant $s$-matrix elements in this case: $s_{21}^{\downarrow \downarrow}$ and $s_{23}^{\downarrow \downarrow}$, are finite, see Eq.~(\ref{eq:66}). For trivial helical edge modes, $\Delta_T^{24}$ also varies quadratically with $\Delta T$, resembling the topological case but with a different overall magnitude. In this case, $\Delta_T^{24;\uparrow \uparrow}$ vanishes, while $\Delta_T^{24;\downarrow \downarrow}$ is finite and proportional to $(1-P)^2 \mathcal{R}\mathcal{T}$. Importantly, the opposite spin-spin correlations $\Delta_T^{24;\uparrow \downarrow}$ and $\Delta_T^{24;\downarrow \uparrow}$ vanish. Thus, although $\Delta_T^{24}$ cannot separate these two helical modes, it can clearly separate topological helical from chiral edge modes.  

Turning now to the autocorrelations, we find that $\Delta_T^{22}$ ($\Delta_T^{44}$) vanishes for chiral edge modes but grows quadratically with $\Delta T$ for topological helical modes, providing a robust distinction between chiral and topological transport, see Fig.~\ref{fig:4}. In this case, $\Delta_T^{22;\uparrow \uparrow}$ vanishes, while $\Delta_T^{22;\downarrow \downarrow}$ is finite and proportional to $\mathcal{R}\mathcal{T}$. This feature is unique to topological helical edge transport. For trivial helical modes, however, $\Delta_T^{22}$ and $\Delta_T^{44}$ also vary quadratically with $\Delta T$, and are proportional to $(1-P)^2 \mathcal{R}\mathcal{T}$, thereby mimicking the topological case with only a different magnitude. Consequently, autocorrelations cannot distinguish topological from trivial helical modes, though they remain effective in separating chiral from topological modes.  

Overall, in setup 1, $\Delta_T$ cross correlations can distinguish both chiral from topological helical modes and topological from trivial helical modes, whereas $\Delta_T$ autocorrelations are limited to distinguishing only chiral from topological helical modes.  

We now turn to setup 2, where the biasing condition is $T_1 = T_3 = T_H$ and $T_2 = T_4 = T_C$, again at zero voltage bias. A complete summary is provided in Table~\ref{Table1}. Similar to setup 1, the cross correlations $\Delta_T^{12}$, $\Delta_T^{34}$, $\Delta_T^{14}$, and $\Delta_T^{23}$ vanish for both chiral as well as helical edge states and thus cannot differentiate between them. For trivial helical states, however, these cross correlations vary quadratically with $\Delta T$ and are proportional to $P(1-P)$, thereby offering a clear distinction between two different helical transport, see Fig.~\ref{fig:7}. As in setup 1, $\Delta_T^{24}$ again distinguishes chiral from topological helical states, as it is proportional to $\mathcal{R}\mathcal{T}$ in the topological case but vanishes in the chiral case. For trivial helical modes, $\Delta_T^{24}$ also varies quadratically, mirroring the topological case and thus failing to distinguish between them.  

Unlike setup 1, however, the autocorrelations in setup 2, namely $\Delta_T^{11}$, $\Delta_T^{22}$, $\Delta_T^{33}$, and $\Delta_T^{44}$, vanish for both chiral as well as topological helical states. This makes them ineffective in separating these two cases. Nevertheless, they acquire finite quadratic dependence for trivial helical modes, being proportional to $P(1-P)$, see Fig.~\ref{fig:8}. Thus, in setup 2, the autocorrelations provide a clean way to distinguish trivial helical modes from topological ones, even though they cannot separate chiral from topological modes.  

{The motivation for introducing two different setups is that each configuration highlights the distinct behavior of correlations, allowing us to distinguish among the edge modes more effectively. As summarized in Table 1, Setup 2 reveals distinct behavior for helical and trivial edge modes through $\Delta_T^{23}$, which is not possible in Setup 1. Likewise, in Setup 2 the $\Delta_T$ noise autocorrelation terms $\Delta_T^{11}, \Delta_T^{22}, \Delta_T^{33}, \Delta_T^{44}$ differentiate helical from trivial transport, whereas these do not provide such distinction in Setup 1. Conversely, $\Delta_T^{22}$ and $\Delta_T^{44}$ in Setup 1 distinguish chiral from helical edge modes, a feature that does not appear in Setup 2. Thus, using two complementary setups ensures that when one configuration fails to differentiate between specific edge modes, the other succeeds. The same reasoning applies to the corresponding shot-noise measurements.}

 In summary, in setup 1, the $\Delta_T$ cross-correlations can differentiate among all three edge modes, whereas the $\Delta_T$ auto-correlations distinguish only chiral from topological helical states. In contrast, for setup 2 $\Delta_T$ autocorrelations can distinguish topological from trivial helical modes, complementing the results from setup 1.

\begin{widetext}

\begin{table}[H]
\centering
\caption{Behavior of $\Delta_T$ noise, including both cross- and autocorrelations, as a function of temperature bias for configurations 1 and 2.}
\begin{tabular}{ll|cccc|cccc|}
\cline{3-10}
           &           & \multicolumn{4}{c|}{Setup 1}                                                    & \multicolumn{4}{c|}{Setup 2}                                                    \\ \hline
\multicolumn{2}{|c|}{Edge modes} & \multicolumn{1}{c|}{$\Delta_T^{12} = \Delta_T^{34}$} & \multicolumn{1}{c|}{$\Delta_T^{14}$} & \multicolumn{1}{c|}{$\Delta_T^{24}$} & \multicolumn{1}{c|}{$\Delta_T^{22} = \Delta_T^{44}$} & \multicolumn{1}{c|}{$\Delta_T^{12} = \Delta_T^{34}$} & \multicolumn{1}{c|}{$\Delta_T^{14} = \Delta_T^{23}$} & \multicolumn{1}{c|}{$\Delta_T^{24} $} & \multicolumn{1}{c|}{$\Delta_T^{11} = \Delta_T^{22} = \Delta_T^{33} = \Delta_T^{44}$} \\ \hline
\multicolumn{2}{|c|}{Chiral} & \multicolumn{1}{c|}{Zero} & \multicolumn{1}{c|}{Zero} & \multicolumn{1}{c|}{Zero} & \multicolumn{1}{c|}{Zero} & \multicolumn{1}{c|}{Zero} & \multicolumn{1}{c|}{Zero} & \multicolumn{1}{c|}{Zero} & \multicolumn{1}{c|}{Zero} \\ \hline
\multicolumn{2}{|c|}{Helical (Topological)} & \multicolumn{1}{c|}{Zero} & \multicolumn{1}{c|}{Zero} & \multicolumn{1}{c|}{Quadratic} & \multicolumn{1}{c|}{Quadratic}  & \multicolumn{1}{c|}{Zero} & \multicolumn{1}{c|}{Zero} & \multicolumn{1}{c|}{Quadratic} & \multicolumn{1}{c|}{Zero} \\ \hline
\multicolumn{2}{|c|}{Helical (Trivial)} & \multicolumn{1}{c|}{Quadratic} & \multicolumn{1}{c|}{Quadratic} & \multicolumn{1}{c|}{Quadratic} & \multicolumn{1}{c|}{Quadratic} & \multicolumn{1}{c|}{Quadratic} & \multicolumn{1}{c|}{Quadratic} & \multicolumn{1}{c|}{Quadratic} & \multicolumn{1}{c|}{Quadratic} \\ \hline
\end{tabular}
\label{Table1}
\end{table}

\end{widetext}

\subsection{Sensitivity to energy-dependent scattering via QPC}

Previously, we have considered energy-independent scattering, however QPC in reality gives rise to energy-dependent scattering. When scattering is energy independent, electron–hole symmetry is preserved and the thermovoltage required to maintain zero charge current in a $\Delta_T$ noise measurement vanishes. In contrast, when the scattering depends on energy, electron–hole symmetry is broken and a finite thermovoltage is then required to ensure zero net charge current.

This means that the condition $V_1 = V_2 = V_3 = V_4 = 0$ can no longer be used when computing $\Delta_T$ noise. A finite thermovoltage must be applied such that net charge current vanishes, and the resulting expression for $\Delta_T$ noise, will be modified accordingly. 

 In setup 1, with contact temperatures: $T_{1}=T_{4}=T_{H}$ and $T_{2}=T_{3}=T_{C}$ with $T_H>T_C$ for both QH and QSH cases, to evaluate $\Delta_{T}$ noise, we consider voltage configuration: $V_{1}=V_{4}=V$ and $V_{2}=V_{3}=0$. In setup 2, with contact temperatures: $T_{1}=T_{3}=T_{H}$ and $T_{2}=T_{4}=T_{C}$, with $T_H > T_C$, and voltage biases are $V_1 = V_3 = V$ and $V_2 = V_4 = 0$. Fermi function in contacts (1,4) and in contacts (2,3) for setup-1 are 
$f_{H}(E)= [1+e^{(E-eV)/k_{B}T_{H}}]^{-1}$ and  
$f_{C}(E)= [1+e^{E/k_{B}T_{C}}]^{-1}$. Using the parametrization:
$T_{H}=\bar{T}+\frac{\Delta T}{2}$ and $T_{C}=\bar{T}-\frac{\Delta T}{2}$, the temperature in contacts 1(4) is then $T_1 = T_H = \bar{T} + \tau_{1(4)}$, where $\tau_{1(4)}$ is the temperature bias in contacts 1 (4), whereas $\tau_{2(3)}$ is the temperature bias in contacts 2 (3). We consider the temperature biases as  
$\tau_{1}=\tau_{4}=\frac{\Delta T}{2}$ with $\tau_{2}=\tau_{3}=-\frac{\Delta T}{2}$ for setup 1.
Similarly, for setup 2, we consider $\tau_{1(3)} = \frac{\Delta T}{2}$ and $\tau_{2(4)}=-\frac{\Delta T}{2}$.

Herein, the QPC is modeled as an energy-dependent scatterer. We first compute the thermovoltage $V_{\text{\text{th}}}$ for which the net charge current vanishes and using it, calculate the $\Delta_T$ noise. 

\subsubsection{$\Delta_T$ noise in QH setup}

To calculate $\Delta_T$ noise in the QH setup as shown in Fig. \ref{fig:1},
using $s$-matrix from Eq. (\ref{eq:12}), but now with modified energy-dependent scattering amplitudes:

\begin{equation} \label{eq:121}
s = \begin{pmatrix}
0 & \sqrt{\mathcal{T}} e^{i\alpha} & 0 & -i \sqrt{1-\mathcal{T}} e^{i\alpha}\\
0 & 0 & e^{i\chi_2} & 0\\
0 & -i \sqrt{1-\mathcal{T}} e^{i\alpha} & 0 & \sqrt{\mathcal{T}} e^{i\alpha}\\
e^{i\chi_1} & 0 & 0 & 0
\end{pmatrix},
\end{equation}

where the QPC transmission (see, Refs. \cite{PhysRevB.41.7906, PhysRevB.57.1838}) is now defined as,
\begin{equation}\label{eq:122}
\mathcal{T}=\frac{1}{1+e^{-2\pi(E-E_{1})/\hbar\omega_{x}}},\,\,\text{with}\,\,
E_{1}=\frac{\hbar\omega_{y}}{2}.
\end{equation}

$\mathcal{T}$ can be derived from the QPC Hamiltonian \cite{PhysRevB.41.7906, PhysRevB.57.1838} with the confinement potential,

\begin{equation}\label{eq:123}
   U(x, y) = \frac{1}{2}m^* \omega_y^2 y^2 - \frac{1}{2}m^* \omega_x^2 x^2 + V_0,
\end{equation}
where $m^*$ is the effective mass of the electron, and $V_0$ is the potential offset, see Refs. \cite{PhysRevB.41.7906, PhysRevB.57.1838} for the derivation of Eq. (\ref{eq:122}). The energy levels in the QPC are quantized as $E_n = V_0 + (n + 1/2)\hbar \omega_y$. We consider the Fermi energy to be zero.

Using the Landauer--Büttiker expression for charge current, Eq.~(\ref{eq:5}), the linear-response charge current in terminal $i$ in the QH setup is,  
\begin{equation}\label{eq:124}
\langle I_{i}\rangle=\sum_{j}(G_{ij}V_{j}+L_{ij}\tau_{j}),
\end{equation}
where $V_{j}$ is the voltage bias and $\tau_{j}$ is the temperature bias at contact $j$. The coefficients $G_{ij}$ and $L_{ij}$ are
\begin{equation}\label{eq:125}
\begin{split}
 G_{ij}&=\frac{2e^{2}}{h}\int_{-\infty}^{\infty}dE\,(\delta_{ij}-T_{ij})\left(\frac{-\partial f}{\partial E}\right),\\
L_{ij}&=\frac{2e}{h\bar{T}}\int_{-\infty}^{\infty}dE\,(\delta_{ij}-T_{ij})\,E\left(\frac{-\partial f}{\partial E}\right),
\end{split}
\end{equation}
where $T_{ij}$ is the transmission probability from terminal $j$ to terminal $i$, as shown in Fig.~\ref{fig:1}, and $f(E)=\bigl(1+e^{E/k_{B}\bar{T}}\bigr)^{-1}$ is the equilibrium Fermi function.

We will first evaluate $\Delta_T^{22}$ in QH setup by imposing $\langle I_2 \rangle = 0$ at thermovoltage $V_{22}^{\mathrm{th}}$. {As derived in Appendix \ref{App_E1},} the charge current in terminal 2 is $
\langle I_{2}\rangle 
=G_{24}V$, which implies $\langle I_2 \rangle = 0$ only at $V = 0$, meaning $V_{\text{th}}^{22} = 0$. We can derive the expression for $\Delta_T^{22}$ at zero thermovoltage and we find that it vanishes completely {as shown in Eq. (\ref{eq:E2}) of Appendix \ref{App_E1}}.
Similarly, we can calculate $\Delta_T^{44}$, imposing $\langle I_4 \rangle = 0$. {As derived in Appendix \ref{App_E1}, $\langle I_4 \rangle$ vanishes at $V_{44}^{\text{th}} = V = 0$. Then,} the expression for $\Delta_T^{44}$ {can also be derived from} Eq. (\ref{eq:11}) {and we realize that it also vanishes as shown in Eq. (\ref{eq:E5}) of Appendix \ref{App_E1}.}

Next, we consider setup 2, wherein we calculate $\Delta_T^{22}$ and $\Delta_T^{44}$. To calculate $\Delta_T^{22}$, we impose $\langle I_2 \rangle = 0$. Using Eqs. (\ref{eq:121}) and (\ref{eq:124}), we get $\langle I_2 \rangle = 0 $ for $V = 0$, thus $V_{\text{th}}^{22} = V = 0$. Similarly, $\langle     I_4 \rangle$ is zero at $V_{\text{th}}^{44} =V= 0$. Thus, as in setup 1, we find that $\Delta_T^{22}$ and $\Delta_T^{44}$ vanish here too using Eq. (\ref{eq:11}).

\subsubsection{$\Delta_T$ noise in QSH setup}

For the QSH setup (Fig.~\ref{fig:2}), the charge current is given as $I_i = \sum_{\eta\in \{\uparrow, \downarrow\}} I_i^{\eta}$, where $I_i^{\eta}$ are individual spin polarized currents. The average spin-polarized current $\langle I_i^{\eta} \rangle$ is \cite{PhysRevB.41.7906},
\begin{equation}\label{eq:126}
\langle I_{i}^{\eta}\rangle
=\frac{e}{h} \sum_{\rho \in \{\uparrow, \downarrow \}} \int dE\,\bigl(\delta_{ij} \delta_{\eta \rho}-  T_{ij}^{\eta \rho}\bigr)\,f_{j}(E-eV_{j}).
\end{equation}
In linear response, we obtain
\begin{equation}\label{eq:127}
\langle I_{i}^{\eta}\rangle=\sum_{j}\sum_{\rho}(G_{ij}^{\eta \rho}V_{j}+L_{ij}^{\eta \rho}\tau_{j}),
\end{equation}
with
\begin{equation}\label{eq:128}
\begin{split}
G_{ij}^{\eta \rho}
&=\frac{e^{2}}{h}\int dE\,(\delta_{ij} \delta_{\eta \rho}-T_{ij}^{\eta \rho}(E))\left(\frac{-\partial f}{\partial E}\right),\,\, \text{and}\\
L_{ij}^{\eta \rho}
&=\frac{e}{h\bar{T}}\int dE\,(\delta_{ij} \delta_{\eta \rho}-T_{ij}^{\eta \rho}(E))\,E\left(\frac{-\partial f}{\partial E}\right).
\end{split}
\end{equation}

Imposing the zero-charge-current condition in terminal 2, i.e., $\langle I_2 \rangle = \langle I_{2}^{\uparrow} + I_2^{\downarrow} \rangle= 0$, we obtain the thermovoltage in order to calculate $\Delta_T^{22}$ noise, i.e., $V_{\text{\text{th}}}^{22}$. Similarly, we can impose charge current in terminal 4, i.e., $\langle I_4 \rangle = \langle I_4^{\uparrow} + I_4^{\downarrow} \rangle= 0$ and find the thermovoltage $V_{\text{\text{th}}}^{44}$ to calculate $\Delta_T^{44}$ noise.

For the helical edge modes, the $s$-matrix of the four terminal QSH setup with a QPC is, 
\begin{widetext}

\begin{equation} \label{eq:129}
s=
\begin{pmatrix}
0 & 0 & \sqrt{\mathcal{T}}\,e^{i\alpha} & 0 & 0 & 0 &
-i\sqrt{1-\mathcal{T}}\,e^{i\alpha} & 0\\
0 & 0 & 0 & 0 & 0 & 0 & 0 & e^{-i\chi_1}\\
0 & 0 & 0 & 0 & e^{i\chi_2} & 0 & 0 & 0\\
0 & \sqrt{\mathcal{T}}\,e^{-i\alpha} & 0 & 0 & 0 &
i\sqrt{1-\mathcal{T}}\,e^{-i\alpha} & 0 & 0\\
0 & 0 & -i\sqrt{1-\mathcal{T}}\,e^{i\alpha} & 0 & 0 & 0 &
\sqrt{\mathcal{T}}\,e^{i\alpha} & 0\\
0 & 0 & 0 & e^{-i\chi_2} & 0 & 0 & 0 & 0\\
e^{i\chi_1} & 0 & 0 & 0 & 0 & 0 & 0 & 0\\
0 & i\sqrt{1-\mathcal{T}}\,e^{-i\alpha} & 0 & 0 & 0 &
\sqrt{\mathcal{T}}\,e^{-i\alpha} & 0 & 0
\end{pmatrix}.
\end{equation}
\end{widetext}

\begin{figure}[H]
 \centering
 \includegraphics[width=1.00\linewidth]{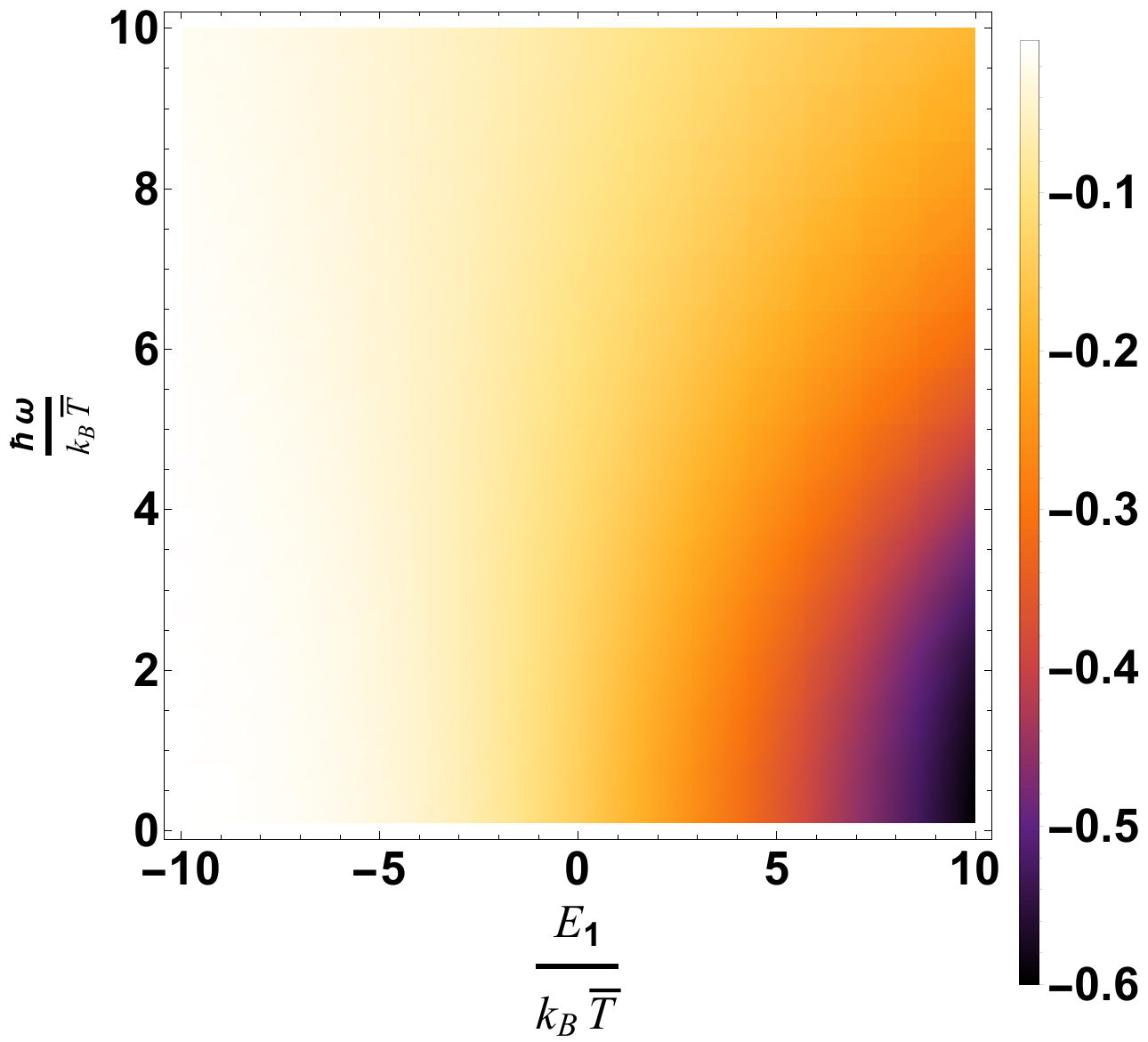}
 \caption{{Density plot of $V_{\text{\text{th}}}^{22}$ (same as $V_{\text{\text{th}}}^{44}$) in QSH setup, with topological helical edge mode, in units of $\frac{k_B \bar{T}}{e}$ in setup 1. We consider $\bar{T} = 10 K$ and $\Delta T = 1K$. (This thermovoltage is exactly same for $V_{\text{th}}^{22}$ and $V_{\text{th}}^{44}$ for trivial helical edge mode (irrespective of the value of $P$).)}}
  \label{fig:14}
 \end{figure}

 {The expression for
 $\langle I_2 \rangle$ can be derived from Eqs. (\ref{eq:127}) and (\ref{eq:128}) and then we can calculate the thermovoltage $V_{th}^{22}$ such that $\langle I_2 \rangle = 0$, which is done in Appendix \ref{App_E2}. As derived in Appendix \ref{App_E2},} $V_{\text{th}}^{22}$ = $\left(\frac{\int dE (1-2 \mathcal{T})E\left(\frac{-\partial f}{\partial E}\right)}{\int dE \mathcal{T} \left(\frac{-\partial f}{\partial E}\right)}\right)\frac{\Delta T}{2e \bar{T}}$. One can then calculate the $\Delta_T$ noise: $\Delta_T^{22}$ at $V_{\text{th}}^{22}$, using Eqs. (\ref{eq:63}) and (\ref{eq:64}).

  Similarly, for the calculation of $\Delta_T^{44}$, we impose $\langle I_4 \rangle = 0$ {to calculate thermovoltage $V_{th}^{44}$}. Using Eqs. (\ref{eq:127}) and (\ref{eq:129}), we derive $\langle I_4 \rangle$, {which implies $\langle I_4 \rangle$ is zero at $V_{th}^{44}$, which is same as $V_{th}^{22}$, see Appendix \ref{App_E2} for this calculation.} We observe that $V^{22}_{\text{th}}$ and $V^{44}_{\text{th}}$ also distinguishes between chiral and helical edge mode as it is finite for helical edge mode, whereas for chiral edge mode, they vanish. We plot $V_{\text{th}}^{22}$ (same as $V_{\text{th}}^{44}$) in Fig. \ref{fig:14} for all the values of QPC parameters $\frac{\hbar \omega}{k_B \bar{T}}$ and $\frac{E_1}{k_B \bar{T}}$.

Thus, $\Delta_T$ noise autocorrelations such as $\Delta_T^{22}, \Delta_T^{44}$ are identical and can be derived from Eqs. (\ref{eq:63}) and (\ref{eq:64}). {This derivation is explained in Appendix \ref{App_E2} and we see that $\Delta_T^{22}$ has non-zero contribution from $\Delta_T^{22; \downarrow \downarrow}$ alone as shown in Eq. (\ref{eq:130}) and is thus given as,}

\begin{equation}\label{eq:130}
 \Delta_T^{22}=   \Delta_T^{22;\downarrow \downarrow} = G_0 \int_{-\infty}^{\infty}dE \mathcal{T}(1-\mathcal{T})(f_H(E) - f_C(E))^2.
\end{equation}

\begin{figure}[H]
\centering
\includegraphics[width=1.00\linewidth]{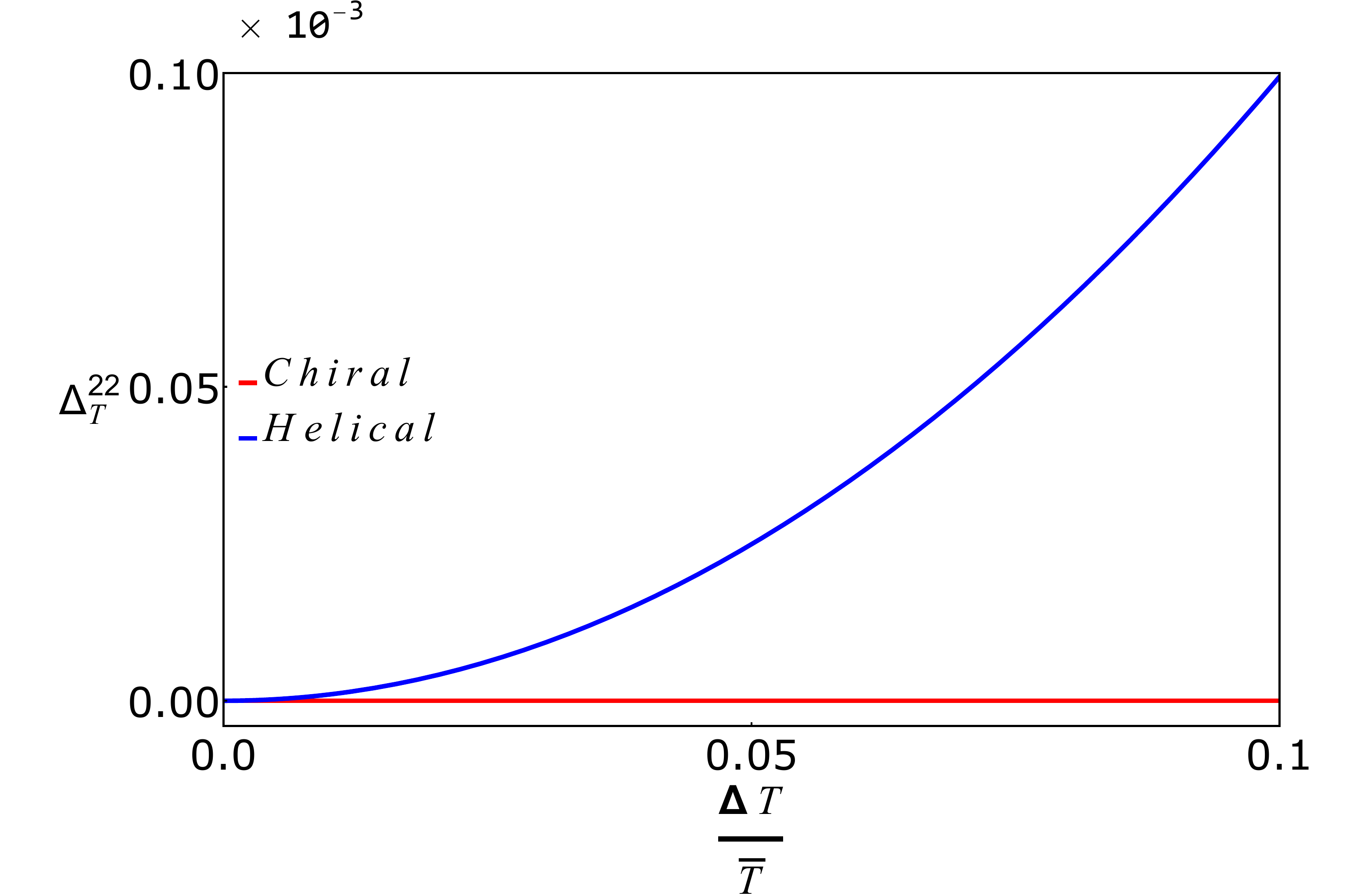}
\caption{{$\Delta_T^{22}$ (or, $\Delta_T^{44}$) in units of $G_0 k_B \bar{T}$ in setup 1 for $\bar{T} = 10 K$ for $\hbar \omega_y = 10 k_B \bar{T}$ and $\hbar \omega_x = 0.9 \hbar \omega_y$. Chiral and helical edge modes can be perfectly distinguished for energy-dependent scattering from QPC also.}}
 \label{fig:15}
\end{figure}

 Similarly, we can derive $\Delta_T^{44}$ wherein $\Delta_T^{44; \uparrow \uparrow}, \Delta_T^{44; \uparrow \downarrow}, \Delta_T^{44; \downarrow \uparrow}$
vanish, whereas $\Delta_T^{44; \downarrow \downarrow}$
is finite and exactly same as $\Delta_T^{22; \downarrow \downarrow}$.
We observe that both $\Delta_T^{22}$ and $\Delta_T^{44}$ clearly distinguish between chiral and helical edge modes, see Fig. \ref{fig:15}. In both the energy-dependent as well as energy-independent case, $\Delta_T^{22}$ and $\Delta_T^{44}$ vary quadratically with $\frac{\Delta T}{\bar{T}}$, but with distinct magnitudes.

Next, we consider setup 2 and calculate $\Delta_T^{22}$ and $\Delta_T^{44}$ by imposing $\langle I_2 \rangle$ as well as $\langle I_4 \rangle$ to be zero. The expression for $\langle I_2 \rangle$ can be obtained from Eq. (\ref{eq:761}) and imposing conditions for setup 2, i.e., $V_1 = V_3 = V, V_2 = V_4 = 0$ and $\tau_1 = \tau_3 = \frac{\Delta T}{2} $ and $\tau_2 = \tau_4 = -\frac{\Delta T}{2} $. {As calculated in Appendix \ref{App_E2}, the thermovoltage $V_{\text{th}}^{22}$ vanishes completely. As also discussed in Appendix \ref{App_E2}, in this setup, $\Delta_T^{22}$ vanishes.} Similarly, $\Delta_T^{44}$ also vanishes at zero $V_{\text{th}}^{44}$. We observe that in setup 2, the distinction between chiral and topological helical edge mode is not possible as in both cases $\Delta_T^{22}$ and $\Delta_T^{44}$ vanish.

 \subsubsection{$\Delta_T$ noise in trivial QSH setup}

Similarly, for trivial helical edge modes in QSH setup (see, Fig. \ref{fig:5}), the modified $s$-matrix with a QPC is given as,
\begin{widetext}

\begin{equation} \label{eq:131}
s=
\begin{pmatrix}
0 & -i\sqrt{ P} e^{i\xi} & \sqrt{\mathcal{T}} e^{i \alpha}x & 0 & 0 & 0 & -i\sqrt{1-\mathcal{T}} e^{i \alpha}x & 0\\
-i\sqrt{ P} & 0 & 0 & 0 & 0 & 0 & 0 & x \, e^{-i\chi_1}\\
0 & 0 & 0 & -i\sqrt{ P} & x \, e^{i\chi_2} & 0 & 0 & 0\\
0 & \sqrt{\mathcal{T}} e^{-i \alpha}x & -i\sqrt{ P}e^{-i\xi} & 0 & 0 & i\sqrt{1-\mathcal{T}} e^{-i \alpha}x & 0 & 0\\
0 & 0 & -i\sqrt{1-\mathcal{T}} e^{i \alpha}x & 0 & 0 & -i\sqrt{ P}e^{i\xi} & \sqrt{\mathcal{T}} e^{i \alpha}x & 0\\
0 & 0 & 0 & x \, e^{-i\chi_2} & -i\sqrt{ P} & 0 & 0 & 0\\
x \, e^{i\chi_1} & 0 & 0 & 0 & 0 & 0 & 0 & -i\sqrt{ P}\\
0 & i\sqrt{1-\mathcal{T}} e^{-i \alpha}x & 0 & 0 & 0 & \sqrt{\mathcal{T}} e^{-i \alpha}x & -i\sqrt{ P}e^{-i\xi} & 0
\end{pmatrix},
\end{equation}
\end{widetext}

where, $x = \sqrt{1 -  P}, \xi = \alpha - \theta$ with $\theta = \text{Tan}^{-1}\left(\frac{\sqrt{1-\mathcal{T}}}{\sqrt{\mathcal{T}}}\right)$, where $\mathcal{T}$ is given in Eq. (\ref{eq:122}). Herein also, we derive $\Delta_T^{22}$ and $\Delta_T^{44}$ imposing $\langle I_2 \rangle$ and $\langle I_4 \rangle$ to zero and at thermovoltages $V_{\text{th}}^{22}$ and $V_{\text{th}}^{44}$ respectively {in Appendix \ref{App_E3}}. We only consider setup 2 for the calculation of $\Delta_T^{22}$ and $\Delta_T^{44}$ with trivial edge modes. Here, we realize that both $V_{\text{th}}^{22}$ and $V_{\text{th}}^{44}$ vanish.

\begin{figure}[H]
\centering
\includegraphics[width=1.00\linewidth]{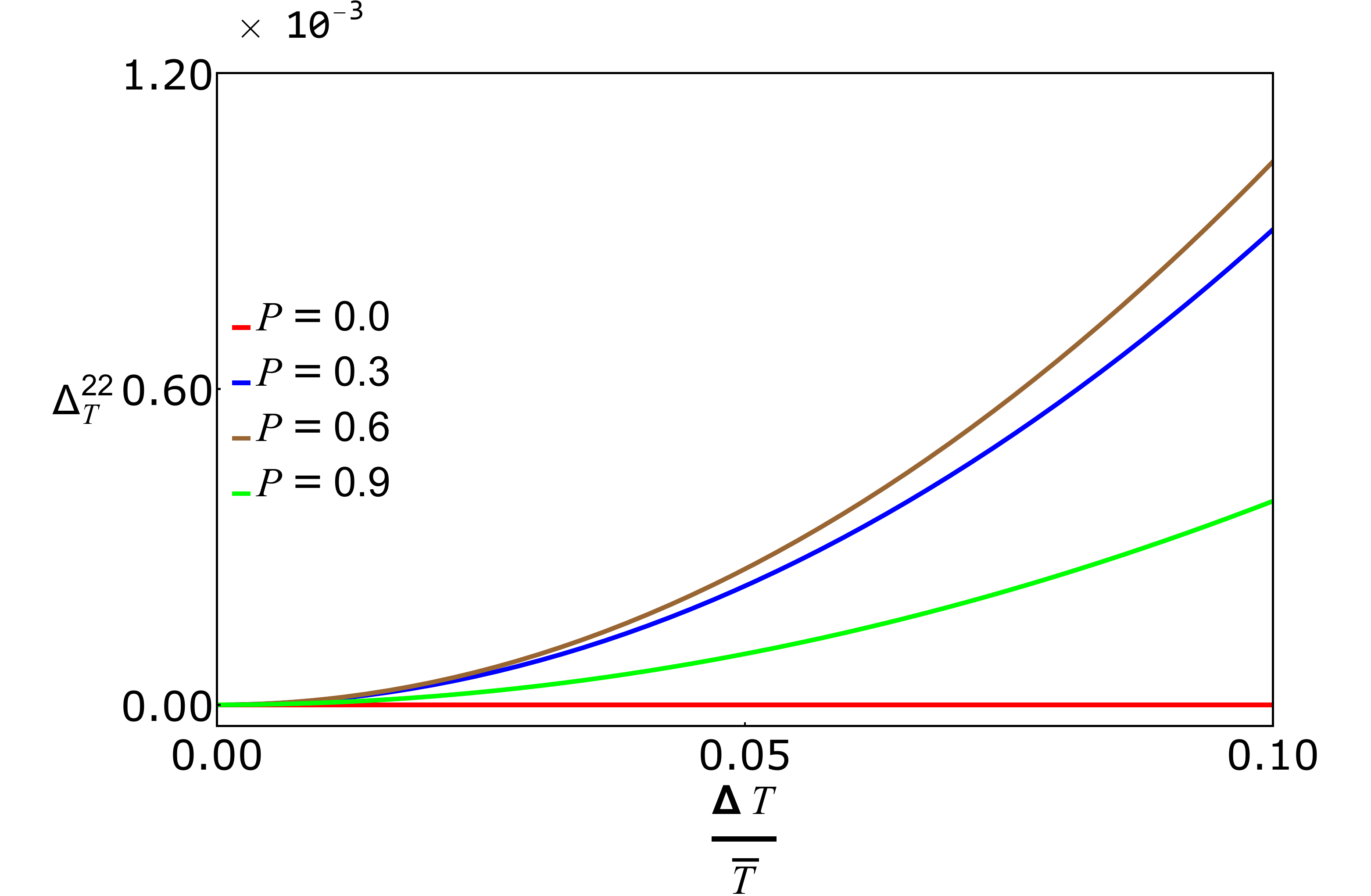}
\caption{{$\Delta_T^{22}$ (or, $\Delta_T^{44}$) in units of $G_0 k_B \bar{T}$ in setup 2 at $\bar{T} = 10 K$ for $\hbar \omega_y = 10 k_B \bar{T}$ and $\hbar \omega_x = 0.9 \hbar \omega_y$. Helical and trivial edge modes can be perfectly distinguished with energy-dependent scattering from QPC.}}
 \label{fig:16}
\end{figure}

Thus, $\Delta_T$ noise such as $\Delta_T^{22}$ and $\Delta_T^{44}$ can be derived from Eqs. (\ref{eq:63}) and (\ref{eq:64}) {and the full derivation is given in Eq. (\ref{eq:133}) in Appendix \ref{App_E3}, and is given as},

\begin{equation}\label{eq:133}
    \Delta_T^{44}=\Delta_T^{22} = \Delta_T^{22; \downarrow \downarrow}  =2G_0 \int_{-\infty}^{\infty} dE \,\,\, P(1-P) (f_H(E) - f_C(E))^2.
\end{equation}
Thus $\Delta_T^{22}$ (or, $\Delta_T^{44}$) clearly distinguish helical ($P = 0 $) from trivial ($P \neq 0$) edge modes, see Fig. \ref{fig:16}. For $P = 0$, $\Delta_T^{22}$ vanishes, while for trivial helical edge mode, $\Delta_T^{22} \neq 0$.

One can also perform this analysis for setup 1, with $T_1 = T_4 = T_H$ and $T_2 = T_3 = T_C$ at $V_1 = V_4 = V$ and $V_2 = V_3 = 0$. However, we do not see any qualitative distinction between helical and trivial helical edge modes using setup 1.

The summary of all the results of $\Delta_T$ noise in presence of QPC distinguishing chiral, helical, and trivial edge modes is given in Table \ref{Table3}

\begin{table}[H]
\centering
\caption{{Setup 1 distinguishes chiral and helical, and setup 2 distinguishes helical from trivial edge modes.}}
{
\begin{tabular}{ll|c|c|}
\cline{3-4}
           &           & \multicolumn{1}{c|}{Setup 1} & \multicolumn{1}{c|}{Setup 2} \\ \hline
\multicolumn{2}{|c|}{Edge modes} 
& \multicolumn{1}{c|}{$\Delta_T^{22} = \Delta_T^{44}$} 
& \multicolumn{1}{c|}{$\Delta_T^{22} = \Delta_T^{44}$} \\ \hline
\multicolumn{2}{|c|}{Chiral} 
& \multicolumn{1}{c|}{Zero} 
& \multicolumn{1}{c|}{Zero} \\ \hline
\multicolumn{2}{|c|}{Helical (Topological)} 
& \multicolumn{1}{c|}{Quadratic}  
& \multicolumn{1}{c|}{Zero} \\ \hline
\multicolumn{2}{|c|}{Helical (Trivial)} 
& \multicolumn{1}{c|}{Quadratic} 
& \multicolumn{1}{c|}{Quadratic} \\ \hline
\end{tabular}
}
\label{Table3}
\end{table}

\section{Experimental realization and conclusion}
\label{expt}

In this work we develop a systematic, experimentally relevant framework to distinguish chiral and two distinct helical edge modes, which can be topological and trivial in nature, by exploiting the temperature bias contribution to quantum noise, $\Delta_T$ noise.

For observing chiral edge-mode transport in the quantum Hall (QH) regime, one typically employs a high-mobility two-dimensional electron gas (2DEG) such as GaAs/AlGaAs heterostructures, see Ref. \cite{PhysRevLett.45.494}. Helical edge-mode transport requires a four-terminal quantum spin Hall (QSH) setup based on HgTe/CdTe quantum wells, where band inversion gives rise to topological helical edge states, see Refs. \cite{doi:10.1126/science.1133734, konig2007quantum, roth2009nonlocal}. Trivial helical edge modes can occur in a such QSH four-terminal setups especially in InAs/GaSb or InSb/GaSb double quantum wells, and one has to use conductance (some other method) to distinguish. As these methods are not very effective, because the charge conductance cannot give a stark distinction between these two distinct helical edge modes, we require better probes for their distinction, which is the motivation of our work.

In the QH setup, strong perpendicular magnetic fields of the order of $1$--$4~\mathrm{T}$ are required to form well-defined Landau levels. Gate voltages are used to tune the carrier density and filling factors, and measurements are typically performed at low temperatures in the $1$--$10~\mathrm{K}$ range to ensure robust quantization. One also requires applied voltage biases too. Our proposal also requires incorporating a quantum point contact (QPC) in the 2DEG to control transmission and backscattering between the chiral edge modes. The fabrication and experimental implementation of QPCs in GaAs-based 2DEGs has been demonstrated extensively, for example in Ref. \cite{PhysRevLett.60.848}.

For helical edge transport in the QSH setup, the HgTe/CdTe quantum well thickness is tuned across the critical thickness $d_{c}$, at which the electronic structure undergoes a transition from a normal to an inverted band ordering. This transition corresponds to a topological phase transition from a trivial insulator to a QSH insulator containing helical edge states, see Refs. \cite{doi:10.1126/science.1133734, konig2007quantum, roth2009nonlocal}. The helical edge modes in the QSH phase are realized when the well thickness satisfies $d > d_{c} \approx 6.3~\mathrm{nm}$, where the band structure becomes inverted. In the original QSH experiments, quantum wells of thickness $d = 7.5$--$9~\mathrm{nm}$ were used, with electron density $n_{s} \sim 3 \times 10^{11}~\mathrm{cm^{-2}}$ and mobility $\sim 1.5 \times 10^{5}~\mathrm{cm^{2}V^{-1}s^{-1}}$. Devices were patterned into Hall-bar geometries using electron-beam lithography followed by Ar ion-beam etching, and gated using a $\mathrm{Si_{3}N_{4}}/\mathrm{SiO_{2}}$ dielectric stack with Ti/Au electrodes. The gate voltage enables continuous tuning between insulating and topological regimes. Transport measurements were typically performed at temperatures in the range $1$--$10~\mathrm{K}$, where four-terminal (same as Fig. \ref{fig:2}) non-local resistance measurements exhibit plateaus of the order of $h/2e^{2}$, consistent with the quantized conductance of a single pair of helical edge states. A QPC can also be implemented in the QSH setup using fabrication techniques analogous to those employed in Ref. \cite{PhysRevLett.60.848}.

Trivial edge-modes in two-dimensional QSH systems occur along with Helical edge modes in InAs/GaSb double quantum wells, see Ref. \cite{nichele2016edge}. Compared to HgTe/CdTe quantum wells, the InAs/GaSb platform offers high mobility, compatibility with semiconductor fabrication, and an electrically tunable band structure controlled independently by top and back gates. By adjusting these gate voltages, a transition is induced between trivial and inverted phases. A key finding is that trivial edge channels exhibit a resistance per unit length comparable to values previously interpreted as signatures of helical edge states, indicating that trivial edge conduction can mimic many transport characteristics of the QSH effect. These observations highlight the need for careful discrimination between trivial and topological edge transport in InAs/GaSb devices. A QPC can also be implemented in this platform using fabrication methods similar to those described in Ref. \cite{PhysRevLett.60.848}.

Our analysis treats both auto- and cross-correlations and applies at zero and finite frequencies. While conventional shot-noise measurements performed at zero temperature under a finite voltage bias and the $\Delta_T$ protocol convey closely related information about transport channels, $\Delta_T$ noise offers a practical advantage, by design it eliminates background fluctuations arising from finite charge currents, yielding a cleaner and more reliable fingerprint of the underlying edge-mode dynamics \cite{shein2022electronic, sivre2019electronic}. The $\Delta_T$ noise evaluation proceeds in three steps. First, compute the full quantum noise correlations at equilibrium (all terminals held at $\bar{T}$ and applied bias voltage is zero), which yields the purely thermal contribution. Second, impose a temperature bias, $ T_{1}=T_{4}=\bar{T}+\frac{\Delta T}{2},$ and $T_{2}=T_{3}=\bar{T}-\frac{\Delta T}{2}$ still at zero voltage bias; the measured correlations now contain both the equilibrium thermal part and an extra non-equilibrium component induced by the temperature gradient. Third, subtract the equilibrium result from the biased result to isolate $\Delta_T$ noise. In this sense $\Delta_T$   
 noise is the finite-temperature analogue of shot noise but crucially without contamination from current-driven (Joule) heating or other bias-induced artifacts. In a similar manner, the $\Delta_T$ noise can also be measured at finite frequencies. The procedure begins by calculating the quantum noise correlation at zero temperature bias, where all reservoirs are maintained at the same temperature. In this case, the total contribution consists of both thermal noise and excess noise. The measurement is then repeated under a finite temperature bias, for which the quantum noise includes thermal noise, excess noise, and the $\Delta_T$ noise. Finally, by subtracting the zero-bias quantum noise correlation from the finite-bias quantum noise correlation, one isolates the $\Delta_T$ noise.

 \begingroup

We have focused on the linear-response regime because our primary objective is to establish the fundamental ability of $\Delta_T$ noise to distinguish between chiral, topological helical, and trivial helical edge modes in a transparent and analytically tractable manner.

We would like to emphasize that for energy-dependent scattering as treated in our work, entering the nonlinear regime involves more than simply considering temperature biases comparable to or larger than the average temperature. In general, nonlinear thermoelectric transport requires a self-consistent treatment of the interaction-induced screening potential, which can modify the transport properties of the system. This becomes particularly relevant for energy-dependent scatterers, such as the quantum point contact (QPC) considered in our work, where the transmission probabilities depend explicitly on energy. In such cases, the interaction-induced potential also affects the scattering characteristics and therefore requires a more elaborate numerical treatment, see Ref. \cite{PhysRevB.110.235430}.

By contrast, Ref. \cite{lumbroso2018electronic} primarily considered energy-independent scattering probabilities, for which nonlinear effects can often be treated more straightforwardly. The situation is therefore considerably simpler than in our work, where both energy-independent and energy-dependent scattering processes, see Sec. \ref{analysis} B, is analyzed.
 \endgroup

 Our results demonstrate that $\Delta_T$ noise provides a robust and unambiguous means of distinguishing between different edge-mode transport regimes in mesoscopic systems. In particular, it enables the differentiation between chiral, topological helical, as well as trivial helical modes with a level of precision unattainable through conventional conductance measurements, because the $\Delta_T$ noise gives the more in-depth knowledge about the system. To the best of our knowledge, this work constitutes the first demonstration that $\Delta_T$ noise can reliably separate distinct edge-mode transport channels. These findings not only advance our fundamental understanding of edge-state dynamics but also hold significant promise for applications in spintronics and quantum information, where the ability to accurately identify and manipulate edge modes is of central importance.

\section*{Appendix}

{The appendix is divided into four parts. Appendix \ref{App_I} derives the expression for $\Delta_T$ noise in QH setup, while Appendix \ref{App_Qn} derives the expression for $\Delta_T$ noise for QSH setup. Appendix \ref{App_C} derives the scattering matrices for the four-terminal QH setup with chiral edge modes, four-terminal QSH setup with helical and trivial edge modes. Appendix \ref{App_E} derives the expressions of $\Delta_T$ noise in QH setup with chiral edge modes, QSH setup with helical and trivial edge modes with energy-dependent scattering in presence of QPC.}

\appendix
\label{appendix}

    \section{Derivation of $\Delta_T$ noise in QH setup}
\label{App_I}

The quantum noise correlations for chiral edge modes at zero voltage bias and finite temperature bias can be expressed, following Eq.~(\ref{eq:8}) for single edge mode transport, as  
\begin{equation} \label{eq:B1}
\begin{split}
S_{i j} &= G_0\sum_{l,p}\int dE \;[A_{l p}(i)A_{p l}(j)] \\&
\times \big[f_{l}(E )(1-f_{p}(E ))+(1-f_{l}(E))f_{p}(E)\big].
\end{split}
\end{equation}
Here, $\mathcal{A}_{l p}(i) = I_{i}\delta_{i l}\delta_{i p} - s_{i l}^{\dagger}s_{i p}$, $G_0 = 2e^2/h$, and $f_l(E) = \frac{1}{1+e^{\frac{E}{k_B T_l}}}$.  
Here, $E$ is the energy measured relative to the chemical potential (or Fermi energy $E_F$), i.e., $E = E' - E_F$, with $E'$ the total electron energy.  

In equilibrium, all contacts are at the same temperature $\bar{T}$ and share a common chemical potential, so the Fermi-Dirac distribution reduces to $f(E) = \frac{1}{1+e^{\frac{E}{k_B \bar{T}}}}$. In this case, Eq.~(\ref{eq:B1}) simplifies to
\begin{equation} \label{eq:B2}
S_{i j}^{eq} = 2 G_0\int dE \, f(E)(1-f(E)) 
\sum_{l, p}[A_{lp}(i)A_{p l}(j)].
\end{equation}
Here, $A_{l p}(i) = \delta_{i l}\delta_{i p} - s_{i l}^{\dagger}s_{i p}$ and 
$A_{p l}(j) = \delta_{j p}\delta_{j l} - s_{j p}^{\dagger}s_{j l}$.  

We now evaluate $S_{i j}^{eq}$ by carrying out the summation over $l$ and $p$:
\begin{equation} \label{eq:B3}
\begin{split}
\sum_{l, p}[A_{l p}(i)A_{p l}(j)] 
= \sum_{l p}\big(\delta_{i l}\delta_{i p}\delta_{j l}\delta_{j p} 
-\delta_{j l}\delta_{j p}s_{i l}^{\dagger}s_{i p}\\
-\delta_{i l}\delta_{i p}s_{j p}^{\dagger}s_{j l} 
+s_{i l}^{\dagger}s_{i p}s_{j p}^{\dagger}s_{j l}\big).
\end{split}
\end{equation}

Using the unitarity condition of the scattering matrix, $\sum_{i} s_{i j}^{\dagger}s_{i l} = I_{j}\delta_{j l}$, Eq.~(\ref{eq:B3}) reduces to  
\begin{equation} \label{eq:B4}
\sum_{l, p}[A_{l p}(i)A_{p l}(j)] 
= 2\delta_{j i}-s_{i j}^{\dagger}s_{i j}-s_{j i}^{\dagger}s_{j i}.
\end{equation}

Substituting \ref{eq:B4} into \ref{eq:B2} gives
\begin{equation} \label{eq:B5}
S_{i j}^{eq} = 2 G_0\int dE \, f(E)(1-f(E)) 
\big[2\delta_{i j}-(s_{i j}^{\dagger}s_{i j}+s_{j i}^{\dagger}s_{j i})\big].
\end{equation} 
This corresponds to the Nyquist-Johnson noise, arising purely from thermal fluctuations.  
Defining $(s_{ij}^{\dagger} s_{ij}) = T_{ij}$ as the transmission probability from terminal $j$ to $i$, Eq.~(\ref{eq:B5}) becomes  
\begin{equation}
S_{i j}^{eq} = 2 G_0\int dE \, f(E)(1-f(E)) 
\big[2\delta_{i j}-(T_{ij} + T_{ji})\big].
\end{equation}

For autocorrelations ($i=j$), we obtain
\begin{equation}  \label{eq:B6}
S_{i i}^{eq} = 4 G_0\int dE \, f(E)(1-f(E)) [1-T_{i i}],
\end{equation}
where $T_{ii}$ is the reflection probability into terminal $i$. For $i \ne j$, the thermal noise cross correlations are
\begin{equation} \label{eq:B7}
S_{i j}^{eq} = -2 G_0\int dE \, f(E)(1-f(E))(T_{i j} + T_{j i}),
\end{equation}
when all terminals have the same chemical potential and temperature. If terminals $i$ and $j$ have different temperatures $T_i$ and $T_j$, but the same chemical potential, $S_{ij}^{\text{th}}$ is generalized as,
\begin{equation} \label{eq:B8}
S_{i j}^{eq} = S_{ij}^{\text{th}}= -2 G_0\int dE \, \Big[T_{i j}f_{j}(E)(1-f_{j}(E))
+T_{j i}f_{i}(E)(1-f_{i}(E))\Big].
\end{equation}

$S_{ij}^{\text{th}}$ is basically the thermal noise component for cross correlation ($i \neq j$). The transport component is obtained by subtracting the equilibrium noise from full quantum noise. In case of autocorrelations ($i=j$), $S_{i i}^{tr} = S_{i i} - S_{i i}^{eq}$, where $S_{ii}$ is defined in (\ref{eq:B1}) and $S_{ii}^{\text{th}}$ is given in (\ref{eq:B6}). Thus,

\begin{flalign} \label{eq:B9}
S_{ii}^{tr} = &G_0\int dE \bigg( \sum_{l,p} [A_{l p}(i)A_{p l}(i)]
\big(f_{l}(E) + f_p(E)-2 f_l(E) f_{p}(E)\big)\notag \\& - 4 f(E)(1-f(E)) [1-T_{i i}]\bigg)
\end{flalign}

In (\ref{eq:B9}), terms like $\sum_{l, p}[A_{l p}(i)A_{p l}(i)] f_p(E)$ and $\sum_{l, p}[A_{l p}(i)A_{p l}(i)] f_l(E)$ appear and they are given as,
\begin{equation}\label{eq:A11}
\begin{split}
&\sum_p f_p(E) \sum_l [A_{l p}(i)A_{p l}(i)]  = \sum_l f_l(E) \sum_l [A_{l p}(i)A_{p l}(i)]\\& = [1-2s_{i i}^{\dagger} s_{i i}]f_{i}(E)+\sum_{p}[s_{i p}s_{i p}^{\dagger}]f_{p}(E)\\&
= [1-2T_{ii}]f_i(E) + \sum_p T_{ip} f_p(E).
\end{split}
\end{equation}


where, we define $s_{ip}s_{ip}^{\dagger} = T_{ip}$. Thus $S_{ii}^{tr}$ is
\begin{equation} \label{eq:B13}
\begin{split}
S_{i i}^{tr} = &2 G_0\int dE \Bigg(\sum_{l}T_{i l}\big(f_{l}(E) - f_{i}(E)\big)+f_{i}(E)^2\\&-\sum_{l p}f_l(E) f_p(E) (s_{i l}s_{i p}s_{i p}s_{i l})\Bigg).
\end{split}
\end{equation}

$S_{ii}^{tr}$ can be further rewritten as

\begin{equation}
    \begin{split}
        S_{ii}^{tr} =& 2G_0 \int dE  \bigg(\sum_{l \neq i}T_{il}f_l(E)(1-f_l(E))-\sum_{l \neq i}T_{il}f_i(E)(1-f_i(E))\bigg) \\& 2G_0\int_{-\infty}^{\infty} dE \Bigg(\sum_{l} T_{i l}f_l(E)^2 
- \sum_{l, p} f_{l}(E)f_{p}(E)\,\!\Big(s_{i l}^{\dagger}s_{i p}s_{i p}^{\dagger}s_{i l}\Big)\Bigg).
    \end{split}
\end{equation}

Here, the terms proportional to $f_i(E) (1-f_i(E))$ in $S_{ii}^{tr}$ contribute to the thermal noise and is added to Eq. (\ref{eq:B6}) and rest contributes to the shot noise. Therefore, the expression for thermal noise for autocorrelation $i = j$ is

\begin{equation}
    \begin{split}
        S_{i i}^{\text{th}} =& 4 G_0\int dE \, f(E)(1-f(E)) [1-T_{i i}] + 2G_0 \int dE \\& \bigg(\sum_{l \neq i}T_{il}f_l(E)(1-f_l(E))-\sum_{l \neq i}T_{il}f_i(E)(1-f_i(E))\bigg),\\
     S_{ii}^{sh} =&   2G_0\int_{-\infty}^{\infty} dE \Bigg(\sum_{l} T_{i l}f_l(E)^2 
- \sum_{l, p} f_{l}(E)f_{p}(E)\,\!\Big(s_{i l}^{\dagger}s_{i p}s_{i p}^{\dagger}s_{i l}\Big)\Bigg).
    \end{split}
\end{equation}

For cross correlations ($i \ne j$), we define $S_{i j}^{sh} = S_{i j} - S_{i j}^{\text{th}}$, where $S_{ij}$ is defined in (\ref{eq:B1}) and $S_{ij}^{\text{th}}$ is given in (\ref{eq:B8}). Thus,
\begin{flalign} \label{eq:B15b}
& S_{ij}^{sh} = G_0\int dE \bigg( \sum_{l,p} [A_{l p}(i)A_{p l}(j)]
\big(f_{l}(E) + f_p(E)-2 f_l(E) f_{p}(E)\big)\notag \\& + 2 \Big[T_{i j}f_{j}(E)(1-f_{j}(E))
+T_{j i}f_{i}(E)(1-f_{i}(E))\Big]\bigg)
\end{flalign}

Now, in Eq. (\ref{eq:B15b}), substituting
\begin{equation} \label{eq:B17}
\begin{split}
&\sum_{l, p}\Big( [A_{l p}(i)A_{p l}(j)]\Big)f_{p}(E) = \sum_{l, p}\Big( [A_{l p}(i)A_{p l}(j)]\Big)f_{l}(E)\\&
= -T_{i j}f_{j}(E) - T_{j i}f_{i}(E),
\end{split}
\end{equation}
we find $S_{ij}^{sh}$ to be,
\begin{equation} \label{eq:B18}
S_{i j}^{sh} = -2 G_0\int dE \sum_{l, p}f_{l}(E)f_{p}(E)[s_{i l}^{\dagger}s_{i p}s_{j p}^{\dagger}s_{j l}].
\end{equation}

It is important to note that Eq.(\ref{eq:B18}) has certain limitations, as highlighted in Ref. \cite{PhysRevB.108.115301} using a two-terminal QH example. In such a setup with a constriction, current conservation requires $\Delta I_1 + \Delta I_2 = 0$, where $\Delta I_{1,2}$ are current fluctuations around their mean values in terminal 1 and 2. Consequently, the quantum noise autocorrelation in terminal 1, $S_{11} = S_{11}^{\text{th}} + S_{11}^{sh} = \langle \Delta I_1^2 \rangle$, and the cross correlation between terminals 1 and 2, $S_{12} = S_{12}^{\text{th}} + S_{12}^{sh} = \langle \Delta I_1 \Delta I_2 \rangle$, must satisfy the relation $S_{11} = -S_{12}$. In order to satisfy this rule, one needs to have $S_{11}^{\text{th}} = -S_{12}^{\text{th}}$ and $S_{11}^{sh} = -S_{12}^{sh}$. The thermal noise component as shown in Eq. (\ref{eq:B8}) satisifies $S_{11}^{\text{th}} = -S_{12}^{\text{th}}$, but the shot noise from Eq. (\ref{eq:B18}) doesnot obey $S_{11}^{sh} = -S_{12}^{sh}$. To resolve this inconsistency, Ref. \cite{PhysRevB.46.12485} proposed a modification introducing two independent functions, $f_a(E)$ and $f_b(E)$, which are energy-dependent, and this restores the required relation. Therefore, the shot noise cross correlation such as $S_{12}^{sh}$ in the QH setup is

\begin{flalign}
S_{12}^{sh} &= -2 G_0\int dE \sum_{l, p}\big(f_{l}(E)-f_a(E)\big)\big(f_{p}(E)-f_b(E)\big)\notag \\
& \times[s_{1 l}^{\dagger}s_{2 p}s_{j p}^{\dagger}s_{j l}].
\end{flalign}

Now, the form of $f_a(E)$ and $f_b(E)$ is not unique, but for a two-terminal QH setup, when one puts both the functions to be the Fermi-Dirac distribution with lower temperature bias, then the relation $S_{11} = -S_{12}$ is satisfied. Here, the expression for both $f_a(E)$ and $f_b(E)$ are same as $f_2(E)$, which is the Fermi-Dirac distribution at temperature $T_2 = \bar{T} - \frac{\Delta T}{2}$, i.e., $f_2(E) = \frac{1}{1+e^{\frac{E}{k_B T_2}}}$, with $\Delta T$ being the temperature bias across the setup. Now, for a multiterminal QH setup, one needs to satisfy $S_{ii}^{sh} = -\sum_j S_{ij}^{sh}$. Therefore, the final and corrected formula for the shot noise for a multiterminal QH setup is therefore \cite{PhysRevB.46.12485},
\begin{flalign} \label{eq:B19}
S_{i j}^{sh} &= -2 G_0\int dE \sum_{l, p}\big(f_{l}(E)-f_a(E)\big)\big(f_{p}(E)-f_b(E)\big)\notag \\
& \times[s_{i l}^{\dagger}s_{i p}s_{j p}^{\dagger}s_{j l}].
\end{flalign}

The $\Delta_T$ noise cross correlation ($i \neq j$) is calculated from shot noise at zero voltage bias but at finite temperature bias using (\ref{eq:B19}) to obtain,
\begin{flalign}\label{eq:A19}
\Delta^{i j}_T =& -2 G_0\int dE \sum_{l, p}\big(f_{l}(E)-f_a(E)\big)\big(f_{p}(E)-f_b(E)\big)\notag \\& \times [s_{i l}^{\dagger}s_{i p}s_{j p}^{\dagger}s_{j l}],
\end{flalign}
and from (\ref{eq:B13}) for autocorrelation ($i = j$), we obtain,
\begin{flalign}\label{eq:A20}
\Delta^{i i}_{T} =& 2G_0\int_{-\infty}^{\infty} dE \Bigg(\sum_{l} T_{i l}f_l(E)^2 
- \sum_{l, p} f_{l}(E)f_{p}(E)\,\!\Big(s_{i l}^{\dagger}s_{i p}s_{i p}^{\dagger}s_{i l}\Big)\Bigg).
\end{flalign}

\section{Derivation of $\Delta_T$ noise in QSH setup}
\label{App_Qn}

The spin-resolved components of quantum noise ($S_{ij}^{\eta \eta'}$) with $\eta, \eta' \in \{\uparrow, \downarrow \}$, can be calculated as  
\begin{equation} \label{eq:C2}
\begin{split}
S_{i j}^{\eta \eta'} = \frac{G_0}{2}\sum_{\rho, \rho' = \uparrow, \downarrow}\sum_{l,p}\int dE \; 
\!\Big[A_{l p}^{\rho \rho'}(i, \eta)
A_{p l}^{\rho' \rho}(j, \eta')\Big] \\
\times \Big(f_{l}(E)(1-f_{p}(E))+(1-f_{l}(E))f_{p}(E)\Big).
\end{split}
\end{equation}
Here, $\mathcal{A}_{l p}^{\rho \rho'}(i, \eta) = \delta_{i l}\delta_{i p}\delta_{\eta \rho}\delta_{\eta \rho'} - s_{i l}^{\eta \rho \dagger}s_{i p}^{\eta \rho'}$ and $f_l(E)$ is the Fermi–Dirac distribution at terminal $l$.  

If all terminals are maintained at the same temperature ($\bar{T}$), i.e., $f_l(E) = f_p(E) = f(E)$, Eq.~(\ref{eq:C2}) reduces to purely thermal noise:  
\begin{flalign} \label{eq:C3}
\! \! \! S_{i j}^{\eta \eta', eq} = G_0\int dE  f(E)(1-f(E)) 
\sum_{l, p}\sum_{\rho, \rho'} 
[A_{lp}^{\rho \rho'}(i, \eta)A_{pl}^{\rho' \rho}(j, \eta')],
\end{flalign}
with $f(E) = \frac{1}{1 + e^{\frac{E - \mu}{k_B \bar{T}}}}$. Substituting the explicit form of $A_{lp}^{\rho \rho'}(i, \eta)$ from Eq.~(\ref{eq:C2}), one obtains  
\begin{flalign} \label{eq:C4}
S_{i j}^{\eta \eta', eq} &= G_0\int dE \, f(E)(1-f(E)) 
\sum_{l,p}\sum_{\rho, \rho'}\mathrm{Tr}\!\Big[(\delta_{i l}\delta_{i p} \delta_{\eta \rho} \delta_{\eta \rho'} \notag \\
& -s_{i l}^{\eta \rho \dagger}s_{i p}^{\eta \rho'}) \times(\delta_{j p}\delta_{j l} \delta_{\eta' \rho'} \delta_{\eta' \rho}-s_{j p}^{\eta' \rho' \dagger}s_{j l}^{\eta' \rho})\Big].
\end{flalign}

As in the chiral case, see Eq.~(\ref{eq:B4}), and using the unitarity of the $s$-matrix, $\sum_{l,p} \sum_{\rho \rho'}s_{il}^{\eta \rho \dagger} s_{ip}^{\eta \rho'} = \delta_{lp} \delta_{\rho \rho'}$, we get  
\begin{equation} \label{eq:C5}
\begin{split}
& \sum_{l,p}\sum_{\rho, \rho'}\mathrm{Tr}\!\Big[(\delta_{i l}\delta_{i p} \delta_{\eta \rho} \delta_{\eta \rho'}   -s_{i l}^{\eta \rho \dagger}s_{i p}^{\eta \rho'}) \times(\delta_{j p}\delta_{j l} \delta_{\eta' \rho'} \delta_{\eta' \rho}\\&-s_{j p}^{\eta' \rho' \dagger}s_{j l}^{\eta' \rho})\Big]= 2\delta_{j i}\delta_{\eta \eta'} 
- (s_{i j}^{\eta \eta' \dagger}s_{i j}^{\eta \eta'}) 
- (s_{j i}^{\eta \eta' \dagger}s_{j i}^{\eta \eta'}) \\& = 2\delta_{ji} \delta_{\eta \eta'} - T_{ij}^{\eta \eta'} - T_{ji}^{\eta \eta'}.
\end{split}
\end{equation}

Where, we define $(s_{ij}^{\eta \eta' \dagger}s_{ij}^{\eta \eta'})  = T_{ij}^{\eta \eta'}$. Thus, for cross-correlations $( i \neq j)$, the thermal noise contribution reads, when all reservoirs are at same chemical potential and temperature, 
\begin{equation} \label{eq:C6}
S_{i j}^{\eta \eta', eq} = -G_0\int dE \, f(E)(1-f(E)) \, \big(T_{i j}^{\eta \eta'} + T_{j i}^{\eta \eta'}\big).
\end{equation}

For the more general case where contacts $i$ and $j$ are at different temperatures but same chemical potential, Eq.~(\ref{eq:C5}) reduces to  
\begin{equation} \label{eq:C7}
\begin{split}
S_{i j}^{\eta \eta', eq} =& -G_0\int dE \Big[T_{i j}^{\eta \eta'} f_{j}(E)(1-f_{j}(E)) \\&
+ T_{j i}^{\eta \eta'} f_{i}(E)(1-f_{i}(E))\Big].
\end{split}
\end{equation}

The spin-summed thermal noise contribution is therefore  
\begin{equation} \label{eq:C8}
S_{i j}^{eq} = \sum_{s, s' = \uparrow,\downarrow} S_{i j}^{\eta \eta', eq}.
\end{equation}

For autocorrelations $(i=j)$ in QSH setup, one obtains,
\begin{equation} \label{eq:C9}
S_{i i}^{\eta \eta', eq} = 2 G_0\int dE \, f_{i}(E)(1-f_{i}(E)) \,
\big(\delta_{\eta \eta'} - T_{i i}^{\eta \eta'}\big).
\end{equation}

\medskip

To extract the {transport} part for the autocorrelation, the equilibrium noise-like contribution ($S_{ii}^{\eta \eta', eq}$) is subtracted from the full quantum noise ($S_{ii}^{\eta \eta'}$), given in Eq. (\ref{eq:C2}), i.e., $S_{ii}^{\eta \eta', tr} = S_{ii}^{\eta \eta'} - S_{ii}^{\eta \eta', th}$ and therefore after simplification, one gets  

\begin{flalign} \label{eq:C2a}
&S_{i i}^{\eta \eta', tr} = G_0\sum_{\rho, \rho' = \uparrow, \downarrow}\sum_{l,p}\int dE \bigg(
\Big[A_{l p}^{\rho \rho'}(i, \eta)
A_{p l}^{\rho' \rho}(i, \eta')\Big] \notag \\& \times \Big(f_{l}(E)  + f_p(E) - 2 f_l(E) f_p(E)\Big)
-2 f_{i}(E)(1-f_{i}(E)) \notag \\& \times 
\big(\delta_{\eta \eta'} - T_{i i}^{\eta \eta'}\big)\bigg),
\end{flalign}

Now, in Eq.~(\ref{eq:C2a}) terms like $\sum_{l, p}[A_{l p}^{\rho \rho'}(i, \eta)A_{p l}^{\rho' \rho}(i, \eta)] f_{p}(E)$ and $\sum_{l,p}[A_{l p}^{\rho \rho'}(i, \eta)A_{p l}^{\rho' \rho}(i, \eta)] f_l(E)$ appear and both of them are identical. Following the procedure used for the chiral case (see, Eq.~(\ref{eq:A11})), these can be simplified as,  
\begin{flalign}\label{eq:C11}
&\sum_{\rho, \rho'} \sum_{l, p}[A_{l p}^{\rho \rho'}(i, \eta)A_{p l}^{\rho' \rho}(i, \eta')] f_{p}(E) \notag \\&= [1 - 2s_{ii}^{\eta \eta' \dagger} s_{ii}^{\eta \eta'}]f_i(E) + \sum_p [s_{ip}^{\eta \eta'} s_{ip}^{\eta \eta' \dagger}]f_p(E)\notag \\&  = [1 - 2T_{ii}^{\eta \eta'}] f_i(E) + \sum_p T_{ip}^{\eta \eta'}f_p(E).
\end{flalign}

Thus, we obtain for $S_{ii}^{\eta \eta', tr}$, 
\begin{flalign} \label{eq:C10}
&S_{i i}^{\eta \eta', tr} = G_0\int dE \Bigg[ \sum_{l}T_{i l}^{\eta \eta'}\big(f_{l}(E) - f_{i}(E)\big) 
+ \delta_{\eta \eta'}f_{i}(E)^2 \notag \\ 
& - \sum_{l p}\sum_{\rho,\rho'} f_l(E) f_p(E)
\big(s_{i l}^{\eta \rho \dagger}s_{i p}^{\eta \rho'}s_{i p}^{\eta' \rho' \dagger}s_{i l}^{\eta' \rho}\big)\Bigg] .
\end{flalign}

Eq. (\ref{eq:C10}) can be further rewritten as

\begin{equation}
\begin{split} \label{eq:C100}
&S_{i i}^{\eta \eta', tr} = G_0\int dE \Bigg[ \sum_{l \neq i}T_{il}^{\eta \eta'}f_{l}(E) (1-f_{l}(E)) \\& - \sum_{l \neq i}T_{il}^{\eta \eta'}f_{i}(E) (1-f_{i}(E) \delta_{\eta \eta'})\Bigg] +  G_0\int dE \Bigg(\sum_{l \neq i} \sum_{\rho \neq \eta'}T_{il}^{\eta \rho}f_i(E)^2\\
& \delta_{\eta \eta'} + \sum_{\rho} T_{ii}^{\eta \rho} f_i(E)^2 \delta_{\eta \eta'}+  \sum_{l \neq i}T_{il}^{\eta \eta'}f_l(E)^2  - \sum_{l, p} \sum_{\rho \rho'} f_l(E) f_p(E) \\&(s_{il}^{\eta \rho \dagger} s_{ip}^{\eta \rho'} s_{ip}^{\eta' \rho' \dagger} s_{i l}^{\eta' \rho})\Bigg).
\end{split}
\end{equation}

Here, the terms proportional to $f_i(E) (1-f_i(E))$ in $S_{ii}^{\eta \eta',tr}$ contribute to the thermal noise and is added to Eq. (\ref{eq:B6}) and rest contributes to the shot noise. Therefore, the expression for thermal noise for autocorrelation $i = j$ is

\begin{equation}
    \begin{split}
         S_{i i}^{\text{th}} =& 2 G_0\int dE \, f_{i}(E)(1-f_{i}(E)) \,
\big(\delta_{\eta \eta'} - T_{i i}^{\eta \eta'}\big)  \\& + G_0 \int dE \bigg(\sum_{l \neq i}T_{il}^{\eta \eta'}f_{l}(E) (1-f_{l}(E)) \\& - \sum_{l \neq i}T_{il}^{\eta \eta'}f_{i}(E) (1-f_{i}(E) \delta_{\eta \eta'})\bigg),\\
     S_{ii}^{sh} =&    G_0\int dE \Bigg(\sum_{l \neq i} \sum_{\rho \neq \eta'}T_{il}^{\eta \rho}f_i(E)^2 \delta_{\eta \eta'} + \sum_{\rho} T_{ii}^{\eta \rho} f_i(E)^2 \delta_{\eta \eta'}\\& + \sum_{l \neq i}T_{il}^{\eta \eta'}f_l(E)^2 - \sum_{l, p} \sum_{\rho \rho'} f_l(E) f_p(E) (s_{il}^{\eta \rho \dagger} s_{ip}^{\eta \rho'} s_{ip}^{\eta' \rho' \dagger} s_{i l}^{\eta' \rho})\Bigg).
    \end{split}
\end{equation} 

For cross-correlations $(i\neq j)$, we define $S_{i j}^{\eta \eta', sh} = S_{i j}^{\eta \eta'} - S_{i j}^{\eta \eta', th}$, where $S_{ij}^{\eta \eta'}$ is defined in Eq. (\ref{eq:C2}) and $S_{ij}^{\eta \eta', th}$ is given in Eq. (\ref{eq:C7}). Thus,
\begin{flalign} \label{eq:C13a}
&S_{i j}^{\eta \eta', sh} = G_0\sum_{\rho, \rho' = \uparrow, \downarrow}\sum_{l,p}\int dE \bigg(
\Big[A_{l p}^{\rho \rho'}(i, \eta)
A_{p l}^{\rho' \rho}(j, \eta')\Big] \notag \\& \times \Big(f_{l}(E)  + f_p(E) - 2 f_l(E) f_p(E)\Big)
-2\Big[T_{i j}^{\eta \eta'} f_{j}(E)(1-f_{j}(E)) \notag \\&
+ T_{j i}^{\eta \eta'} f_{i}(E)(1-f_{i}(E))\Big],
\end{flalign}

The expression in Eq.~(\ref{eq:C13}) contains terms like $\sum_{l, p}[A_{l p}^{\rho \rho'}(i, \eta)A_{p l}^{\rho' \rho}(j, \eta')] f_{p}(E)$, and $\sum_{l,p}[A_{l p}^{\rho \rho'}(i, \eta)A_{p l}^{\rho' \rho}(j, \eta')] f_l(E)$. We find,  
\begin{flalign} \label{eq:C12}
&\sum_{l, p}[A_{l p}^{\rho \rho'}(i, \eta)A_{p l}^{\rho' \rho}(j, \eta')] f_{p}(E)
= \sum_{l,p}[A_{l p}^{\rho \rho'}(i, \eta)A_{p l}^{\rho' \rho}(j, \eta')] f_l(E) \notag  \\
&= -T_{i j}^{\eta \eta'}f_{j}(E) - T_{j i}^{\eta \eta'}f_{i}(E).
\end{flalign}

Substituting \ref{eq:C12} in \ref{eq:C13a} yields the shot noise cross-correlation ($S_{ij}^{\eta \eta', sh}$) to be,
\begin{flalign} \label{eq:C13}
\! \! S_{i j}^{\eta \eta', sh} = -G_0\int dE \sum_{l, p} f_{l}(E)f_{p}(E) 
\big(s_{i l}^{\eta \rho \dagger}s_{i p}^{\eta \rho'}s_{j p}^{\eta' \rho \dagger}s_{j l}^{\eta' \rho'}\big).
\end{flalign}

A calculation in a two-reservoir QSH setup in presence of a constriction shows that the shot noise cross correlation such as $S_{12}^{\eta \eta', sh}$ derived from \ref{eq:C13} is inconsistent unless two energy-dependent distribution functions, $f_a(E)$ and $f_b(E)$, are introduced, following the same reasoning was made for chiral case. Without this, constraints such as $S_{11}^{\eta \eta'} = -S_{12}^{\eta \eta'}$ are not satisfied. With $f_a(E)$ and $f_{b}(E)$ included, the cross-correlation $S_{12}^{\eta \eta', sh}$ becomes

\begin{flalign} 
S_{12}^{\eta \eta', sh} = &-G_0\int dE \sum_{l, p} \big(f_{l}(E)-f_a(E)\big)\big(f_{p}(E)-f_b(E)\big) \notag \\& \times
\big(s_{1 l}^{\eta \rho \dagger}s_{1 p}^{\eta \rho'}s_{2 p}^{\eta' \rho \dagger}s_{2 l}^{\eta' \rho'}\big).
\end{flalign}

Now, when we set both $f_a(E)$ and $f_b(E)$ equal to $f_2(E) = \dfrac{1}{1+e^{\tfrac{E}{k_B T_2}}}$, the above consistency condition $S_{11}^{\eta \eta', sh} = -S_{12}^{\eta \eta', sh}$ is satisfied. Here, $T_2 = \bar{T} - \tfrac{\Delta T}{2}$ with $\Delta T$ being the temperature bias applied across the system. Now, in a generic multiterminal QSH setup, $S_{ii}^{\eta \eta', sh} = -\sum_{j\neq i} S_{ij}^{\eta \eta', sh}$ should be satisfied. Therefore, the modified expression for $S_{ij}^{\eta \eta', sh}$ is,

\begin{flalign} \label{eq:C14}
S_{i j}^{\eta \eta', sh} = &-G_0\int dE \sum_{l, p} \big(f_{l}(E)-f_a(E)\big)\big(f_{p}(E)-f_b(E)\big)\notag  \\& \times
\big(s_{i l}^{\eta \rho \dagger}s_{i p}^{\eta \rho'}s_{j p}^{\eta' \rho \dagger}s_{j l}^{\eta' \rho'}\big).
\end{flalign}

The {$\Delta_T$ noise} cross correlation ($i \neq j$) is calculated from shot noise at zero voltage bias but at finite temperature bias, using (\ref{eq:C14}), to obtain, 
\begin{flalign}\label{eq:C15}
\Delta_T^{i j, \eta \eta'} = &-G_0\int dE \sum_{l, p}\sum_{\rho,\rho'} 
\big(f_{l}(E)-f_a(E)\big)\big(f_{p}(E)-f_b(E)\big) \notag \\
&\times \big(s_{i l}^{\eta \rho \dagger}s_{i p}^{\eta \rho'}s_{j p}^{\eta' \rho' \dagger}s_{j l}^{\eta' \rho}\big),
\end{flalign}
and for auto-correlation $(i = j)$ using (\ref{eq:C10}), we obtain,
\begin{equation}
\begin{split}\label{eq:C16}
\Delta_T^{i i, \eta \eta'} = &\, G_0\int dE \Bigg(\sum_{l \neq i} \sum_{\rho \neq \eta'}T_{il}^{\eta \rho}f_i(E)^2 \delta_{\eta \eta'} + \sum_{\rho} T_{ii}^{\eta \rho} f_i(E)^2 \delta_{\eta \eta'}\\&+  \sum_{l \neq i}T_{il}^{\eta \eta'}f_l(E)^2 - \sum_{l, p} \sum_{\rho \rho'} f_l(E) f_p(E) (s_{il}^{\eta \rho \dagger} s_{ip}^{\eta \rho'} s_{ip}^{\eta' \rho' \dagger} s_{i l}^{\eta' \rho})\Bigg).
\end{split}
\end{equation}

\begingroup

\section{Derivation of $s$-matrices for the QH and QSH setups}
\label{App_C}

In this section, we derive the $s-$matrices for the QH setup with chiral edge mode and QSH setup with both helical and trivial edge modes. First, we discuss about the QH setup, then we focus on the QSH setup.

\subsection{Derivation of $s$-matrix for the QH setup with chiral edge modes}
\label{App_C1}

The general $s$-matrix for a generic 4-terminal QH setup is
\begin{equation}
s = \begin{pmatrix}
s_{11} & s_{12} & s_{13} & s_{14}\\
s_{21} & s_{22} & s_{23} & s_{24}\\
s_{31} & s_{32} & s_{33} & s_{34}\\
s_{41} & s_{42} & s_{43} & s_{44}
\end{pmatrix}.
\end{equation}

Here, $s_{ij}$ denotes the scattering amplitude for an electron entering from terminal $j \in \{1,2,3,4\}$ and exiting through terminal $i \in \{1,2,3,4\}$. A constriction is introduced in the QH bar, characterized by transmission and reflection amplitudes $\tau$ and $\rho$, as shown in Fig.~\ref{fig:1}. From Fig.~\ref{fig:1}, it is clear that an electron entering terminal 1 cannot return to terminal 1, which implies $s_{11}=0$. In the same way, $s_{22}, s_{33}, s_{44}$ also vanish. For the other amplitudes such as $s_{13}$, i.e., the amplitude for an electron to scatter from terminal 3 to terminal 1, is zero because there is no physical path for such a process. Similarly $s_{21}, s_{24}, s_{31}, s_{42}, s_{43}$ are zero as well because there is no path for such a process, as there are chiral edge modes.

The remaining nonzero scattering amplitudes are $s_{12}, s_{14}, s_{23}, s_{32}, s_{34}, s_{41}$. $s_{41}$ is finite even without scattering from the constriction, because an electron traveling from terminal 1 to terminal 4 does not encounter the constriction at all and reaches terminal 4 with probability 1. $s_{41}$ is thus independent of $\tau$ and $\rho$ and we write $s_{41}=e^{i\chi_1}$, where $\chi_1$ is the propagation phase acquired along the trajectory from terminal 1 to terminal 4. Similarly, an electron traveling from terminal 3 to terminal 2 acquires a propagation phase $\chi_2$, so $s_{23}=e^{i\chi_2}$, and it also doesnot encounter the constriction.

We denote the remaining amplitudes as $s_{12}=\tau e^{i\theta_t}$, $s_{14}=\rho e^{i\theta_r}$, $s_{32}=\rho e^{i\theta_r}$, and $s_{34}=\tau e^{i\theta_t}$. For example, $s_{12}$ corresponds to transmission from terminal 2 to terminal 1 through the constriction, i.e., $\tau e^{i\theta_t}$, with $\theta_t$ the transmission phase. Likewise, $s_{14}$ describes reflection from terminal 4 to terminal 1, giving $\rho e^{i\theta_r}$, where $\theta_r$ is the reflection phase. Thus, the $s$-matrix for the four-terminal QH setup with a constriction is

\begin{equation}
s = \begin{pmatrix}
0 & \tau e^{i\theta_t} & 0 & \rho e^{i\theta_r}\\
0 & 0 & e^{i\chi_2} & 0\\
0 & \rho e^{i\theta_r} & 0 & \tau e^{i\theta_t}\\
e^{i\chi_1} & 0 & 0 & 0
\end{pmatrix}.
\end{equation}

Unitarity of the $s$-matrix, $s^\dagger s = I$, requires $\theta_r = \theta_t - \frac{\pi}{2}$. Choosing $\theta_t=\alpha$ gives $\theta_r=\alpha - \frac{\pi}{2}$ and this reduces the $s$-matrix to be

\begin{equation} \label{eq:C12}
s = \begin{pmatrix}
0 & \tau e^{i\alpha} & 0 & -i \rho e^{i\alpha}\\
0 & 0 & e^{i\chi_2} & 0\\
0 & -i \rho e^{i\alpha} & 0 & \tau e^{i\alpha}\\
e^{i\chi_1} & 0 & 0 & 0
\end{pmatrix}.
\end{equation}

\subsection{Derivation of $s-$matrix for the QSH setup}
\label{App_C2}

The general form of the $s$-matrix for the QSH setup shown in Fig.~\ref{fig:2} is

\begin{equation}
s = \begin{pmatrix}
s_{11}^{\uparrow \uparrow} & s_{11}^{\uparrow \downarrow} & s_{12}^{\uparrow \uparrow} & s_{12}^{\uparrow \downarrow} & s_{13}^{\uparrow \uparrow} & s_{13}^{\uparrow \downarrow} & s_{14}^{\uparrow \uparrow} & s_{14}^{\uparrow \downarrow} \\
s_{11}^{\downarrow \uparrow} & s_{11}^{\downarrow \downarrow} & s_{12}^{\downarrow \uparrow} & s_{12}^{\downarrow \downarrow} & s_{13}^{\downarrow \uparrow} & s_{13}^{\downarrow \downarrow} & s_{14}^{\downarrow \uparrow} & s_{14}^{\downarrow \downarrow} \\
s_{21}^{\uparrow \uparrow} & s_{21}^{\uparrow \downarrow} & s_{22}^{\uparrow \uparrow} & s_{22}^{\uparrow \downarrow} & s_{23}^{\uparrow \uparrow} & s_{23}^{\uparrow \downarrow} & s_{24}^{\uparrow \uparrow} & s_{24}^{\uparrow \downarrow} \\
s_{21}^{\downarrow \uparrow} & s_{21}^{\downarrow \downarrow} & s_{22}^{\downarrow \uparrow} & s_{22}^{\downarrow \downarrow} & s_{23}^{\downarrow \uparrow} & s_{23}^{\downarrow \downarrow} & s_{24}^{\downarrow \uparrow} & s_{24}^{\downarrow \downarrow} \\
s_{31}^{\uparrow \uparrow} & s_{31}^{\uparrow \downarrow} & s_{32}^{\uparrow \uparrow} & s_{32}^{\uparrow \downarrow} & s_{33}^{\uparrow \uparrow} & s_{33}^{\uparrow \downarrow} & s_{34}^{\uparrow \uparrow} & s_{34}^{\uparrow \downarrow} \\
s_{31}^{\downarrow \uparrow} & s_{31}^{\downarrow \downarrow} & s_{32}^{\downarrow \uparrow} & s_{32}^{\downarrow \downarrow} & s_{33}^{\downarrow \uparrow} & s_{33}^{\downarrow \downarrow} & s_{34}^{\downarrow \uparrow} & s_{34}^{\downarrow \downarrow} \\
s_{41}^{\uparrow \uparrow} & s_{41}^{\uparrow \downarrow} & s_{42}^{\uparrow \uparrow} & s_{42}^{\uparrow \downarrow} & s_{43}^{\uparrow \uparrow} & s_{43}^{\uparrow \downarrow} & s_{44}^{\uparrow \uparrow} & s_{44}^{\uparrow \downarrow} \\
s_{41}^{\downarrow \uparrow} & s_{41}^{\downarrow \downarrow} & s_{42}^{\downarrow \uparrow} & s_{42}^{\downarrow \downarrow} & s_{43}^{\downarrow \uparrow} & s_{43}^{\downarrow \downarrow} & s_{44}^{\downarrow \uparrow} & s_{44}^{\downarrow \downarrow} 
\end{pmatrix}.
\end{equation}

Here, $s_{ij}^{\eta \eta'}$ denotes the scattering amplitude for an electron entering from terminal $j\in \{1,2,3,4\}$ with spin $\eta' \in {\uparrow,\downarrow}$ to exit at terminal $i\in \{1,2,3,4\}$ with spin $\eta \in {\uparrow,\downarrow}$. As in the QH case, we introduce a constriction characterized by transmission amplitude $\tau$ and reflection amplitude $\rho$.

From Fig.~\ref{fig:2}, we see that $s_{11}^{\uparrow\uparrow}, s_{11}^{\downarrow\uparrow}$ vanish because an electron incident from terminal 1 cannot reflect back to the same terminal with the same spin. Likewise, $s_{11}^{\uparrow\downarrow}$ and $s_{11}^{\downarrow\downarrow}$ vanish because the helical edge modes forbid spin-flip scattering. By the same reasoning, all diagonal elements $s_{22}^{\eta\eta'}, s_{33}^{\eta\eta'}$ and $s_{44}^{\eta\eta'}$ are zero.

Amplitudes such as $s_{14}^{\downarrow\downarrow}$, $s_{23}^{\uparrow\uparrow}$, $s_{32}^{\downarrow\downarrow}$ and $s_{41}^{\uparrow\uparrow}$ remain finite since these correspond to propagation without encountering the constriction. We assign
$s_{23}^{\uparrow\uparrow} = e^{i\chi_2}$ and $s_{41}^{\uparrow\uparrow} = e^{i\chi_1}$, where $\chi_1$ ($\chi_2$) is the phase acquired by an up-spin electron travelling from terminal 1 to 4 (from 3 to 2).
Similarly, $s_{14}^{\downarrow\downarrow} = e^{-i\chi_1}$ and $s_{32}^{\downarrow\downarrow}=e^{-i\chi_2}$, where the minus sign arises because the down-spin mode propagates in the opposite direction, reversing the sign of the wavevector and thus the phase. All spin-flip amplitudes vanish due to the absence of spin-flip scattering in the helical QSH edges. The remaining finite amplitudes are
$s_{12}^{\uparrow\uparrow}, s_{14}^{\uparrow\uparrow}, s_{21}^{\downarrow\downarrow}, s_{23}^{\downarrow\downarrow}, s_{32}^{\uparrow\uparrow}, s_{34}^{\uparrow\uparrow}, s_{41}^{\downarrow\downarrow}, s_{43}^{\downarrow\downarrow}$.
We take
$s_{12}^{\uparrow\uparrow} = s_{21}^{\downarrow\downarrow} = s_{34}^{\uparrow\uparrow} = s_{43}^{\downarrow\downarrow} = \tau e^{i\theta_t}$,
and
$s_{14}^{\uparrow\uparrow} = s_{23}^{\downarrow\downarrow} = s_{32}^{\uparrow\uparrow} = s_{41}^{\downarrow\downarrow} = \rho e^{i\theta_r}$.

Therefore, the $s$-matrix for the QSH setup with helical edge modes is given by:

\begin{equation}
s = \begin{pmatrix}
0 & 0 & \tau e^{i \theta_t} & 0 & 0 & 0 & \rho e^{i \theta_r} & 0\\
0 & 0 & 0 & 0 & 0 & 0 & 0 & e^{-i \chi_1}\\
0 & 0 & 0 & 0 & e^{i \chi_2} & 0 & 0 & 0\\
0 & \tau e^{-i \theta_t} & 0 & 0 & 0 & \rho e^{-i \theta_r} & 0 & 0\\
0 & 0 & \rho e^{i \theta_r} & 0 & 0 & 0 & \tau e^{i \theta_t} & 0\\
0 & 0 & 0 & e^{-i \chi_2} & 0 & 0 & 0 & 0\\
e^{i \chi_1} & 0 & 0 & 0 & 0 & 0 & 0 & 0\\
0 & \rho e^{-i \theta_r} & 0 & 0 & 0 & \tau e^{-i \theta_t} & 0 & 0
\end{pmatrix}
\end{equation}

The unitarity condition $s^\dagger s = I$ imposes
$\theta_r = \theta_t - \frac{\pi}{2}$.
We set $\theta_t = \alpha$, which gives $\theta_r = \alpha - \frac{\pi}{2}$.

The scattering matrix for the setup in Fig.~\ref{fig:2} reduces to:

\begin{equation} \label{eq:C66}
\resizebox{1.00\hsize}{!}{$s=
\begin{pmatrix}
0 & 0 & \tau e^{i\alpha} & 0 & 0 & 0 & -i\rho e^{i\alpha} & 0\\
0 & 0 & 0 & 0 & 0 & 0 & 0 & e^{-i\chi_1}\\
0 & 0 & 0 & 0 & e^{i\chi_2} & 0 & 0 & 0\\
0 & \tau e^{-i\alpha} & 0 & 0 & 0 & i\rho e^{-i\alpha} & 0 & 0\\
0 & 0 & -i\rho e^{i\alpha} & 0 & 0 & 0 & \tau e^{i\alpha} & 0\\
0 & 0 & 0 & e^{-i\chi_2} & 0 & 0 & 0 & 0\\
e^{i\chi_1} & 0 & 0 & 0 & 0 & 0 & 0 & 0\\
0 & i\rho e^{-i\alpha} & 0 & 0 & 0 & \tau e^{-i\alpha} & 0 & 0
\end{pmatrix}.$}
\end{equation}

Similarly, the $s$-matrix for the QSH setup with trivial edge modes as in Fig. \ref{fig:5}, with spin-flip probability $P$, is given by:
\begin{widetext}

{
\begin{equation} \label{eq:C108}
s= \begin{pmatrix}
s_{11}^{\uparrow \uparrow} & s_{11}^{\uparrow \downarrow} & s_{12}^{\uparrow \uparrow} & s_{12}^{\uparrow \downarrow} & s_{13}^{\uparrow \uparrow} & s_{13}^{\uparrow \downarrow} & s_{14}^{\uparrow \uparrow} & s_{14}^{\uparrow \downarrow} \\
s_{11}^{\downarrow \uparrow} & s_{11}^{\downarrow \downarrow} & s_{12}^{\downarrow \uparrow} & s_{12}^{\downarrow \downarrow} & s_{13}^{\downarrow \uparrow} & s_{13}^{\downarrow \downarrow} & s_{14}^{\downarrow \uparrow} & s_{14}^{\downarrow \downarrow} \\
s_{21}^{\uparrow \uparrow} & s_{21}^{\uparrow \downarrow} & s_{22}^{\uparrow \uparrow} & s_{22}^{\uparrow \downarrow} & s_{23}^{\uparrow \uparrow} & s_{23}^{\uparrow \downarrow} & s_{24}^{\uparrow \uparrow} & s_{24}^{\uparrow \downarrow} \\
s_{21}^{\downarrow \uparrow} & s_{21}^{\downarrow \downarrow} & s_{22}^{\downarrow \uparrow} & s_{22}^{\downarrow \downarrow} & s_{23}^{\downarrow \uparrow} & s_{23}^{\downarrow \downarrow} & s_{24}^{\downarrow \uparrow} & s_{24}^{\downarrow \downarrow} \\
s_{31}^{\uparrow \uparrow} & s_{31}^{\uparrow \downarrow} & s_{32}^{\uparrow \uparrow} & s_{32}^{\uparrow \downarrow} & s_{33}^{\uparrow \uparrow} & s_{33}^{\uparrow \downarrow} & s_{34}^{\uparrow \uparrow} & s_{34}^{\uparrow \downarrow} \\
s_{31}^{\downarrow \uparrow} & s_{31}^{\downarrow \downarrow} & s_{32}^{\downarrow \uparrow} & s_{32}^{\downarrow \downarrow} & s_{33}^{\downarrow \uparrow} & s_{33}^{\downarrow \downarrow} & s_{34}^{\downarrow \uparrow} & s_{34}^{\downarrow \downarrow} \\
s_{41}^{\uparrow \uparrow} & s_{41}^{\uparrow \downarrow} & s_{42}^{\uparrow \uparrow} & s_{42}^{\uparrow \downarrow} & s_{43}^{\uparrow \uparrow} & s_{43}^{\uparrow \downarrow} & s_{44}^{\uparrow \uparrow} & s_{44}^{\uparrow \downarrow} \\
s_{41}^{\downarrow \uparrow} & s_{41}^{\downarrow \downarrow} & s_{42}^{\downarrow \uparrow} & s_{42}^{\downarrow \downarrow} & s_{43}^{\downarrow \uparrow} & s_{43}^{\downarrow \downarrow} & s_{44}^{\downarrow \uparrow} & s_{44}^{\downarrow \downarrow} 
\end{pmatrix}=
\begin{pmatrix}
0 & -i\sqrt{ P} e^{i\xi} & \tau e^{i \alpha}x & 0 & 0 & 0 & -i\rho e^{i \alpha}x & 0\\
-i\sqrt{ P} & 0 & 0 & 0 & 0 & 0 & 0 & x \, e^{-i\chi_1}\\
0 & 0 & 0 & -i\sqrt{ P} & x \, e^{i\chi_2} & 0 & 0 & 0\\
0 & \tau e^{-i \alpha}x & -i\sqrt{ P}e^{-i\xi} & 0 & 0 & i\rho e^{-i \alpha}x & 0 & 0\\
0 & 0 & -i\rho e^{i \alpha}x & 0 & 0 & -i\sqrt{ P}e^{i\xi} & \tau e^{i \alpha}x & 0\\
0 & 0 & 0 & x \, e^{-i\chi_2} & -i\sqrt{ P} & 0 & 0 & 0\\
x \, e^{i\chi_1} & 0 & 0 & 0 & 0 & 0 & 0 & -i\sqrt{ P}\\
0 & i\rho e^{-i \alpha}x & 0 & 0 & 0 & \tau e^{-i \alpha}x & -i\sqrt{ P}e^{-i\xi} & 0
\end{pmatrix},
\end{equation}}
\end{widetext}

where, $x = \sqrt{1 -  P}, \xi = \alpha - \theta$ with $\theta = \text{Tan}^{-1}\left(\frac{\rho}{\tau}\right)$. {The derivation of the $s$-matrix with the trivial helical edge modes can be explained by one simple case, this explanation holds for other cases too. We see that the $s-$matrix element $s_{14}^{\uparrow\uparrow}$ takes the form $-i\rho\, e^{i\alpha} x$ in Eq. (\ref{eq:C108}). The physical scattering path associated with this process is illustrated by the blue dashed line connecting terminal~4 to terminal~1 in Fig.~\ref{fig:5}. First the spin-up electron has to reflect from the constriction with amplitude $-i \rho e^{i \alpha}$ and the amplitude for no spin-flip scattering is $x = \sqrt{1 - P}$. Therefore, one can multiply both the amplitudes in order to derive $s_{14}^{\uparrow \uparrow}$. One can understand other scattering processes in a similar way.}

\section{Derivation of $\Delta_T$ noise with energy-dependent scattering via QPC in QH and QSH setups}
\label{App_E}

In this section, we derive the $\Delta_T$ noise with energy-dependent scattering via QPC in QH and QSH setups in detail. First, we focus on the $\Delta_T$ noise for QH setup, then focus on the same derivation with QSH setup.

\subsection{Derivation of $\Delta_T$ noise in QH setup}
\label{App_E1}

In this section, we derive the $\Delta_T$ noise in QH setup as shown in Fig. \ref{fig:1} with energy-dependent scattering via QPC at their respective thermovoltages, which $s$-matrix is already given in Eq. (\ref{eq:121}) in Sec. \ref{analysis} B 1. We will first evaluate $\Delta_T^{22}$ in QH setup by imposing $\langle I_2 \rangle = 0$.  
The total charge current entering through terminal 2 is $\langle I_2 \rangle = \sum_{j} G_{2j} V_j + L_{2j} \tau_j$. From the definition of $L_{ij}$ in Eq. (\ref{eq:125}), we note that $L_{ij}$ is nonzero only when $T_{ij}$ is an asymmetric function of $E$. We observe that $L_{22}$, $L_{23}$ and $L_{24}$ vanish as for $L_{22}$ the term inside brackets in Eq. (\ref{eq:125}) reduces to $1-T_{22} =1$, as $T_{22} = 0$. In $L_{23}$, term in brackets is $-T_{23} = -1$ as $T_{23} = 1$ from Eq. (\ref{eq:121}), and for $L_{24}$, it vanishes, as term in bracket is: $T_{24} = 0$. This follows directly from the $s$-matrix (\ref{eq:121}). Therefore, the charge current in terminal 2 is $
\langle I_{2}\rangle 
=G_{22}V_2 +G_{24}V_4$. We consider $V_2 = 0$ and $V_4 = V$ for setup 1, which implies $\langle I_2 \rangle = G_{24} V$. This implies $\langle I_2 \rangle = 0$ only at $V = 0$, which implies the thermovoltage $V_{\text{th}}^{22} = 0$. We can derive the expression for $\Delta_T^{22}$ at zero thermovoltage and we find that it vanishes completely. The expression for $\Delta_T^{22}$, using Eq. (\ref{eq:11}), is 

\begin{equation}\label{eq:135}
\begin{split}
    \Delta_T^{22} =& 2G_0\int_{-\infty}^{\infty} dE \Bigg(\sum_{l} T_{2 l}f_{l}(E)^2
- \sum_{l,p} f_{l}(E)f_{p}(E)\\&\times\Big(s_{2 l}^{\dagger}s_{2 p}s_{2 p}^{\dagger}s_{2 l}\Big)\Bigg).
\end{split}
\end{equation}

We realize in Eq. (\ref{eq:135}), the terms with $l,p=1,2,4$ do not survive because from Eq. (\ref{eq:121}), it is evident that $s_{21}$, $s_{22}$ and $s_{24}$ are zero meaning the respective scattering probabilities such as $T_{21}$, $T_{22}$ and $T_{24}$ vanish, thus Eq. (\ref{eq:135}) reduces to,
\begin{equation}\label{eq:E2}
     \Delta_T^{22} = 2G_0 \int_{-\infty}^{\infty}dE \left(T_{23}f_3(E)^2 - (|s_{23}|^2 |s_{23}|^2) f_3(E)^2\right).
\end{equation}

Since $|s_{23}|^2 = T_{23} = 1$, $\Delta_T^{22}$ vanishes.
Similarly, we can calculate $\Delta_T^{44}$, imposing $\langle I_4 \rangle = 0$. The expression for $\langle I_4 \rangle$ is $\sum_{j} G_{4j} V_j + L_{4j} \tau_j$. From Eqs. (\ref{eq:121}) and (\ref{eq:125}), the expression for $L_{41}$ is,

\begin{equation}
    L_{41} = \frac{2e}{h\bar{T}}\int_{-\infty}^{\infty}dE\,(-T_{41})\,E\left(\frac{-\partial f}{\partial E}\right).
\end{equation}
From Eq. (\ref{eq:121}), we see that $T_{41} = 1$, which makes the integrand of $L_{41}$ an odd function in $E$ and therefore, $L_{41}$ vanishes. Similarly, $L_{42}, L_{43}$ vanish as $T_{42} = T_{43} = 0$ and for $L_{44} $, it is 1 as $T_{44} = 0$. Therefore, the expression for $\langle I_4 \rangle$ reduces to $G_{41} V + G_{44} V$, as $G_{42}$ and $G_{43}$ vanish, which follows directly from Eq. (1) and this implies $\langle I_4 \rangle = 0$ only when thermovoltage is zero, i.e., $V_{44}^{\text{th}} = V = 0$. The expression for $\Delta_T^{44}$ from Eq. (\ref{eq:11}) is then given as:

\begin{equation}
\begin{split}
    \Delta_T^{44} =& 2G_0\int_{-\infty}^{\infty} dE \Bigg(\sum_{l} T_{4 l}f_{l}(E)^2
- \sum_{l,p} f_{l}(E)f_{p}(E)\\&\times\Big(s_{4 l}^{\dagger}s_{4 p}s_{4 p}^{\dagger}s_{4 l}\Big)\Bigg),
\end{split}
\end{equation}

The terms with $l,p=2,3,4$ do not survive because from Eq. (\ref{eq:121}), we see that $s_{42} = s_{43} = s_{44} = 0$ meaning the corresponding probabilities are $T_{42}= T_{43} = T_{44} = 0$. The final expression for $\Delta_T^{44}$ is
\begin{equation}\label{eq:E5}
     \Delta_T^{44} = 2G_0 \int_{-\infty}^{\infty}dE \left(T_{41}f_1(E)^2 - (|s_{41}|^2 |s_{41}|^2) f_1(E)^2\right)
\end{equation}
Since, $|s_{41}|^2 = T_{41} = 1$, $\Delta_T^{44}$ also vanishes for QH case.

Next, we consider setup 2, wherein we calculate $\Delta_T^{22}$ and $\Delta_T^{44}$. To calculate $\Delta_T^{22}$, we impose $\langle I_2 \rangle = 0$. Using Eqs. (\ref{eq:121}) and (\ref{eq:124}), we get $\langle I_2 \rangle = G_{21} V_1 + G_{22} V_2 + G_{23} V_3 + G_{24} V_4$, since the coefficients $L_{21}, L_{22}, L_{23}, L_{24}$ are zero, as discussed in the previous paragraph, and as $V_1 = V_3 = V$, $V_2 = V_4 = 0$ with $G_{21} = 0$ implying $\langle I_2 \rangle = 0 $ for $V = 0$, thus $V_{\text{th}}^{22} = V = 0$. Similarly, $\langle     I_4 \rangle$ = $G_{41} V$, which is zero at $V_{\text{th}}^{44} =V= 0$. Thus, as in setup 1, we find that $\Delta_T^{22}$ and $\Delta_T^{44}$ vanish here too using Eq. (\ref{eq:11}).

\subsection{Derivation of $\Delta_T$ noise with QSH setup.}
\label{App_E2}

In this section, we derive the $\Delta_T$ noise in QSH setup with helical edge modes as shown in Fig. \ref{fig:2} with energy-dependent scattering via QPC at their respective thermovoltages, which $s$-matrix is already given in Eq. (\ref{eq:129}) in Sec. \ref{analysis} B 2. Similar to QH setup, we first derive $\Delta_T^{22}$ by imposing $\langle I_2 \rangle = 0$. According to Eq. (\ref{eq:127}), the charge current in terminal 2 is $\langle  I_2^{\uparrow} + I_2^{\downarrow} \rangle = \sum_{j} \sum_{\rho} (G_{2j}^{\eta \rho}V_j + L_{2j}^{\eta \rho} \tau_j)$. The expression for $\langle I_2 \rangle$ using Eq. (\ref{eq:128}) is,

\begin{equation}
\label{eq:761}
\begin{split}
\langle I_2 \rangle &= \langle I_2^{\uparrow} + I_2^{\downarrow} \rangle =  G_{21}^{\uparrow \uparrow}V_1 + G_{22}^{\uparrow \uparrow}V_2 + G_{23}^{\uparrow \uparrow}V_3 + G_{24}^{\uparrow \uparrow}V_4 \\&+ L_{21}^{\uparrow \uparrow}\tau_1 + L_{22}^{\uparrow \uparrow}\tau_2 + L_{23}^{\uparrow \uparrow}\tau_3 + L_{24}^{\uparrow \uparrow}\tau_4 \\&+ G_{21}^{\downarrow \downarrow}V_1 + G_{22}^{\downarrow \downarrow}V_2 + G_{23}^{\downarrow \downarrow}V_3 + G_{24}^{\downarrow \downarrow}V_4 \\&+ L_{21}^{\downarrow \downarrow}\tau_1 + L_{22}^{\downarrow \downarrow}\tau_2 + L_{23}^{\downarrow \downarrow}\tau_3 + L_{24}^{\downarrow \downarrow}\tau_4.
\end{split}
\end{equation}
Using setup 1, wherein, $V_1 = V_4 = V, V_2 = V_3 = 0$ and $\tau_1 = \tau_4 = \Delta T/2$, while $\tau_2 = \tau_3 = -\Delta T/2$, we realize from Eqs. (\ref{eq:129}), that $G_{21}^{\uparrow \uparrow} = G_{24}^{\uparrow \uparrow} = L_{21}^{\uparrow \uparrow} = L_{22}^{\uparrow \uparrow} = L_{23}^{\uparrow \uparrow} = L_{24}^{\uparrow \uparrow} = G_{24}^{\downarrow \downarrow} = L_{22}^{\downarrow \downarrow} = L_{24}^{\downarrow \downarrow}$ are zero as $T_{21}^{\uparrow \uparrow} = T_{24}^{\uparrow \uparrow}, T_{22}^{\uparrow \uparrow} = 0, T_{24}^{\downarrow \downarrow}$ are zero. However, $L_{23}^{\uparrow \uparrow}$ is zero even though $T_{23}^{\uparrow \uparrow}$ is finite and equal to 1. From the definition of $L_{ij}^{\eta \rho}$ as in Eq. (\ref{eq:128}), the expression for $L_{23}^{\uparrow \uparrow}$ is $\frac{e}{h\bar{T}}\int_{-\infty}^{\infty}dE\,\,E\left(\frac{-\partial f}{\partial E}\right)$. The integrand is an odd function, which is why $L_{23}^{\uparrow \uparrow}$ is zero. Therefore,
 $\langle I_2 \rangle$ reduces to,

 \begin{equation}
\langle I_2 \rangle =  V G_{21}^{\downarrow \downarrow }+\frac{\Delta T L_{21}^{\downarrow \downarrow }}{2}-\frac{\Delta T L_{23}^{\downarrow \downarrow }}{2},
 \end{equation}
  which implies that thermovoltage is $V_{\text{th}}^{22} = V=  \frac{\Delta T L_{23}^{\downarrow \downarrow }-\Delta T L_{21}^{\downarrow \downarrow }}{2 G_{21}^{\downarrow \downarrow }}$. Using Eq. (\ref{eq:128}) and (\ref{eq:129}), $V_{\text{th}}^{22}$ can be written as $\left(\frac{\int dE (1-2 \mathcal{T})E\left(\frac{-\partial f}{\partial E}\right)}{\int dE \mathcal{T} \left(\frac{-\partial f}{\partial E}\right)}\right)\frac{\Delta T}{2e \bar{T}}$. We now calculate the $\Delta_T$ noise $\Delta_T^{22}$ at $V_{\text{th}}^{22}$, using Eqs. (\ref{eq:63}) and (\ref{eq:64}).

  Similarly, for the calculation of $\Delta_T^{44}$, we impose $\langle I_4 \rangle = 0$. Using Eqs. (\ref{eq:127}) and (\ref{eq:129}), we find that $\langle I_4 \rangle = V \left(G_{41}^{\downarrow \downarrow }+G_{41}^{\uparrow \uparrow }\right)+\frac{\Delta T L_{41}^{\downarrow \downarrow }}{2}-\frac{\Delta T L_{43}^{\downarrow \downarrow }}{2}+V \left(G_{44}^{\downarrow \downarrow }+G_{44}^{\downarrow \uparrow }+G_{44}^{\uparrow \downarrow }+G_{44}^{\uparrow \uparrow }\right)$, which implies $\langle I_4 \rangle$ is zero at thermovoltage, $V_{\text{th}}^{44} = \frac{\Delta T L_{43}^{\downarrow \downarrow }-\Delta T L_{41}^{\downarrow \downarrow }}{2 \left(G_{41}^{\downarrow \downarrow }+G_{41}^{\uparrow \uparrow }+G_{44}^{\downarrow \downarrow }+G_{44}^{\downarrow \uparrow }+G_{44}^{\uparrow \downarrow }+G_{44}^{\uparrow \uparrow }\right)}$. Using Eqs. (\ref{eq:128}) and (\ref{eq:129}), we realize that $V_{\text{th}}^{44}$ is exactly same as $V_{\text{th}}^{22}$. We observe that $V^{22}_{\text{th}}$ and $V^{44}_{\text{th}}$ also distinguishes between chiral and helical edge mode as it is finite for helical edge mode, whereas for chiral edge mode, they vanish. We plot $V_{\text{th}}^{22}$ (same as $V_{\text{th}}^{44}$) in Fig. \ref{fig:14} for all the values of QPC parameters $\frac{\hbar \omega}{k_B \bar{T}}$ and $\frac{E_1}{k_B \bar{T}}$.

Thus, $\Delta_T$ noise autocorrelations such as $\Delta_T^{22}, \Delta_T^{44}$ are identical and can be derived from Eqs. (\ref{eq:63}) and (\ref{eq:64}). For $i = 2$ and $\eta = \eta' = \uparrow$, we have,
\begin{equation} \label{eq:165}
\begin{split}
\Delta_T^{22,\uparrow \uparrow} =& G_0\int dE \Bigg(\sum_{l \neq i} \sum_{\rho \neq \uparrow}T_{2l}^{\uparrow \rho}f_2(E)^2  + \sum_{\rho} T_{22}^{\uparrow \rho} f_2(E)^2 \\& + \sum_{l \neq 2}T_{2l}^{\uparrow \uparrow}f_l(E)^2 - \sum_{l, p} \sum_{\rho \rho'} f_l(E) f_p(E) (s_{2l}^{\uparrow \rho \dagger} s_{2p}^{\uparrow \rho'} s_{2p}^{\uparrow \rho' \dagger} s_{2 l}^{\uparrow \rho})\Bigg).
\end{split}
\end{equation}

We see that the terms with $l,p = \{1,2,4 \}$ do not contribute because from Eq. (\ref{eq:129}), $s_{21}^{\uparrow \uparrow} = s_{22}^{\uparrow \uparrow} = s_{24}^{\uparrow \uparrow} = 0$ and only $s_{23}^{\uparrow \uparrow}$ is finite, which implies $T_{23}^{\uparrow \uparrow} = 1$. Therefore,

\begin{equation}
\label{eq:791}
    \Delta_T^{22; \uparrow \uparrow} = G_0 \int dE \left(T_{23}^{\uparrow \uparrow}  f_3(E)^2 - |s_{23}|^2 |s_{23}|^2 f_{3}(E)^2\right). 
\end{equation}

Since, $|s_{23}^{\uparrow \uparrow}|^2 = T_{23}^{\uparrow \uparrow} = 1$, this makes $\Delta_T^{22; \uparrow \uparrow} = 0$. Similarly, $\Delta_T^{22; \uparrow \downarrow}$ and $\Delta_T^{22; \downarrow \uparrow}$ are zero since there is no spin-flip scattering present with topological helical edge modes. The expression for $\Delta_T^{22;\downarrow \downarrow}$ is then given as,

\begin{equation} \label{eq:166}
\begin{split}
\Delta_T^{22,\downarrow \downarrow} =& G_0\int dE \Bigg(\sum_{l \neq i} \sum_{\rho \neq \downarrow}T_{2l}^{\downarrow \rho}f_2(E)^2  + \sum_{\rho} T_{22}^{\downarrow \rho} f_2(E)^2 \\& + \sum_{l \neq 2}T_{2l}^{\downarrow \downarrow}f_l(E)^2 - \sum_{l, p} \sum_{\rho \rho'} f_l(E) f_p(E) (s_{2l}^{\downarrow \rho \dagger} s_{2p}^{\downarrow \rho'} s_{2p}^{\downarrow \rho' \dagger} s_{2 l}^{\downarrow \rho})\Bigg).
\end{split}
\end{equation}

We see from Eq. (\ref{eq:129}) that $s_{21}^{\downarrow \downarrow}$ and $s_{23}^{\downarrow \downarrow}$ are finite and $s_{22}^{\downarrow \downarrow}$ and $s_{24}^{\downarrow \downarrow}$ vanish. $\Delta_T^{22; \downarrow \downarrow}$ reduces to, using Eq. (\ref{eq:129}),

\begin{equation} 
\label{eq:811}
\begin{split}
    \Delta_T^{22; \downarrow \downarrow} &= G_0 \int dE (T_{21}^{\downarrow \downarrow} f_1(E)^2 - T_{21}^{\downarrow \downarrow 2} f_1(E)^2 + T_{23}^{\downarrow \downarrow} f_3(E)^2\\& - T_{23}^{\downarrow \downarrow 2} f_3(E)^2 - 2 T_{21}^{\downarrow \downarrow}T_{23}^{\downarrow \downarrow}f_1(E)f_3(E) ).
    \end{split}
\end{equation}

From Eq. (\ref{eq:129}), we see that $T_{21}^{\downarrow \downarrow} = \mathcal{T}$, $T_{23}^{\downarrow \downarrow} = 1 - \mathcal{T}$, and with $f_1(E) = f_H(E)$, $f_3(E) = f_C(E)$, we get, 

\begin{equation}\label{eq:130}
 \Delta_T^{22}=   \Delta_T^{22;\downarrow \downarrow} = G_0 \int_{-\infty}^{\infty}dE \mathcal{T}(1-\mathcal{T})(f_H(E) - f_C(E))^2.
\end{equation}

Therefore, $\Delta_T^{22}$ = $\Delta_T^{22; \downarrow \downarrow}$. Similarly, we can derive $\Delta_T^{44}$, wherein $\Delta_T^{44; \uparrow \uparrow}, \Delta_T^{44; \uparrow \downarrow}$, $\Delta_T^{44; \downarrow \uparrow}$ vanish, whereas $\Delta_T^{44; \downarrow \downarrow}$ is finite and exactly same as $\Delta_T^{22; \downarrow \downarrow}$.
We observe that both $\Delta_T^{22}$ and $\Delta_T^{44}$ clearly distinguish between chiral and helical edge modes, see Fig. \ref{fig:15}. In both the energy-dependent as well as energy-independent case, $\Delta_T^{22}$ and $\Delta_T^{44}$ vary quadratically with $\frac{\Delta T}{\bar{T}}$, but with distinct magnitudes.

Next, we consider setup 2 and calculate $\Delta_T^{22}$ and $\Delta_T^{44}$ by imposing $\langle I_2 \rangle$ as well as $\langle I_4 \rangle$ to be zero. The expression for $\langle I_2 \rangle$ can be obtained from Eq. (\ref{eq:761}) and imposing conditions for setup 2, i.e., $V_1 = V_3 = V, V_2 = V_4 = 0$ and $\tau_1 = \tau_3 = \frac{\Delta T}{2} $ and $\tau_2 = \tau_4 = -\frac{\Delta T}{2} $. Therefore, $\langle I_2 \rangle$ is $V G_{21}^{\downarrow \downarrow }+\frac{\Delta T L_{21}^{\downarrow \downarrow }}{2}+V \left(G_{23}^{\downarrow \downarrow }+G_{23}^{\uparrow \uparrow }\right)+\frac{\Delta T L_{23}^{\downarrow \downarrow }}{2}$, which gives thermovoltage $V_{\text{th}}^{22} =  \frac{-\Delta T L_{21}^{\downarrow \downarrow }-\Delta T L_{23}^{\downarrow \downarrow }}{2 \left(G_{21}^{\downarrow \downarrow }+G_{23}^{\downarrow \downarrow }+G_{23}^{\uparrow \uparrow }\right)}$. Following Eq. (\ref{eq:128}), the term inside brackets of $L_{21}^{\downarrow \downarrow}$ is $-T_{21}^{\downarrow \downarrow} = -\mathcal{T}$. Similarly, for $L_{23}^{\downarrow \downarrow}$, the term inside bracket is $-T_{23} = -(1-\mathcal{T})$. $L_{21}^{\downarrow \downarrow}$ and $L_{23}^{\downarrow \downarrow}$ are thus given as:

\begin{equation}
    \begin{split}
        L_{21}^{\downarrow \downarrow} &= \frac{e}{h\bar{T}}\int_{-\infty}^{\infty}dE\,(-\mathcal{T})\,E\left(\frac{-\partial f}{\partial E}\right), \\
        L_{23}^{\downarrow \downarrow} &= \frac{e}{h\bar{T}}\int_{-\infty}^{\infty}dE\,(-1)(1-\mathcal{T})\,E\left(\frac{-\partial f}{\partial E}\right).
    \end{split}
\end{equation}

The numerator of $V_{\text{th}}^{22}$ i.e., $-\Delta T (L_{21}^{\downarrow \downarrow} + L_{23}^{\downarrow \downarrow}) = (\Delta T) \frac{e}{h\bar{T}}\int_{-\infty}^{\infty}dE\,\,E\left(\frac{-\partial f}{\partial E}\right)$ is zero as integrand vanishes. Thus, $V_{\text{th}}^{22} = 0$. The general expression for $\Delta_T^{22; \uparrow \uparrow}$ is same as in Eq. (\ref{eq:165}) and Eq. (\ref{eq:791}), which is why $\Delta_T^{22; \uparrow \uparrow}$ is zero in setup 2 also. $\Delta_T^{22; \uparrow \downarrow}$ and $\Delta_T^{22; \downarrow \uparrow}$ are always zero as there is no spin-flip scattering present with helical edge modes. The expression for $\Delta_T^{22; \downarrow \downarrow}$ is exactly same as Eq. (\ref{eq:811}) and since in setup 2, we consider $f_1 (E) = f_3(E)$, the integrand of Eq. (\ref{eq:811}) becomes $(T_{21}^{\downarrow \downarrow}- T_{21}^{\downarrow \downarrow 2} + T_{23}^{\downarrow \downarrow} - T_{23}^{\downarrow \downarrow 2} - 2 T_{21}^{\downarrow \downarrow}T_{23}^{\downarrow \downarrow})f_1(E)^2$. Using $T_{21}^{\downarrow \downarrow} = 1 - T_{23}^{\downarrow \downarrow}$, we see that $\Delta_T^{22; \downarrow \downarrow}$ too vanishes. Therefore, in setup 2, $\Delta_T^{22}$ vanishes. Similarly, $\Delta_T^{44}$ also vanishes at zero $V_{\text{th}}^{44}$. We observe that in setup 2, the distinction between chiral and topological helical edge mode is not possible as in both cases $\Delta_T^{22}$ and $\Delta_T^{44}$ vanish.

\subsection{Derivation of $\Delta_T$ noise in trivial QSH setup}
\label{App_E3}

In this section, we derive the $\Delta_T$ noise in QSH setup with trivial edge modes as shown in Fig. \ref{fig:5} with energy-dependent scattering via QPC at their respective thermovoltages, which $s$-matrix is already given in Eq. (\ref{eq:131}) in Sec. \ref{analysis} B 3. Herein also, we calculate $\Delta_T^{22}$ and $\Delta_T^{44}$ imposing $\langle I_2 \rangle$ and $\langle I_4 \rangle$ to zero and at thermovoltages $V_{\text{th}}^{22}$ and $V_{\text{th}}^{44}$ respectively. We only consider setup 2 for the calculation of $\Delta_T^{22}$ and $\Delta_T^{44}$ with trivial edge modes. Using Eq.~(\ref{eq:127}), the expression for $\langle I_2 \rangle$ is 

\begin{equation}
    \begin{split}
        \langle I_2 \rangle &= V_1 \bigg(G_{21}^{\downarrow \downarrow }+G_{21}^{\downarrow \uparrow }+G_{21}^{\uparrow \downarrow }+G_{21}^{\uparrow \uparrow }\bigg)+\tau _1 \bigg(L_{21}^{\downarrow \downarrow }+L_{21}^{\downarrow \uparrow } \\&+L_{21}^{\uparrow \downarrow }+L_{21}^{\uparrow \uparrow }\bigg)+V_2 \bigg(G_{22}^{\downarrow \downarrow }+G_{22}^{\downarrow \uparrow }+G_{22}^{\uparrow \downarrow }+G_{22}^{\uparrow \uparrow }\bigg)\\&+\tau _2 \bigg(L_{22}^{\downarrow \downarrow }+L_{22}^{\downarrow \uparrow }+L_{22}^{\uparrow \downarrow }+L_{22}^{\uparrow \uparrow }\bigg)+V_3 \bigg(G_{23}^{\downarrow \downarrow }+G_{23}^{\downarrow \uparrow }\\&+G_{23}^{\uparrow \downarrow }+G_{23}^{\uparrow \uparrow }\bigg)+\tau _3 \bigg(L_{23}^{\downarrow \downarrow }+L_{23}^{\downarrow \uparrow }+L_{23}^{\uparrow \downarrow }+L_{23}^{\uparrow \uparrow }\bigg)\\&+V_4 \bigg(G_{24}^{\downarrow \downarrow }+G_{24}^{\downarrow \uparrow }+G_{24}^{\uparrow \downarrow }+G_{24}^{\uparrow \uparrow }\bigg)+\tau _4 \bigg(L_{24}^{\downarrow \downarrow }+L_{24}^{\downarrow \uparrow }\\&+L_{24}^{\uparrow \downarrow }+L_{24}^{\uparrow \uparrow }\bigg).
    \end{split}
\end{equation}

Using setup 2, wherein, $V_1 = V_3 = V, V_2 = V_4 = 0$ and $\tau_1 = \tau_3 = \frac{\Delta T}{2}$ and $\tau_2 = \tau_4 = \frac{-\Delta T}{2}$, aforementioned equation for $\langle I_2 \rangle$ reduces to

\begin{equation}
    \begin{split}
        \langle I_2 \rangle &= V \bigg(G_{21}^{\downarrow \downarrow }+G_{21}^{\downarrow \uparrow }+G_{21}^{\uparrow \downarrow }+G_{21}^{\uparrow \uparrow }\bigg)+\frac{\Delta T}{2} \bigg(L_{21}^{\downarrow \downarrow }+L_{21}^{\downarrow \uparrow }\\&+L_{21}^{\uparrow \downarrow }+L_{21}^{\uparrow \uparrow }\bigg)-\frac{\Delta T}{2} \bigg(L_{22}^{\downarrow \downarrow }+L_{22}^{\downarrow \uparrow }+L_{22}^{\uparrow \downarrow }+L_{22}^{\uparrow \uparrow }\bigg)\\&+V \bigg(G_{23}^{\downarrow \downarrow }+G_{23}^{\downarrow \uparrow }+G_{23}^{\uparrow \downarrow }+G_{23}^{\uparrow \uparrow }\bigg)+\frac{\Delta T}{2} \bigg(L_{23}^{\downarrow \downarrow }+L_{23}^{\downarrow \uparrow }\\&+L_{23}^{\uparrow \downarrow }+L_{23}^{\uparrow \uparrow }\bigg)-\frac{\Delta T}{2} \bigg(L_{24}^{\downarrow \downarrow }+L_{24}^{\downarrow \uparrow }+L_{24}^{\uparrow \downarrow }+L_{24}^{\uparrow \uparrow }\bigg).
    \end{split}
\end{equation}
From Eqs. (\ref{eq:128}) and (\ref{eq:131}), we realize that $G_{21}^{\uparrow \uparrow}= G_{21}^{\uparrow \downarrow}= G_{21}^{\downarrow \uparrow}= L_{21}^{\uparrow \uparrow}= L_{21}^{\uparrow \downarrow}= L_{21}^{\downarrow \uparrow}  = G_{23}^{\uparrow \downarrow}= G_{23}^{\downarrow \uparrow}= L_{23}^{\uparrow \uparrow}= L_{23}^{\uparrow \downarrow}= L_{23}^{\downarrow \uparrow} = L_{22}^{\uparrow \uparrow} = L_{22}^{\uparrow \downarrow} = L_{22}^{\downarrow \uparrow} = L_{22}^{\downarrow \downarrow}= L_{24}^{\uparrow \uparrow} = L_{24}^{\uparrow \downarrow} = L_{24}^{\downarrow \uparrow} = L_{24}^{\downarrow \downarrow}=0 $, since $T_{21}^{\uparrow \uparrow} = T_{21}^{\uparrow \downarrow} = T_{21}^{\downarrow \uparrow} = T_{22}^{\uparrow \uparrow} = T_{22}^{\downarrow \downarrow} = T_{23}^{\uparrow \downarrow} = T_{23}^{\downarrow \uparrow} = T_{24}^{\uparrow \uparrow} = T_{24}^{\uparrow \downarrow} = T_{24}^{\downarrow \uparrow} = T_{24}^{\downarrow \downarrow} = 0$. The expression for $\langle I_2 \rangle$ is $V G_{21}^{\downarrow \downarrow }+\frac{\Delta T L_{21}^{\downarrow \downarrow }}{2}+V \left(G_{23}^{\downarrow \downarrow }+G_{23}^{\uparrow \uparrow }\right)+\frac{\Delta T L_{23}^{\downarrow \downarrow }}{2}$, which implies thermovoltage $V_{\text{th}}^{22}: \frac{-\Delta T L_{21}^{\downarrow \downarrow }-\Delta T L_{23}^{\downarrow \downarrow }}{2 \left(G_{21}^{\downarrow \downarrow }+G_{23}^{\downarrow \downarrow }+G_{23}^{\uparrow \uparrow }\right)}$. From Eq. (\ref{eq:128}), the term inside brackets of $L_{21}^{\downarrow \downarrow} = -T_{21}^{\downarrow \downarrow} = - \mathcal{T}x^2$ and for $L_{23}^{\downarrow \downarrow}$, it is $-T_{23}^{\downarrow \downarrow} = -(1 - \mathcal{T})x^2$. Hence, the expression for $L_{21}^{\downarrow \downarrow}$ and $L_{23}^{\downarrow \downarrow}$ are given as

\begin{equation}
    \begin{split}
        L_{21}^{\downarrow \downarrow} &= \frac{e}{h\bar{T}}\int_{-\infty}^{\infty}dE\,(-\mathcal{T})x^2\,E\left(\frac{-\partial f}{\partial E}\right), \\
        L_{23}^{\downarrow \downarrow} &= \frac{e}{h\bar{T}}\int_{-\infty}^{\infty}dE\,(-1)(1-\mathcal{T})x^2\,E\left(\frac{-\partial f}{\partial E}\right),
    \end{split}
\end{equation}

The numerator of $V_{\text{th}}^{22}$ is $-\Delta T (L_{21}^{\downarrow \downarrow} + L_{23}^{\downarrow \downarrow}) = x^2 (\Delta T) \frac{e}{h\bar{T}}\int_{-\infty}^{\infty}dE\,\,E\left(\frac{-\partial f}{\partial E}\right)$, which vanishes as integrand is zero. Therefore, $V_{\text{th}}^{22} = 0$. Similarly, for the calculation of $\Delta_T^{44}$, we impose $\langle I_4 \rangle  =0$ to determine $V_{\text{th}}^{44}$. In setup 2, the expression for $\langle I_4 \rangle$ is

\begin{equation}
    \begin{split}
        \langle I_4 \rangle &= V \bigg(G_{41}^{\downarrow \downarrow }+G_{41}^{\downarrow \uparrow }+G_{41}^{\uparrow \downarrow }+G_{41}^{\uparrow \uparrow }\bigg)+\frac{\Delta T}{2} \bigg(L_{41}^{\downarrow \downarrow }+L_{41}^{\downarrow \uparrow }\\&+L_{41}^{\uparrow \downarrow }+L_{41}^{\uparrow \uparrow }\bigg)-\frac{\Delta T}{2} \bigg(L_{42}^{\downarrow \downarrow }+L_{42}^{\downarrow \uparrow }+L_{42}^{\uparrow \downarrow }+L_{42}^{\uparrow \uparrow }\bigg)\\&+V \bigg(G_{43}^{\downarrow \downarrow }+G_{43}^{\downarrow \uparrow }+G_{43}^{\uparrow \downarrow }+G_{43}^{\uparrow \uparrow }\bigg)+\frac{\Delta T}{2} \bigg(L_{43}^{\downarrow \downarrow }+L_{43}^{\downarrow \uparrow }\\&+L_{43}^{\uparrow \downarrow }+L_{43}^{\uparrow \uparrow }\bigg)-\frac{\Delta T}{2} \bigg(L_{44}^{\downarrow \downarrow }+L_{44}^{\downarrow \uparrow }+L_{44}^{\uparrow \downarrow }+L_{44}^{\uparrow \uparrow }\bigg) 
    \end{split}
\end{equation}

From Eqs. (\ref{eq:128}) and (\ref{eq:131}), we realize that $G_{41}^{\uparrow \downarrow}= G_{41}^{\uparrow \downarrow}= L_{42}^{\uparrow \uparrow}= L_{42}^{\uparrow \downarrow}= L_{42}^{\downarrow \uparrow}= L_{42}^{\downarrow \downarrow}= G_{43}^{\uparrow \uparrow}  = G_{43}^{\uparrow \downarrow}= G_{43}^{\downarrow \uparrow}= L_{43}^{\uparrow \uparrow}  = L_{43}^{\uparrow \downarrow}= L_{43}^{\downarrow \uparrow} = L_{44}^{\uparrow \uparrow}= L_{44}^{\uparrow \downarrow}= L_{44}^{\downarrow \uparrow}= L_{44}^{\downarrow \downarrow}=0 $, since $T_{41}^{\uparrow \downarrow} = T_{41}^{\downarrow \uparrow} = T_{42}^{\uparrow \uparrow} = T_{42}^{\uparrow \downarrow}= T_{42}^{\downarrow \uparrow} = T_{42}^{\downarrow \downarrow} = T_{43}^{\uparrow \uparrow} =  T_{43}^{\uparrow \downarrow}=T_{43}^{\downarrow \uparrow} = T_{44}^{\uparrow \uparrow} = T_{44}^{\downarrow \downarrow} = T_{24}^{\downarrow \uparrow}$. The final expression for $\langle I_4 \rangle$ is $V G_{43}^{\downarrow \downarrow }+\frac{\Delta T L_{41}^{\downarrow \downarrow }}{2}+V \left(G_{41}^{\downarrow \downarrow }+G_{41}^{\uparrow \uparrow }\right)+\frac{\Delta T L_{43}^{\downarrow \downarrow }}{2}$, which implies thermovoltage $V_{\text{th}}^{44}= \frac{-\Delta T L_{41}^{\downarrow \downarrow }-\Delta T L_{43}^{\downarrow \downarrow }}{2 \left(G_{41}^{\downarrow \downarrow }+G_{43}^{\downarrow \downarrow }+G_{41}^{\uparrow \uparrow }\right)}$. Using Eq. (\ref{eq:128}), the expression for $L_{41}^{\downarrow \downarrow}$ and $L_{43}^{\downarrow \downarrow}$ are given as,
\begin{equation}
    \begin{split}
        L_{41}^{\downarrow \downarrow} &= \frac{e}{h\bar{T}}\int_{-\infty}^{\infty}dE\,(-1)(1-\mathcal{T})x^2\,E\left(\frac{-\partial f}{\partial E}\right), \\
        L_{43}^{\downarrow \downarrow} &= \frac{e}{h\bar{T}}\int_{-\infty}^{\infty}dE\,(-\mathcal{T})x^2\,E\left(\frac{-\partial f}{\partial E}\right).
    \end{split}
\end{equation}
The numerator of $V_{\text{th}}^{44}$ is $-\Delta T (L_{41}^{\downarrow \downarrow} + L_{43}^{\downarrow \downarrow}) = x^2 (\Delta T) \frac{e}{h\bar{T}}\int_{-\infty}^{\infty}dE\,\,E\left(\frac{-\partial f}{\partial E}\right)$, which again vanishes as integrand is zero.

Thus, $\Delta_T$ noise  $\Delta_T^{22}$ can be derived from Eqs. (\ref{eq:63}) and (\ref{eq:64}). For $i = 2$ and $\eta = \eta' = \uparrow$, $\Delta_T^{22; \uparrow \uparrow}$ for trivial helical edge modes is:
\begin{equation} \label{eq:1651}
\begin{split}
\Delta_T^{22,\uparrow \uparrow} =& G_0\int dE \Bigg(\sum_{l \neq i} \sum_{\rho \neq \uparrow}T_{2l}^{\uparrow \rho}f_2(E)^2  + \sum_{\rho} T_{22}^{\uparrow \rho} f_2(E)^2 \\& + \sum_{l \neq 2}T_{2l}^{\uparrow \uparrow}f_l(E)^2 - \sum_{l, p} \sum_{\rho \rho'} f_l(E) f_p(E) (s_{2l}^{\uparrow \rho \dagger} s_{2p}^{\uparrow \rho'} s_{2p}^{\uparrow \rho' \dagger} s_{2 l}^{\uparrow \rho})\Bigg).
\end{split}
\end{equation}

Using Eq. (\ref{eq:131}), $\Delta_T^{22; \uparrow \uparrow}$ reduces to

\begin{equation}
    \begin{split}
        \Delta_T^{22; \uparrow \uparrow} =&G_0 \int dE (T_{22}^{\uparrow \downarrow} f_2(E)^2 - T_{22}^{\uparrow \downarrow 2} f_2(E)^2 + T_{23}^{\uparrow \uparrow} f_3(E)^2\\& - T_{23}^{\uparrow \uparrow 2} f_3(E)^2 - 2 T_{22}^{\uparrow \downarrow}T_{23}^{\uparrow \uparrow}f_2(E)f_3(E) ).
    \end{split}
\end{equation}

From Eq. (\ref{eq:131}), we realize that $T_{22}^{\uparrow \downarrow} = P$ and $T_{23}^{\uparrow \uparrow} = 1-P$. In setup 2, we have considered $f_2(E) = f_C(E)$ and $f_3(E) = f_H(E)$, thus $\Delta_T^{22; \uparrow \uparrow}$ becomes,

\begin{equation}
    \Delta_T^{22;\uparrow \uparrow} = G_0 P(1-P)\int_{-\infty}^{\infty}dE (f_H(E) - f_C(E))^2.
\end{equation}

Similarly, the expression for $\Delta_T^{22; \uparrow \downarrow}$ and $\Delta_T^{22; \downarrow \uparrow}$ using Eq. (\ref{eq:64}), are

\begin{equation} \label{eq:1651}
\begin{split}
\Delta_T^{22,\uparrow \downarrow} =& G_0\int dE \Bigg( \sum_{l \neq 2}T_{2l}^{\uparrow \downarrow}f_l(E)^2 - \sum_{l, p} \sum_{\rho \rho'} f_l(E) f_p(E)\\&\times (s_{2l}^{\uparrow \rho \dagger} s_{2p}^{\uparrow \rho'} s_{2p}^{\downarrow \rho' \dagger} s_{2 l}^{\downarrow \rho})\Bigg),\\
\Delta_T^{22,\downarrow \uparrow} =& G_0\int dE \Bigg( \sum_{l \neq 2}T_{2l}^{\downarrow \uparrow}f_l(E)^2 - \sum_{l, p} \sum_{\rho \rho'} f_l(E) f_p(E)\\&\times (s_{2l}^{\downarrow \rho \dagger} s_{2p}^{\downarrow \rho'} s_{2p}^{\uparrow \rho' \dagger} s_{2 l}^{\uparrow \rho})\Bigg),
\end{split}
\end{equation}

Using Eq. (\ref{eq:131}), we realize that both $\Delta_T^{22; \uparrow \downarrow}$ and $\Delta_T^{22; \downarrow \uparrow}$ vanish. The expression for $\Delta_T^{22; \downarrow \downarrow}$ is,

\begin{equation} \label{eq:1651}
\begin{split}
\Delta_T^{22,\downarrow \downarrow} =& G_0\int dE \Bigg(\sum_{l \neq 2} \sum_{\rho \neq \downarrow}T_{2l}^{\downarrow \rho}f_2(E)^2  + \sum_{\rho} T_{22}^{\downarrow \rho} f_2(E)^2 \\& + \sum_{l \neq 2}T_{2l}^{\downarrow \downarrow}f_l(E)^2 - \sum_{l, p} \sum_{\rho \rho'} f_l(E) f_p(E) (s_{2l}^{\downarrow \rho \dagger} s_{2p}^{\downarrow \rho'} s_{2p}^{\downarrow \rho' \dagger} s_{2 l}^{\downarrow \rho})\Bigg).
\end{split}
\end{equation}

From Eq. (\ref{eq:131}), we see that $s_{21}^{\downarrow \downarrow}$, $s_{22}^{\downarrow \uparrow}$ and $s_{23}^{\downarrow \downarrow}$ are non-zero and therefore, $\Delta_T^{22; \downarrow \downarrow}$ reduces to,

\begin{equation}
    \begin{split}
        \Delta_T^{22; \downarrow \downarrow} = & G_0 \int dE \bigg(T_{22}^{\downarrow \uparrow }f_2(E)^2 + T_{21}^{\downarrow \downarrow} f_1(E)^2 + T_{23}^{\downarrow \downarrow } f_3(E)^2 \\& -(|s_{21}^{\downarrow \downarrow}|^2 |s_{21}^{\downarrow \downarrow}|^2 f_1(E)^2 + |s_{23}^{\downarrow \downarrow}|^2 |s_{23}^{\downarrow \downarrow}|^2 f_3(E)^2\\&+|s_{22}^{\downarrow \uparrow}|^2 |s_{21}^{\downarrow \uparrow}|^2 f_2(E)^2 + 2|s_{21}^{\downarrow \downarrow}|^2 |s_{23}^{\downarrow \downarrow}|^2 f_1(E) f_3(E) \\&+ 2|s_{21}^{\downarrow \downarrow}|^2 |s_{22}^{\downarrow \uparrow}|^2 f_1(E) f_2(E) +  2|s_{22}^{\downarrow \uparrow}|^2 |s_{23}^{\downarrow \downarrow}|^2 f_2(E) f_3(E))\bigg).
    \end{split}
\end{equation}

Since in setup 2, we consider $f_1(E) = f_3(E) = f_H(E)$ and $f_2(E) = f_C(E)$, we get $\Delta_T^{22; \downarrow \downarrow}$,
\begin{equation}
    \Delta_T^{22;\downarrow \downarrow} = G_0 P(1-P)\int_{-\infty}^{\infty}dE (f_H(E) - f_C(E))^2.
\end{equation}

Therefore, $\Delta_T^{22}$ reduces to,
\begin{equation}\label{eq:133}
    \Delta_T^{22} = \Delta_T^{22; \downarrow \downarrow}  =2G_0 \int_{-\infty}^{\infty} dE \,\,\, P(1-P) (f_H(E) - f_C(E))^2.
\end{equation}
Via a similar procedure, one can also calculate $\Delta_T^{44}$ and one sees that it is exactly same as $\Delta_T^{22}$.
\endgroup

\vspace{2cm}

\bibliography{apssamp}

\end{document}